\documentclass[11pt]{article}
\usepackage{amssymb}
\usepackage[sectionbib]{natbib}
\usepackage{array,epsfig, rotating}
\usepackage{amsmath}
\usepackage{amsfonts}
\usepackage{multirow}
\usepackage{amsthm}
\usepackage{array}
\usepackage{booktabs}
\usepackage{graphicx}
\usepackage{graphics}
\usepackage{epsfig,latexsym, verbatim}
\usepackage{color}
\usepackage{multicol}
\usepackage{threeparttable}
\usepackage{float}
\usepackage{cancel}
\usepackage{url}
\usepackage{rotating}
\usepackage[normalem]{ulem}
\usepackage{xr}
\usepackage{secdot}
\usepackage{subcaption}
\usepackage{algorithm}
\usepackage{algpseudocode}
\usepackage{cancel}
\usepackage{titlesec}
\usepackage{enumitem}
\usepackage{hyperref}

\titleformat{\section}
{\normalfont\fontsize{14}{15}\bfseries}{\thesection}{1em}{}

\definecolor{red}{rgb}{1,0,0}
\definecolor{green}{rgb}{0,0,0}
\definecolor{blue}{rgb}{0,0,1}

\definecolor{gw}{RGB}{10,101,181}

\DeclareMathOperator*{\argmin}{arg\,min}

\def\bs{\boldsymbol}

\floatstyle{ruled}
\newtheorem{theorem}{Theorem}
\newtheorem{lemma}{Lemma}
\newtheorem{corollary}{Corollary}
\newtheorem{assump}{Assumption}
\newtheorem{assumpB}{Assumption}

\theoremstyle{definition}

\def\paragraph#1{\vspace*{0.5em}\noindent\textbf{#1} }

\def\boxit#1{\vbox{\hrule\hbox{\vrule\kern6pt
\vbox{\kern6pt#1\kern6pt}\kern6pt\vrule}\hrule}}

\newcommand{\lsim}{\raisebox{-0.13cm}{~\shortstack{$<$ \\[-0.07cm]
$\sim$}}~}

\begin{document}
\renewcommand{\baselinestretch}{1.2}
\markboth{\hfill{\footnotesize\rm Kunal Das, Shan Yu, Guannan Wang, and Li Wang}\hfill}
{\hfill {\footnotesize\rm Density Estimation over Complex Domains} \hfill}
\renewcommand{\thefootnote}{}
$\ $\par \fontsize{10.95}{14pt plus.8pt minus .6pt}\selectfont
\vspace{0.8pc} 
\centerline{\large\bf Nonparametric Density Estimation for  Data Scattered on Irregular}
\vspace{0.8pc} 
\centerline{\large\bf Spatial Domains: A Likelihood-Based Approach Using}
\vspace{0.8pc}
\centerline{\large\bf Bivariate Penalized Spline Smoothing}
\vspace{.4cm} 
\centerline{ 
Kunal Das$^{a}$, 
Shan Yu$^{b}$,  
Guannan Wang$^{c}$ 
and  
Li Wang$^{d}$
\footnote{\emph{Address for correspondence}: 
Li Wang, Department of Statistics, George Mason University, Fairfax, VA 22030, USA. Email: lwang41@gmu.edu}} \vspace{.4cm} 
\centerline{\it 
$^{a}$Iowa State University, 
$^{b}$University of Virginia,
$^{c}$College of William \& Mary} 
\centerline{\it 
and
$^{d}$George Mason University} \vspace{.55cm}
\fontsize{9}{11.5pt plus.8pt minus .6pt}\selectfont


\begin{quotation}
\noindent {\it Abstract:}
Accurately estimating data density is crucial for making informed decisions and modeling in various fields. This paper presents a novel nonparametric density estimation procedure that utilizes bivariate penalized spline smoothing over triangulation for data scattered over irregular spatial domains. The approach is likelihood-based with a regularization term that addresses the roughness of the logarithm of density based on a second-order differential operator. The proposed method offers greater efficiency and flexibility in estimating density over complex domains and has been theoretically supported by establishing the asymptotic convergence rate under mild natural conditions. Through extensive simulation studies and a real-world application that analyzes motor vehicle theft data from Portland City, Oregon, we demonstrate the advantages of the proposed method over existing techniques detailed in the literature.

\vspace{9pt}
\noindent {\it Key words and phrases:}
Bivariate splines; 
Complex domain; 
Density estimation;
Penalized splines; 
Triangulations.
\end{quotation}

\fontsize{10.95}{14pt plus.8pt minus .6pt}\selectfont
\thispagestyle{empty}
\setcounter{equation}{0}
\section{Introduction}
\label{SEC:intro}

Density estimation for data scattered over a spatial domain is a critical component in analysis and modeling, as it quantifies the distribution of a set of geographic events or observations. When integrated with ancillary data, this information identifies patterns and trends in the underlying process, such as high-density areas, clusters, and hotspots.

\paragraph{}Let ${\bs{x}_1, \ldots, \bs{x}_n}$ denote $n$ independent and identically distributed (i.i.d.) samples from an unknown probability density $f$, defined on a two-dimensional (2D) domain $\Omega \subseteq \mathbb{R}^2$. A fundamental goal in statistics is to estimate $f$ from these samples. In a parametric approach, $f$ is assumed to belong to a known parametric family of distributions $\mathcal{P}_{\bs{\theta}} = \{f({\bs{\theta}}): \bs{\theta} \in \mathrm{\Theta}\}$, characterized by a finite-dimensional parameter $\bs{\theta}$. In this case, the estimate $\bs{\theta}$ serves as a proxy to estimate the density. Maximum likelihood estimation (MLE), with its desirable properties such as consistency and efficiency, can be employed. However, in the absence of known parametric forms, one might consider a naive ML estimator as a sum of delta function spikes at the sample points. This estimator is problematic because it violates the intrinsic constraints of a density estimator: non-negativity ($f \geq 0$) and unit integral ($\int_{\Omega}f = 1$). In addition, this naive estimator becomes impractical when $\Omega$ is continuous. The nonparametric approach to density estimation strikes a balance between these two extremes.

\paragraph{}Traditional nonparametric density estimators, including histogram density estimations \citep{scott2015multivariate}, kernel density estimators (KDE) \citep{wand1994kernel}, and Parzen windows estimators \citep{Parzen}, have been widely used to estimate the underlying distribution of spatial data. These methods are straightforward to implement but can be sensitive to the choice of bin widths and kernels, and they suffer from edge effects near boundaries.  Another popular method, the $k$-nearest neighbor density estimator \citep{Loftsgaarden:Quesenberry:65}, estimates the density by counting the number of $k$-nearest neighbors. However, like KDE and histogram estimators, this method also struggles to estimate densities over complex domains. Since these methods rely on Euclidean distances, they fail to provide efficient and flexible estimations when the shape of spatial domains influences the occurrence of observations. 

\paragraph{}Recent advancements in nonparametric density estimation under shape restrictions \citep{carando2009nonparametric} and log-concave density estimation via maximum likelihood \citep{cule} also face limitations when applied to complex spatial domains. These domains, characterized by non-trivial geometries such as irregular boundaries, interior holes, or sharp concavities, present unique challenges due to the intrinsic heterogeneity of point patterns across subregions, which is driven by the underlying density. The case of motor vehicle thefts in Portland, Oregon, exemplifies the importance of considering such complex spatial domains in density estimation. Figure \ref{fig:realdata} (a) illustrates the spatial distribution of reported motor vehicle thefts in Portland during June 2023. The Willamette River, bisecting the city, acts as a natural barrier significantly influencing the spatial pattern of criminal activities. Notably, the data reveals a higher concentration of reported thefts on the eastern side of the river compared to the western side, indicating the potential heterogeneity of the underlying true unknown data-generating density. Traditional density estimation methods, which rely on Euclidean distances, would incorrectly suggest a continuous high-density area of vehicle thefts across the river, particularly where the eastern and western urban cores are in close proximity. This misrepresentation fails to account for the river's role as a physical and psychological barrier to criminal movement.     


\paragraph{}These challenges require flexible and sophisticated density estimation methods that also increase computational complexity, particularly for real-world scenarios with large datasets and complex models. Penalized likelihoods, first introduced by \cite{goodd1971nonparametric} for univariate density estimation, offer a perspective to address these challenges in the nonparametric context. The main idea is to minimize the negative logarithmic likelihood functional of $f$, with the addition of a roughness penalty term $\mathcal{E}(f) = -n^{-1}\sum_{i = 1}^{n}\log(f(X_i)) + \lambda_n \mathcal{E}(f)$, where the log-likelihood term quantifies the conformity of the estimate to the sample data, while the penalty functional $\mathcal{E}(f)$ (such as $\int_{\Omega}(f^{''})^2$) and the smoothing parameter $\lambda_n$ control the smoothness of the estimate.

\paragraph{}\cite{leonard1978density} extended the work in \cite{goodd1971nonparametric} by addressing the non-negativity and unity constraints through a logistic density transformation. They introduced $f = \exp(g)/\int \exp(g)$ and estimated $g$ by minimizing the constraint-free equation:
\begin{equation}
\label{eq1}
-\frac{1}{n}\sum_{i = 1}^{n}g(X_i) + \log\int \exp(g) + \lambda_n \mathcal{E}(g). 
\end{equation}

\paragraph{}However, $f$ only determines $g$ up to a constant function in the null space of $\mathcal{E}(\cdot)$, which potentially produces nonunique solutions. To resolve this, \cite{Silverman1982} proposed estimating the log density $g = \log f$ instead of $f$, bypassing the non-negativity constraint, and replaced $\log \int \exp(g)$ with $\int \exp(g)$ in \eqref{eq1} to enforce the unity constraint.
 
\paragraph{}Penalized density estimation for multivariate domains remained largely unexplored until the work of \cite{cox1990asymptotic}, which established general asymptotic theories for penalized likelihood estimators. \cite{Gu1993} formulated the density estimation problem in a Reproducing Kernel Hilbert Space (RKHS) and provided a unique solution by imposing a one-dimensional constraint on the constant space of $g$ after enforcing a one-to-one logistic transformation $g = \log(f)$. \cite{Gu1993a} further developed the discussion using smoothing splines. However, these methods are incompatible with density estimation on complex domains with irregular shapes or boundaries, since they are developed within the framework of smooth rectangular domains. 

\paragraph{}Recent advancements in density estimation for complex 2D domains have shown promise, notably the work of \cite{Ferraccioli2021}, which employs finite element methods (FEMs) to discretize a penalized negative log-likelihood functional with a Laplacian differential operator regularization term. While \cite{Ferraccioli2021} established consistency of their estimators in terms of the symmetrized Kullback-Leibler distance, following \cite{Gu1993}, this approach has some limitations. FEM typically utilizes piecewise polynomial functions over a triangulated domain, often lacking smoothness at element boundaries. This discontinuity can be problematic where gradient continuity is crucial. Moreover, FEM's rigid mesh structures may not adapt well to highly irregular or dynamic domains, and computational complexity increases substantially with domain intricacy.

\paragraph{}This article addresses the aforementioned challenges by developing a flexible nonparametric density estimation method for data scattered over a 2D spatial domain with complex and irregular shapes. Our method, based on using bivariate penalized spline over a triangulation \citep[BPST; see][]{lai2013bivariate}, incorporates a differential operator to balance the estimator's smoothness and avoid unbounded likelihoods. In comparison to the FEM method proposed by \cite{Ferraccioli2021}, our approach offers enhanced smoothness and continuity across the domain—a critical feature for applications requiring gradient continuity, such as medical imaging and geographic information systems. The bivariate spline method excels in flexibility and adaptability, accommodating irregular shapes and dynamic domain changes without necessitating remeshing. Moreover, it achieves high computational efficiency while maintaining superior accuracy, thereby reducing the computational overhead associated with traditional methods. 

\paragraph{}Theoretically, we establish the asymptotic convergence rate of our density estimator in terms of $L_2$ and $L_{\infty}$ norms under mild conditions. Compared to the consistency in symmetrized Kullback-Leibler distance established by \cite{Ferraccioli2021}, our convergence in $L_2$ and $L_{\infty}$ norms provides stronger, more comprehensive theoretical guarantees. These results offer tighter bounds on estimation error across the entire domain and are more directly interpretable in practical applications. This theoretical advancement, coupled with the method's practical benefits, represents a significant contribution to density estimation for complex bivariate domains. 

\paragraph{}The remainder of this paper is structured as follows. Section \ref{SEC:estimation} provides an overview of bivariate spline smoothing over triangulation and the proposed nonparametric likelihood density estimator. Section \ref{SEC:results} analyzes the estimator's theoretical properties and convergence rate. The implementation details are presented in Section \ref{SEC:imp}. Section \ref{SEC:simulations} evaluates the performance of the proposed method against competitors through simulation studies. An application for data on motor vehicle theft from Portland, Oregon, is demonstrated in Section \ref{SEC:application}. Finally, Section \ref{SEC:conclusions} discusses potential directions for future research.

\section{Density Estimation via Bivariate Splines over Triangulation} 
\renewcommand{\theequation}{2.\arabic{equation}} \setcounter{equation}{0} 
\label{SEC:estimation}

This section introduces our method for estimating the density function $f$ on a spatial domain $\Omega \subseteq \mathbb{R}^2$ with irregular shapes, including complex boundaries, interior holes, and sharp concavities. We prioritize these domain-shape constraints to develop a methodology sufficiently flexible to handle real-world challenges in spatial data density estimation, where data are seldom distributed over regularly shaped regions with smooth boundaries.

\subsection{Penalized likelihood density estimator} 
\label{ssec:plikelihood}

Let $\{\bs{X}_i\}_{i=1}^n$ denote the locations of $n$ observations scattered throughout the domain $\Omega$, assumed to be generated from the underlying density function $f$. Following \cite{Silverman1982}, we propose estimating the logarithm of the density function, $g = \log(f)$, instead of $f$ itself, to circumvent the non-negativity ($f \geq 0$) constraint of density functions. Specifically, we estimate $g$ by minimizing the penalized negative log-likelihood functional:
\begin{equation}
\label{eq2.1}
\mathcal{L}_{pen}(g) = -\frac{1}{n}\sum_{i=1}^n g(\bs{X}_i) + \int_{\Omega} \exp{(g)} + \lambda_n \mathcal{E}_q(g),
\end{equation}
where the penalty functional $\mathcal{E}_q(g)$, in generic notation for a $d$-dimensional case (i.e., where the domain $\Omega$ and the density $f$ are $d$-dimensional), is defined as: $\mathcal{E}_q(g) = \sum_{|\bs{\alpha}| = q} c_{\bs{\alpha}}\mathcal{E}_{\bs{\alpha}}(g)$ with $c_{\bs{\alpha}} = \frac{q!}{\alpha_1! \alpha_2! \cdots \alpha_d!}$ and $\mathcal{E}_{\bs{\alpha}}(g) = \int_{\Omega} |D^{\bs{\alpha}}g(\cdot)|^2 \,d\Omega$. Here, $\bs{\alpha} = (\alpha_1, \alpha_2, \cdots, \alpha_d)$ denotes a multi-index with $D^{\bs{\alpha}} = D^{\alpha_1} D^{\alpha_2} \cdots D^{\alpha_d}$ and $|\bs{\alpha}| = \alpha_1 + \alpha_2 + \cdots + \alpha_d$. Since we focus on 2D domains, we set $d = 2$ throughout the article and choose $q = 2$ to employ a second-order penalty functional.
 
\paragraph{}Consequently, the penalty functional takes the form: 
\begin{align*}
\mathcal{E}_2(g(\bs{u})) &= \frac{2!}{2!0!}\int_{\Omega}|D_{u_1}^2 g(\bs{u})|^2 \,d\Omega + \frac{2!}{1!1!}\int_{\Omega}|D_{u_1} D_{u_2} g(\bs{u})|^2 \,d\Omega + \frac{2!}{0!2!}\int_{\Omega}|D_{u_2}^2 g(\bs{u})|^2 \,d\bs{u} \\
&= \int_{\Omega}\{|D_{u_1}^2 g(\bs{u})|^2 + 2 |D_{u_1} D_{u_2} g(\bs{u})|^2 + |D_{u_2}^2 g(\bs{u})|^2\}\,d\bs{u}, \hspace{0.4 in} \text{for} \hspace{0.1 in} \bs{u} \in \Omega.
\end{align*}
Equation \eqref{eq2.1} comprises three terms. The first term computes the negative log-likelihood at the data points. The second term, inspired by Theorem 3.1 of \cite{Silverman1982}, ensures that the estimated density integrates to unity, a necessary condition for our density estimation. The third term, a penalized regularization term, serves two essential purposes: (1) preventing an unbounded likelihood and (2) controlling the estimator's roughness. The regularization parameter $\lambda_n$ balances two competing objectives: adapting the estimate to the sample data through the negative log-likelihood and defining the estimator's smoothness via regularization. As $\lambda_n$ increases, the influence of regularization increases, yielding a smoother, more uniform density estimate with fewer local variations.

\subsection{Estimation via Bivariate Penalized Splines over Triangulation (BPST)}
\label{ssec:BPST}

The direct minimization of \eqref{eq2.1} leads to an infinite-dimensional problem without a readily available analytical solution. To address this issue on complex domains, we approximate the functional $\mathcal{L}_{pen}(g)$ and function $g$ using bivariate splines on triangulation. This approach effectively handles data distributed over irregularly shaped domains, as discussed by \cite{lai2013bivariate} and \cite{yu_wang}. This section provides a brief overview of the triangulation method and highlights how bivariate splines on triangulation can be applied to density estimation over complex domains.

\paragraph{}Triangulations are essential tools for numerically solving partial differential equations and analyzing complex two-dimensional geometric designs. Their utility stems from the fact that any polygonal domain with an arbitrary shape in two dimensions can be partitioned into a finite number of triangles. A triangle $T$ can be considered the convex hull of three noncollinear points. A triangulation $\triangle = \{T_1, \cdots, T_N\}$ of a domain $\Omega$ is a collection of $N$ triangles such that $\Omega = \bigcup\limits_{j = 1}^{N}T_j$ and any nonempty intersection between a pair of triangles in $\triangle$ is either a shared edge or a shared vertex.

\paragraph{}The ratio $\pi_T = |T|/\rho_T$ is used to define a shape parameter for a triangle $T$, where $|T|$ denotes the longest edge length of $T$, and $\rho_T$ is the radius of the largest inscribed disk in $T$. If $\pi_T$ is small for all $T$, the triangulation can be considered relatively uniform, implying that all angles in the triangles $T \in \triangle$ are relatively similar. It is desirable to construct a triangulation that is as uniform as possible and to avoid triangles with acute angles, as they can lead to numerical instabilities. To this end, we introduce the concept of the $\beta$-quasi-uniform triangulation family, which requires $|\triangle|/\rho_T \leq \beta$, for all $T \in \triangle$, where $|\triangle|=\max\{|T|: T \in \triangle\}$; $|T|$ is the longest edge of a triangle $T$ and $\rho_T$ is the radius of the largest disk that can be inscribed in $T$. This quasi-uniformity requirement is satisfied if the smallest angles in the triangulation are bounded away from zero by a positive constant \citep{lai_schumaker_2007}. 

\paragraph{}For an integer, $r \geq 0$, denote by $\mathcal{C}^r(\Omega)$ the collection of all functions that are $r$ times continuously differentiable on $\Omega$. Given a triangulation $\triangle$, let $\mathcal{S}^r_m(\triangle) = \{s \in \mathcal{C}^r(\Omega) : s|_T \in \mathbb{P}_m(T), T \in \triangle\}$ be the spline space of degree $m$ and smoothness $r$ on $\triangle$, and $s|_T$ denote the polynomial piece of the spline $s$ when restricted to triangle $T$ with $\mathbb{P}_m$ being the space generated by all polynomials in 2D of degree less than or equal to $m$.

\paragraph{}The idea is to approximate the function $g(\cdot)$ using Bernstein basis polynomials that are defined as piecewise polynomials over a 2D triangulated domain. Compared to the FEM employed by \cite{Ferraccioli2021}, our proposed Bivariate Penalized Splines on Triangulations (BPST) approach offers enhanced flexibility. The spline functions allow greater flexibility in choosing the smoothness degree $r \geq 0$ and the polynomial degree $m$, instead of restricting to linear approximations ($m = 1$) as in the FEM method. 

\paragraph{}To introduce Bernstein basis polynomials, we first discuss barycentric coordinates, which are more useful than Cartesian coordinates for working with polynomials on triangles. Consider a non-degenerate triangle $T$ (with nonzero area) in $\mathbb{R}^2$ with vertices $\bs{u}_1, \bs{u}_2, \bs{u}_3$, where $\bs{u}_i = (u_{i_1}, u_{i_2}) \in \mathbb{R}^2$, for $i = 1, 2, 3$, and the vertices are numbered in counterclockwise order. Then, every point $\bs{u} = (u_1, u_2) \in \mathbb{R}^2$ can be uniquely represented as: $\bs{u} = b_1\bs{u}_1 + b_2\bs{u}_2 + b_3\bs{u}_3$ with $b_1 + b_2 + b_3 = 1$, where $b_1, b_2, b_3$ are the barycentric coordinates of the point $\bs{u}$ relative to the triangle $T$. Furthermore, for each $i = 1, 2, 3$, $b_i$ can be expressed as a linear polynomial in $u_1, u_2$, taking the value 1 at the vertex $\bs{u}_i$ and vanishing at all points on the edge of $T$ opposite to $\bs{u}_i$. Given $b_1, b_2, b_3$, for nonnegative integers $i, j, k$ summing to $m$, we define: $B_{ijk}^m := \frac{m !}{i ! j ! k !}b_1^ib_2^jb_3^k$.

\paragraph{}Given that each $b_i$ is a linear polynomial in $u_1, u_2$, it follows that $B_{ijk}^m(\bs{u}) = B_{ijk}^m(u_1, u_2)$ is a polynomial in $u_1, u_2$ of degree $m$. We refer to these polynomials as the Bernstein basis polynomials of degree $m$ relative to $T$. More explicit details on Bernstein basis polynomials and their intriguing properties can be found in \cite{lai_schumaker_2007}.

\paragraph{}We employ these Bernstein basis polynomials to represent the bivariate splines. Let $\{B_h\}_{h\in\mathcal{H}}$ denote the set of degree-$m$ bivariate Bernstein basis polynomials for the space $\mathcal{S}^r_m(\triangle)$, where $\mathcal{H}$ is the index set of Bernstein basis polynomials with cardinality $|\mathcal{H}| = N(m + 1)(m + 2)/2$. Let $\mathbf{B}$ be the $n \times H$ evaluation matrix of the $H$ Bernstein basis polynomials $B_1, B_2, \ldots, B_H$ evaluated at the $n$ data locations $\bs{x}_1, \bs{x}_2, \ldots, \bs{x}_n$, i.e.,
 \begin{align*}
 \mathbf{B} = \begin{bmatrix}
 B_1(\bs{x}_1) & \cdots & B_H(\bs{x}_1)\\
 B_1(\bs{x}_2) & \cdots & B_H(\bs{x}_2)\\
 \vdots & \ddots & \vdots\\
 B_1(\bs{x}_n) & \cdots & B_H(\bs{x}_n)
 \end{bmatrix}.
 \end{align*}

\paragraph{}Hence, using the evaluation matrix structure of the basis polynomials $\mathbf{B}$ and the spline coefficient vector $\bs{\gamma} = (\gamma_1, \cdots, \gamma_H)$, we can approximate the log density function $g$ at location $\bs{x}$ as $g(\bs{x}) \approx \mathbf{B}(\bs{x})^{\top}\bs{\gamma}$ with $\mathbf{B}(\bs{x}) = \{B_1(\bs{x}), \ldots, B_H(\bs{x})\}^{\top}$. This leads to the discretization of the first term in \eqref{eq2.1} as $-n^{-1} \bs{1}_{n \times 1}^{\top}\mathbf{B}\bs{\gamma}$. To discretize the second term in \eqref{eq2.1}, $\int_{\Omega} \exp{(g)}$, we employ a standard Gaussian quadrature rule with $n_g = 9$ nodes. Let $\bs{\omega} \in \mathbb{R}^{n_g}$ denote the associated vector of quadrature weights. Then, for each triangle $T \in \triangle$, let $\mathbf{B}_{T}$ be the $n_g \times H$ evaluation matrix of the $H$ basis functions at the $n_g$ quadrature nodes in triangle $T$, leading to the discretization of the second term as $\sum_{T \in \triangle} \bs{\omega}^\top \exp{(\mathbf{B}_T \bs{\gamma})}$.

\paragraph{}The regularization or roughness penalty can therefore be discretized as $\bs{\gamma}^\top \mathbf{K} \bs{\gamma}$, where $K$ is the diagonal block penalty matrix that satisfies $\mathcal{E}_q(g) = \mathcal{E}_q(\mathbf{B}\bs{\gamma}) = \bs{\gamma}^{\top} \mathbf{K} \bs{\gamma}$. To satisfy the smoothness condition of the splines fitted over the triangulated domain, we impose linear constraints on the spline coefficients $\bs{\gamma}$ such that $\mathbf{H}\gamma = \bs{0}$. Then, the negative penalized estimating equation \eqref{eq2.1} can be discretized as the following constrained estimating equation:
\begin{equation} 
\label{eq2.2}
\mathcal{L}_{pen}(\bs{\gamma}) = - \bs{1}^{\top}\mathbf{B}\bs{\gamma} + \sum_{T \in \triangle}  \bs{\omega}^{\top}\exp{(\mathbf{B}_{T}\bs{\gamma})} + \lambda_n \bs{\gamma}^{\top}\mathbf{K}\bs{\gamma} \text{~subject~to~} \mathbf{H}\bs{\gamma} = \bs{0}. 
\end{equation}
To remove the linear constraint on the spline coefficients $\bs{\gamma}$, QR decomposition is applied on $\mathbf{H}^{\top}$, resulting $\mathbf{H}^{\top} = \mathbf{Q}\mathbf{R} = (\mathbf{Q}_1  \mathbf{Q}_2) \binom{\mathbf{R}_1}{\mathbf{R}_2}$, where $\mathbf{Q}$ is an orthogonal matrix and $\mathbf{R}$ is an upper triangular matrix. The submatrix $\mathbf{Q}_1$ represents the first $r$ columns of $\mathbf{Q}$, where $r$ is the rank of $\mathbf{H}$ and $\mathbf{R}_2$ is a matrix of zeros. We use the reparameterization $\bs{\gamma} = \mathbf{Q}_2 \bs{\theta}$, for some $\bs{\theta}$. Hence the estimating equation \eqref{eq2.2} reduces to
\begin{equation} 
\label{eq2.3}
\mathcal{L}(\bs{\theta}) = - \bs{1}^{\top}\mathbf{BQ_2}\bs{\theta} + \sum_{T \in \triangle}  \bs{\omega}^{\top}\exp{(\mathbf{B}_T\mathbf{Q}_2\bs{\theta})} + \lambda_n \bs{\theta}^{\top}\mathbf{Q_2^{\top} K Q_2}\bs{\theta}.
\end{equation}

\paragraph{}Another major advantage of this reparameterization is that  the coefficients to estimate from \eqref{eq2.3}, which is $\bs{\theta}$, are of much lower dimension compared to $\bs{\gamma}$, which leads to a much faster computational task.

\section{Theoretical Results}
\label{SEC:results}
\renewcommand{\theequation}{3.\arabic{equation}} 
\setcounter{equation}{0} 

This section investigates the asymptotic convergence rate of the proposed BPST estimator for density. In order to achieve this, useful notations are initially introduced along with some general assumptions. 


\paragraph{}For any $a, b \in \mathbb{R}$, we denote $a \wedge b$ as the minimum of the two and $a \vee b$ as the maximum of the two. Given two sequences of positive numbers, say $a_n$ and $b_n$, if the ratio $a_n/b_n$ is bounded for all $n$, we write $b_n \gtrsim a_n$ or $a_n \lsim b_n$. We denote $a_n \asymp b_n$ if and only if $a_n \lsim b_n$ and $b_n \lsim a_n$. Furthermore, $a_n \prec b_n$ or $b_n \succ a_n$ if $a_n/b_n \rightarrow 0$ as $n \rightarrow \infty$. Given a sequence of random variables $V_n, n \geq 1$, we write $V_n = O_p(b_n)$ if $\lim\limits_{c\to \infty}\limsup\limits_{n \to \infty}P[|V_n| \geq cb_n] = 0$ and $V_n = o_p(b_n)$ if $\lim\limits_{n\to \infty}P[|V_n| \geq cb_n] = 0$ for any positive constant $c$. 

\paragraph{}For any function $h$ on the closure of the domain $\Omega$, the $L_2$ norm is defined as $\|h\|_2^2 = \int_{\bs{x}\in\Omega}h^2(\bs{x})\,d\bs{x}$, while the $L_{\infty}$ or the supremum norm is given by $\|h\|_{\infty} = \sup_{\bs{x}\in \Omega}|h(\bs{x})|$. Further, define $|h|_{k, \infty, \Omega} = \sup_{i + j =k}\|D_{z_1}^iD_{z_2}^j h\|_{\infty, \Omega}$ as the maximum norm of all $k$-th order derivatives of $h$ over $\Omega$, where $z_1$ and $z_2$ denote the directions along the horizontal and vertical axes of $\mathbb{R}^2$, respectively.

\paragraph{}Following the discussions related to the methodology in Section \ref{SEC:estimation}, in the following, we list some generic assumptions for density estimations for the convenience of the readers.

\begin{assump}
\label{assmp1}
 The observations $\{\bs{x}_{i}\}_{i=1}^n$ over the domain $\Omega$ are $n$ i.i.d. points generated from an unknown true underlying density $f_0$.
\end{assump}

\begin{assump}
\label{assmp2}
 The triangulation $\triangle_n = \{T_1, T_2, \cdots, T_{N_n}\}$ represents a $\beta$-quasi triangulation of the domain $\Omega$.
\end{assump}


\begin{assump}
\label{assmp3}
 The logarithm of the true density $g_0 \in W_{\infty}^{m + 1}(\Omega)$, where $W_{\infty}^{m + 1}(\Omega) = \left\{g: \|g\|_{m + 1, \infty, \Omega} < \infty\right\}$ and 
 $\|g\|_{m + 1, \infty, \Omega} = \sum_{k = 0}^{m + 1}|g|_{k, \infty, \Omega} = \sum_{k = 0}^{m + 1}\sup_{i + j = k} \|D_{z_1}^iD_{z_2}^j g\|_{\infty}$.
\end{assump}

\begin{assump}
\label{assmp4}
The true log-density function $g_0$ is bounded from zero and infinity in $\Omega$.
\end{assump}


\begin{assump}
\label{assmp5}
The number of triangles generated in the triangulation process, i.e. $N_n$ and the sample size, $n$, satisfy condition $N_n = Cn^{\eta}$, for some constants $C > 0$ and $\eta < 1$. 
\end{assump}

\paragraph{}Assumptions \ref{assmp1}--\ref{assmp5} establish the framework for a BPST-based density estimation method. Assumption \ref{assmp1} is standard in density estimation, assuming i.i.d. observations. Assumptions \ref{assmp2} and \ref{assmp5} are specific to triangulation methods, defining the approximation structure and fineness, related to the sample size. Assumption \ref{assmp3} imposes smoothness conditions on the log-density via Sobolev spaces, which is more general than common assumptions of specific differentiability orders. Assumption \ref{assmp4} ensures the density is bounded away from zero and infinity, similar to typical boundedness conditions.

\paragraph{}Next, we define empirical and theoretical inner products for functions $s_1(\bs{x})$ and $s_2(\bs{x})$, $\bs{x}\in \Omega$, as:
\[
\langle s_1, s_2\rangle_n = {\rm E}_n[s_1(\bs{X}) s_2(\bs{X})] = \frac{1}{n}\sum_{i=1}^{n}s_1(\bs{X}_i)s_2(\bs{X}_i), ~
\langle s_1, s_2 \rangle = {\rm E}[s_1(\bs{X}) s_2(\bs{X})].
\]
The corresponding empirical and theoretical norms are $\|s\|_{n}^{2} = \langle s, s\rangle_n$ and $\|s\|^2 = \langle s, s \rangle$, respectively. Under Assumption \ref{assmp4}, the theoretical norm $\|.\|$ is equivalent to the usual $L_2$ norm with respect to the Lebesgue measure, $\|.\|_2$. Specifically, there exist constants $C_1$ and $C_2$ such that $C_1\|s\|_2 \leq \|s\| \leq C_2 \|s\|_2$ for all square-integrable functions $s$. 





\subsection{Convergence rates of the estimator}
\label{ssec:convergence}

This section establishes the convergence rate of the BPST estimator to estimate the true unknown density. The asymptotic convergence rates are derived using two main theorems, which essentially provide convergence results for the approximation error and estimation error components as discussed in the following.

\paragraph{}The degree of the spline, $m$, and the order of the penalty functional $\mathcal{E}_q$, $q$, significantly influence the asymptotic convergence rate of the BPST estimator. Both $m$ and $q$ are assumed to be integers. The relationship $0 \leq q \leq m$ is a natural constraint, as the $m$-th derivative of a polynomial spline function of degree $m$ is piecewise constant, while its $(m+1)$-th derivative contains Dirac delta functions, rendering the $(m+1)$-th order penalty functional ill-defined. Additionally, $q > d/2$ is necessary to ensure the desired convergence rate of the penalty functional's eigenvalues, as detailed in Lemma \ref{prop4} in the Appendix. In the theoretical studies below, we set $q = 2$ and $d = 2$.

\paragraph{}Note that the estimating equation, in the form of a negative penalized log-likelihood in \eqref{eq2.1}, can be expressed as:
\begin{align*}
\mathcal{L}_{pen}(g) &= -\frac{1}{n}\sum_{i=1}^n g(\bs{X}_i) + \int_{\Omega} \exp{(g)} + \lambda_n \mathcal{E}_q(g)= L(g) + \lambda_n\mathcal{E}_q(g),
\end{align*}
where $L(g) =  -n^{-1}\sum_{i=1}^n g(\bs{X}_i) + \int_{\Omega} \exp{(g)}$. This formulation can be viewed as a convex extended linear model, analogous to the concave extended linear models discussed by \cite{huang2021asymptotic}, as the functional $L(g)$ is convex over the domain $\Omega$. For a proof of the convexity of $L(\cdot)$, refer to Section B of the Supplementary Material in \cite{Ferraccioli2021}. 

\paragraph{}The expected value of the penalized log-likelihood, $\mathcal{L}_{pen}(g)$, is given by:
\[
\Lambda_{pen}(g) = {\rm E}(L(g)) + \lambda_n\mathcal{E}_q(g) = \Lambda(g) + \lambda_n\mathcal{E}_q(g).
\]

\begin{assumpB}\label{cond1}
Whenever $\|h\|_{\infty} \leq C$ for a constant $C>0$, there exist constants $C_1, C_2 > 0$, such that $C_1\|h\|^2 \leq \Lambda(g_0 + h) - \Lambda(g_0) \leq C_2 \|h\|^2.$
\end{assumpB}

\paragraph{}This assumption ensures that the expected log-likelihood, $\Lambda(\cdot)$, behaves like a quadratic functional around its minimal point $g_0$. 

\paragraph{}Let $\hat{g}_n$ and $\tilde{g}_n$ be the minimizers of the penalized estimation equations $\mathcal{L}_{pen}(g)$ and $\Lambda_{pen}(g)$, respectively, that is, 
\begin{equation}
\hat{g}_n = \argmin\limits_{g \in \mathcal{S}^r_m(\triangle_n)}\mathcal{L}_{pen}(g), ~~ \tilde{g}_n = \argmin\limits_{g \in \mathcal{S}^r_m(\triangle_n)}\Lambda_{pen}(g).
\end{equation}
For the logarithm of the true underlying density $g_0$, we consider the following decomposition of the estimation error:
\[
\hat{g}_n - g_0 = \hat{g}_n - \tilde{g}_n + \tilde{g}_n - g_0,
\]
where $\hat{g}_n - \tilde{g}_n$ and $\tilde{g}_n - g_0$ represent the estimation and approximation errors, respectively. To achieve asymptotic convergence results for the BPST estimator, it is essential to determine appropriate bounds for these two types of error in suitable norms.

\paragraph{}A measure of the complexity of the bivariate spline space is defined as:
\[
U_n = \sup\limits_{s \in \mathcal{S}^r_m(\triangle_n), \|s\|_2 \neq 0}\left\{{\|s\|_{\infty}}/{\|s\|_{2}}\right\},
\]
which is a significant quantity for the asymptotic analysis. Under the assumption of $\beta$-quasi triangulation of the domain $\Omega$, $U_n \asymp |\triangle_n|^{-1}$ (see Lemma 1, p. 250, \cite{huang1998projection}). 



\paragraph{}The following theorem establishes the existence and asymptotic properties of the minimizer $\tilde{g}_n$, providing convergence rates for its estimation error under specific assumptions.

\begin{theorem}\label{theo1}
Suppose that Assumption \ref{assmp1}--\ref{assmp5} and Assumption \ref{cond1} hold. If $\lim_n |\triangle_n| \vee \lambda_n = 0$ and 
$\lim_n U_n^2\{|\triangle_n|^{2(m + 1)} \vee (\lambda_n|\triangle_n|^{\{2(m + 1 - q)\}})\} = 0$,  then $\tilde{g}_n$ exists and $\|\tilde{g}_n\|_{\infty} = O_p(1)$. Also, $\|\tilde{g}_n - g_0\|_{\infty} = o_p(1)$ and 
\[
\|\tilde{g}_n - g_0\|^2 + \lambda_n \mathcal{E}_q(\tilde{g}_n) = O_p(|\triangle_n|^{2(m + 1)} \vee (\lambda_n|\triangle_n|^{\{2(m + 1 - q) \wedge 0\}})).
\]
\end{theorem}



\paragraph{}Note that $L(\tilde{g}_n + \alpha s)$ is a convex function of $\alpha$; thus, it admits left and right derivatives and is differentiable at all but countably many points. The directional derivative of $L(\tilde{g}_n + \alpha s)$ at $\tilde{g}_n$, along the direction of $s$, can be denoted as
\[
L^{\prime}(\tilde{g}_n; s) = \frac{\,d}{\,d\alpha}L(\tilde{g}_n + \alpha s)\bigg\rvert_{\alpha = 0^{+}}.
\]
Assuming mild exchangeability between expectation and differentiation, we have ${\rm E}[L^{\prime}(\tilde{g}_n; s)] = \frac{\,d}{\,d\alpha}{\rm E}[L(\tilde{g}_n + \alpha s)]\bigg\rvert_{\alpha = 0^{+}}$. As a quadratic functional, $\mathcal{E}_q(\cdot)$ takes the form $\mathcal{E}_q(\tilde{g}_n + \alpha s) = \mathcal{E}_q(\tilde{g}_n) + 2\alpha \mathcal{E}_q(\tilde{g}_n, s) + \alpha^2 \mathcal{E}_q(s)$. Since $\tilde{g}_n$ minimizes the convex functional $\Lambda_{pen}(\cdot)$ over $\mathcal{S}^r_m(\triangle_n)$, by the first derivative condition, 
\[
\frac{\,d}{\,d \alpha}\Lambda_{pen}(\tilde{g}_n + \alpha s)\bigg\rvert_{\alpha = 0^{+}} = \frac{\,d}{\,d \alpha}\Lambda(\tilde{g}_n + \alpha s)\bigg\rvert_{\alpha = 0^{+}} + 2 \lambda_n \mathcal{E}_q(\tilde{g}_n, s) = 0, ~ s \in \mathcal{S}^r_m(\triangle_n).
\]
Consequently, for any $s \in \mathcal{S}^r_m(\triangle_n)$,
\begin{align*}
\frac{\,d}{\,d\alpha}L(\tilde{g}_n + \alpha s)\bigg\rvert_{\alpha = 0^{+}} + 2 \lambda_n \mathcal{E}_q(\tilde{g}_n, s) &= \frac{\,d}{\,d\alpha}L(\tilde{g}_n + \alpha s)\bigg\rvert_{\alpha = 0^{+}} - \frac{\,d}{\,d\alpha}\Lambda(\tilde{g}_n + \alpha s)\bigg\rvert_{\alpha = 0^{+}}\\
&= ({\rm E}_n - {\rm E})L^{\prime}(\tilde{g}_n; s).
\end{align*}

\begin{assumpB}\label{cond2}
The difference between ${\rm E}_nL^{\prime}(\tilde{g}_n; s)$ and ${\rm E}L^{\prime}(\tilde{g}_n; s)$ satisfies that
\[
\sup\limits_{s \in \mathcal{S}_m^q(\triangle_n)} \frac{|({\rm E}_n - {\rm E})L^{\prime}(\tilde{g}_n; s)|^2}{\|s\|^2 + \lambda_n \mathcal{E}_q(s)} = O_p\left(\frac{1}{n \lambda_n^{d/(2q)}} \wedge \frac{1}{n |\triangle_n|^d}\right).
\]
\end{assumpB}


\begin{assumpB}\label{cond3}
There exist constants $C_1, C_2 > 0$, such that, for all $s \in \mathcal{S}^r_m(\triangle_n)$ with $\|s\|_{\infty} \leq C_1$, 
\[
\frac{\,d}{\,d\alpha}L(\tilde{g}_n + \alpha s)\bigg\rvert_{\alpha = 1^{+}} - \frac{\,d}{\,d\alpha}L(\tilde{g}_n + \alpha s)\bigg\rvert_{\alpha = 0^{+}} \geq C_2 \|s\|^2,
\]
with probability tending to one as $n \xrightarrow{}\infty$.
\end{assumpB}

\paragraph{}In Appendix \ref{App4}, we verify that the BPST density estimator satisfies Assumptions \ref{cond1}, \ref{cond2}, and \ref{cond3}.

\paragraph{}Theorem \ref{theo2} below characterizes the asymptotic behavior of the difference between the empirical and theoretical minimizers $\hat{g}_n$ and $\tilde{g}_n$, providing convergence rates for the estimation error and the difference in their energy functionals.

\begin{theorem}\label{theo2}
Under Assumptions \ref{cond2}--\ref{cond3} and the fundamental Assumptions \ref{assmp1}--\ref{assmp5}, if $\lim_n |\triangle_n| \vee \lambda_n = 0$ and $\lim\limits_{n}U_n^2\left(\frac{1}{n\lambda_n^{d/2q}} \wedge \frac{1}{n|\triangle_n|^d}\right) = 0$, then $\|\hat{g}_n - \tilde{g}_n\|_{\infty} = o_p(1)$ and
\[
\|\hat{g}_n - \tilde{g}_n\|^2 + \lambda_n\mathcal{E}_q(\hat{g}_n - \tilde{g}_n) = O_p\left(\frac{1}{n\lambda_n^{d/2q}} \wedge \frac{1}{n|\triangle_n|^d}\right).
\]
\end{theorem}

\paragraph{}Combining the results of Theorem \ref{theo1} and Theorem \ref{theo2}, the final result on the convergence rate for $\|\hat{g}_n - g_0\|^2$ can be obtained as described in the following corollary.

\begin{corollary}\label{corr1}
Suppose, Assumptions \ref{cond1}, \ref{cond2} and \ref{cond3} hold. If $\lim_n |\triangle_n| \vee \lambda_n = 0$ along with 
\begin{align}
	\label{eq3.2}
	\lim_n U_n^2\left\{|\triangle_n|^{2(m + 1)} \vee (\lambda_n|\triangle_n|^{\{2(m + 1 - q) \wedge 0\}}) + \frac{1}{n\lambda_n^{d/2q}} \wedge \frac{1}{n|\triangle_n|^d}\right\} = 0,	
\end{align}
also hold, then,
$\|\hat{g}_n - g_0\|_{\infty} = o_p(1)$
and,
\begin{align}
	\label{eq3.3}
	\|\hat{g}_n - g_0\|^2 + \lambda_n \mathcal{E}_q(\hat{g}_n) = O_p\left(|\triangle_n|^{2(m + 1)} \vee (\lambda_n|\triangle_n|^{\{2(m + 1 - q) \wedge 0\}}) + \frac{1}{n\lambda_n^{d/2q}} \wedge \frac{1}{n|\triangle_n|^d}\right).
\end{align} 
\end{corollary}

\paragraph{}A significant point to note is that the result in Corollary \ref{corr1} not only provides a bound for the $L_2$ norm $\|\hat{g}_n - g_0\|^2$ but can also bound the penalty functional $\mathcal{E}_q(\hat{g}_n)$ and therefore can be interpreted as a stronger result than the results regarding bound for $\|\hat{g}_n - g_0\|^2$ solely.

\paragraph{}Note that behavior of the asymptotic result for penalized BPST estimator in \eqref{eq3.3} depend on the interplays among the smoothness of the unknown function$(m + 1)$, degree of the bivariate splines$(m)$, order of the penalty functional$(q)$, penalty parameter$(\lambda_n)$ and the triangulation parameter$(|\Delta_n|)$, making it hard to interpret. An attempt to simplify things needs to be considered. By structure, we are in a setting where $q \leq m < m + 1.$ And, under this setup, the expression $\left(|\triangle_n|^{2(m + 1)} \vee (\lambda_n|\triangle_n|^{\{2(m + 1 - q) \wedge 0\}}) + \frac{1}{n\lambda_n^{d/2q}} \wedge \frac{1}{n|\triangle_n|^d}\right)$, in both \eqref{eq3.2} and \eqref{eq3.3} reduces to $\left(|\Delta_n|^{2(m+1)} \vee \lambda_n + \frac{1}{n\lambda_n^{1/q}} \wedge \frac{1}{n|\Delta_n|^d}\right)$, making the asymptotic behavior of the penalized BPST estimator to depend only on $\lambda_n$ and $|\Delta_n|$ and their corresponding exponents. To further investigate the asymptotic behavior corresponding to different choices of values for these two quantities, let us consider the following scenarios:

\paragraph{Case 1:} $\lambda_n \lsim |\Delta_n|^{2(m+1)}.$

\paragraph{}In this scenario, the LHS of \eqref{eq3.2} translates to a more simplified expression as follows,
\begin{align*}
		& \lim_n U_n^2\left\{|\triangle_n|^{2(m + 1)} \vee (\lambda_n|\triangle_n|^{\{2(m + 1 - q) \wedge 0\}}) + \frac{1}{n\lambda_n^{d/2q}} \wedge \frac{1}{n|\triangle_n|^d}\right\}\\
		&= \lim_n U_n^2 \left(|\Delta_n|^{2(m+1)} \vee \lambda_n + \frac{1}{n\lambda_n^{1/q}} \wedge \frac{1}{n|\Delta_n|^d}\right)\\
		&= \lim_n |\Delta_n|^{-2} \left(|\Delta_n|^{2(m+1)}+  \frac{1}{n|\Delta_n|^d}\right), \hspace{0.2 in} \text{as} \hspace{0.1 in} U_n \asymp |\Delta_n|^{-1}\\
		&= \lim_n\left(|\Delta_n|^{2m} + \frac{1}{n|\Delta_n|^{d+2}}\right),
\end{align*}
and, the condition in \eqref{eq3.2} simplifies to $n|\Delta_n|^{d+2} \rightarrow \infty.$ Note that using a small $\lambda_n$ with $\lambda_n \lsim |\Delta_n|^{2(m+1)}$ indicates a light penalization, and the behavior of the penalized BPST estimator in these situations is quite similar to that of the unpenalized polynomial spline estimators in \cite{huang2003asymptotics}. Furthermore, specifically, if we choose the tuning parameter $|\Delta_n|$ such that $|\Delta_n| \asymp n^{-1/(2(m+1)+d)}$, then the asymptotic rate of convergence in the RHS of \eqref{eq3.3} boils down to $O_p(n^{-(m+1)/(m+2)})$, which happens to be the Stone's optimal rate of convergence as described in \cite{stone1982optimal}. 

\paragraph{Case 2:} $|\Delta_n|^{2(m+1)} \lsim \lambda_n \lsim |\Delta_n|^{2q}.$

\paragraph{}The LHS of \eqref{eq3.2} boils down to, $\lim_n U_n^2 \left(\lambda_n + \frac{1}{n|\Delta_n|^d}\right)$ and condition in \eqref{eq3.2} holds if $ \lambda_n |\Delta_n|^{-2} \rightarrow 0$ and, $n|\Delta_n|^{-(d+2)} \rightarrow \infty$. And, Stone's optimal rate of convergence($O_p(n^{-(m+1)/(m+2)})$) can be achieved in the RHS of \eqref{eq3.3} if we choose both the tuning parameters $\lambda_n$ and $|\Delta_n|$ such that $\lambda_n \asymp |\Delta_n|^{2(m+1)}$ and $|\Delta_n| \asymp n^{-1/(2(m+1) + d)}.$ 

\paragraph{Case 3:} $\lambda_n \gtrsim |\Delta_n|^{2q}.$

\paragraph{}Here the LHS of \eqref{eq3.2} takes the form $\lim_n U_n^2\left(\lambda_n + \frac{1}{n\lambda_n^{1/q}}\right)$ and the condition in \eqref{eq3.2} holds if $\lambda_n |\Delta_n|^{-2} \rightarrow 0$ and $n\lambda_n^{1/q}|\Delta_n|^2 \rightarrow \infty$. Note that this scenario can be referred to as a case of high penalization, and the best rate of convergence in the RHS of \eqref{eq3.3}is found to be $O_p(n^{-2q/(2q+d)})$ when $\lambda_n \asymp n^{-2q/(2q+d)}$, and this convergence rate is slower than the Stone's optimal rate of convergence $O_p(n^{-(m+1)/(m+2)})$.

\section{Algorithm and implementation details}
\label{SEC:imp}
\renewcommand{\theequation}{4.\arabic{equation}} 
\setcounter{equation}{0} 

\subsection{Estimation Algorithm} 
\label{ssec:algorithms}

The minimization of the objective function in Equation \eqref{eq2.3} with respect to $\bs{\theta}$ can be achieved using either gradient descent or quasi-Newton methods. Given the strict convexity of the function in \eqref{eq2.3}, convergence is guaranteed for both algorithms. For our numerical studies in simulations and real data analyses, we employ the \texttt{nlm()} function in \texttt{R}, which implements a Newton-type algorithm for nonlinear minimization. This function prioritizes a line search algorithm (detailed in Algorithm \ref{algo1}) over the Newton-Raphson method. The line search approach seeks a local minimum of a multidimensional nonlinear function using its gradients, determining both the search direction and an appropriate step length that satisfies standard conditions. This method is widely applied in solving unconstrained optimization problems across various fields, including machine learning and game theory.

\begin{algorithm}[!ht]
\caption{Line search algorithm}\label{algo1}
\begin{itemize}
 \item Pick the starting point $\bs{\theta}_0$.
 \item Repeat the steps until $\mathcal{L}(\bs{\theta}_k) = \mathcal{L}_k$ converges to a local minimum. 
 \begin{enumerate}
\item Choose $\bs{p}_k$, a descent direction starting at $\bs{\theta}_k$, such that,\newline $\nabla \mathcal{L}_k^{\top}\bs{p}_k < 0$ for $\nabla \mathcal{L}_k \neq 0$
\item Find $\alpha_k > 0$, the step length, such that $\mathcal{L}_k(\bs{\theta}_k + \alpha_k\bs{p}_k) < \mathcal{L}_k(\bs{\theta}_k)$
\item Fix $\bs{\theta}_{k+1} = \bs{\theta}_{k} + \alpha_k\bs{p}_k$.
 \end{enumerate}
\end{itemize}    
\end{algorithm}



\paragraph{}Initiating the optimization algorithm described in Equation \eqref{eq2.3} requires a meaningful initial estimator for the density $f$. A well-chosen initial value can potentially accelerate the algorithm's convergence, requiring fewer iterations compared to an arbitrarily selected initial value, such as a constant vector. Moreover, this initial density estimate also sets the preliminary choices for $g$ and $\bs{\theta}$.

\paragraph{}In one-dimensional settings, the initial step for estimating the density typically involves building a histogram of the data distribution. Extending this approach to our 2D scenario, we have developed a histogram-like density estimator. Following the notation from Section \ref{SEC:estimation}, let $\nu_k$ represent the number of points within triangle $T_k$ (i.e., $\nu_k = \sum_{i = 1}^n I(\bs{X}_i \in T_k)$). Then, for any point $\bs{u} = (u_1, u_2) \in T_k$, we define the initial estimated density at $\bs{u}$ as $\hat{f}_{\mathrm{initial}}(\bs{u}) = {\nu_k}/(n A_k)$, where $A_k$ is the area of triangle $T_k$. This approach essentially provides the coarsest possible distribution of the data across the domain. Thus, the initial density estimator at any point $\bs{u} \in \Omega$ is given by
\[
\hat{f}_{\mathrm{initial}}(\bs{u}) = \sum_{j = 1}^{N_n}I(\bs{u} \in T_j)\frac{\nu_j}{n A_j}.
\]

\paragraph{}While this coarse initial estimator suffices to begin optimization, it requires refinement for low sample size (LSS) data. In such cases, many triangles may contain few or no data points, resulting in inaccurate zero or near-zero density estimates. To address this issue, we propose a modified initial estimator that incorporates data from neighboring triangles. Let $\mathbb{V} = \{V_1, V_2, \cdots, V_{N_n}\}$ be the collection of vertex sets for the $N$ triangles in the triangulation $\triangle_n$. The vertices of the $k$th triangle, $T_k$ (for $k = 1, \cdots, N$), are denoted by the set $V_k = \{\bs{u}_{k1}, \bs{u}_{k2}, \bs{u}_{k3}\}$, with $\bs{u}_{kl} \in \mathbb{R}^2(l  = 1, 2, 3)$. The neighborhood of triangle $T_k$, denoted $N_k$, is defined as $N_k = \{\cup_{j} T_j : \cup_{l = 1}^3 \bs{u}_{jl} \in \{\bs{u}_{k1}, \bs{u}_{k2}, \bs{u}_{k3}\}$, $j = 1, 2, \cdots, N\}$, which is the polygon formed by merging $T_k$ with all triangles that share a vertex or an edge with it. In addition, let $\mathcal{N}_k = \sum_{i = 1}^n I(\bs{X}_i \in N_k)$ represent the total number of data points within region $N_k$, and let $\mathcal{A}_k(=\sum_{j \in N_k}A_j)$ be the area of $N_k$. The adjusted initial density estimator for an LSS scenario at any point $\bs{u} \subseteq \mathbb{R}^2$ is then given by:
\[
\hat{f}_{\mathrm{LSS}}(\bs{u}) = \sum_{j = 1}^{N_n}I(\bs{u} \in T_j)\frac{\mathcal{N}_j}{n \mathcal{A}_j}.
\]
Given the initial estimate of $f$, the initial estimate of $g$ can be obtained using a logarithmic transformation, and the initial guess for $\bs{\theta}$ can be obtained considering penalized smoothing of the initial density using the \texttt{gam()} function from \texttt{R} package \texttt{mgcv}.

\paragraph{}In Algorithm \ref{algo1}, the step length $\alpha_k$ is crucial. Its selection requires balancing two factors: substantially reducing the function value in the subsequent iteration and efficiently determining the step length. An excessively large step length may lead to overshooting, while a very small one may result in slow convergence. To address this, either exact or inexact search procedures can be employed. The steepest descent method exemplifies an exact search, utilizing the negative gradient $-\nabla\mathcal{L}_k$ as an efficient search direction for minimizing the objective function. The step length $\alpha_k$ is obtained by minimizing a single-variable objective function in each iteration until convergence:
$\alpha_k = \argmin\limits_{\alpha}\mathcal{L}(\bs{\theta}_k - \alpha \nabla\mathcal{L}(\bs{\theta}_k))$.

\subsection{Penalty parameter selection}
\label{ssec:penalty}

The choice of the smoothing parameter $\lambda_n$ is crucial in balancing the degree of smoothness of the estimator and the adaptability of the estimator to the data. For this purpose, we have chosen the smoothing parameter using a $k$-fold cross-validation (CV) approach, which assesses the $L_2$ norm distance between the true and the estimated density, i.e.,
\begin{equation}
\label{eq4.1}
\int_{\Omega}(\hat{f} - f)^2 = \int_{\Omega}\hat{f}^2 - 2\int_{\Omega}\hat{f}f + \int_{\Omega}f^2,
\end{equation}
where the third term on the right-hand side (RHS) of \eqref{eq4.1}, $\int_{\Omega} f^2 $, is independent of $ \hat {f}$ and is thus excluded from the calculation of $\lambda_n$. 

\paragraph{}Following \cite{sain1994cross}, we set up the CV with $k = 10$ folds, denoting the $k$-th fold of the data as $\bs{x}^{[k]}, k = 1, \cdots, 10$, which contains $|\bs{x}^{[k]}|$ observations. The first term on the RHS in \eqref{eq4.1} is computed based on the training dataset $\bs{x}^{[-k]} = \{\bs{x}^{[j]}; j = 1, \cdots, 10; j \neq k\}$ as $\int_{\Omega}(\hat{f}^{[-k]}_{\lambda_n}(\bs{x}))^2$. The second term in \eqref{eq4.1}, $-2\int_{\Omega}\hat{f}f$, represents the expectation of $\hat{f}$ with respect to the true density $f$, and is empirically computed using the estimator $\hat{f}_{\lambda_n}^{[-k]}$ from the training set evaluated at the test data points $\bs{x}^{[k]}$. By aggregating the results from all folds, we determine the optimal value of $\lambda_n$ by minimizing the cumulative CV error:
\begin{equation}\label{eq4.2}
\mathrm{E}_{\mathrm{CV}}(\lambda_n)  = \frac{1}{10} \sum_{k = 1}^{10}\left[\int_{\Omega}(\hat{f}_{\lambda_n}^{[-k]}(\bs{x}))^2 - \frac{2}{|\bs{x}^{[k]}|}\sum_{\bs{u} \in \bs{x}^{[k]}}\hat{f}_{\lambda_n}^{[-k]}(\bs{u})\right].
\end{equation}

\subsection{Smoothness, degree and triangulation selection}
\label{ssec:parameter-selction}

The implementation and execution of the proposed procedure in both simulation studies and real data analysis primarily utilize the \texttt{R} packages \texttt{Triangulation} and \texttt{BPST} \citep{BPST}. For optimization, mainly the \texttt{nlm()} function from package \texttt{stats} is used as described in Section \ref{ssec:algorithms}. In most cases, we recommend setting the smoothness parameter for the spline space at $r = 1$, and correspondingly, the degree of polynomials at $m \geq 3r + 2 (=5)$, to fulfill the theoretical requirements for the full approximation power of the spline space [Chapter 10, \cite{lai_schumaker_2007}]. However, satisfactory performance has been observed even with $m = 3$ in both simulation studies and real data analysis.

\paragraph{}The determination of an optimal triangulation is largely dependent on the shape and size of the domain. Complex domains with sharp concavities require finer triangulation compared to more regular-shaped domains, such as rectangles or squares. Ideally, the triangles should be as uniform as possible across the domain to satisfy the assumption of $\beta$-quasi-uniform triangulation for theoretical convergence. An optimal triangulation criterion also takes into account the shape, size, and number of triangles generated. A 'good' triangulation, in terms of shape and size, involves triangles that are neither too acute nor too obtuse, thus minimizing the risk of numerical instabilities. According to \cite{lai_schumaker_2007} and \cite{lindgren2011explicit}, the optimal triangulation for a given number of triangles can be identified using the max-min criterion, which maximizes the minimum angle among all angles of the triangle. Our analyses of both numerical and real-data applications indicate that a minimum number of triangles is essential to capture the features of the rapidly changing underlying density functions. However, further refinement of this triangulation has little to almost no effect on lowering the MISE or CV errors for estimation. Consequently, to conserve space, we have opted to omit the extensive comparison tables for varying triangle sizes and numbers.  

\section{Simulations}
\label{SEC:simulations}
\renewcommand{\theequation}{5.\arabic{equation}} 
\setcounter{equation}{0} 

This section presents three simulation studies considering different settings based on the complexity of domains and densities. The simulations are performed under the following scenarios: nontrivial density on a regular domain (Simulation 1, Section \ref{ssec:sim1}); simple density on a complex domain (Simulation 2, Section \ref{ssec:sim2}); and complex density on a complex domain (Simulation 3, Section \ref{ssec:sim3}). These scenarios cover various complicated situations that might arise in real-life density estimation applications.

\paragraph{}In each scenario, we evaluate the performance of the BPST method using the Mean Integrated Squared Error (MISE), calculated as MISE = $\int_{\Omega}(\hat{f} - f)^2$, computed by an approximation over a regular fine grid covering the entire domain. The performance of BPST is compared with several existing methods, including kernel density estimation (KDE), (partial) differential equation (DE-PDE), or FEM, which employs finite element basis functions for partial differential regularization \citep{Ferraccioli2021}, and the HEAT estimator \citep{chaudhuri1999sizer}, which serves as the initial estimate in the DE-PDE method.

\paragraph{}For the first simulation study, we generate $100$ samples of $200$ observations each from the corresponding true probability density function, and for the second and third one, $100$ samples of size $600$ each are considered to resemble the situation in real data application. The KDE, DE-PDE, and HEAT estimates are obtained using the \texttt{ks} package, which employs anisotropic Gaussian kernels with a bandwidth matrix determined by $k$-fold CV  \citep{chacon2018multivariate} and the \texttt{fdaPDE} package \citep{fdaPDE} in \texttt{R}. The average MISEs of each method for the 100 replications are then compared through boxplots to obtain a quantitative summary of the fits. The white asterisk mark ($\ast$) within each boxplot denotes the average of the MISE values from 100 replications for each method considered.


\subsection{Simulation study 1}
\label{ssec:sim1}
The aim of this simulation study is to evaluate the adaptability of our proposed method on data characterized by multiple modes with varying directions and magnitudes of spread, as well as on scenarios where the original probability density function changes abruptly with varying degrees of anisotropy across the domain. The simulated density is a mixture of four Gaussian distributions, properly normalized to integrate to 1, on the domain $\Omega = [-6,6] \times [-6, 6]$ with the following means, variances, and mixing coefficients:
\begin{itemize}
    \item Means: \(\mu_1 = (-2, -1.5)'\), \(\mu_2 = (2, -2)'\), \(\mu_3 = (-2, 1.5)'\), \(\mu_4 = (2, 2)'\);
    \item Variances: \(\Sigma_1 = \begin{bmatrix} 0.8 & -0.5 \\ -0.5 & 1 \end{bmatrix}\), \(\Sigma_2 = \begin{bmatrix} 1.5 & 0 \\ 0 & 1.5 \end{bmatrix}\), \(\Sigma_3 = \begin{bmatrix} 0.6 & 0 \\ 0 & 0.6 \end{bmatrix}\), \(\Sigma_4 = \begin{bmatrix} 1 & 0.9 \\ 0.9 & 1 \end{bmatrix}\);
    \item Mixture weights: \(\pi = \left(1,1,1,1\right)/4\).
\end{itemize}

\begin{figure}[!ht]
\centering
\begin{subfigure}[c]{.27\linewidth}
\centering
\includegraphics[height = 1.6in, width = 1.45in]{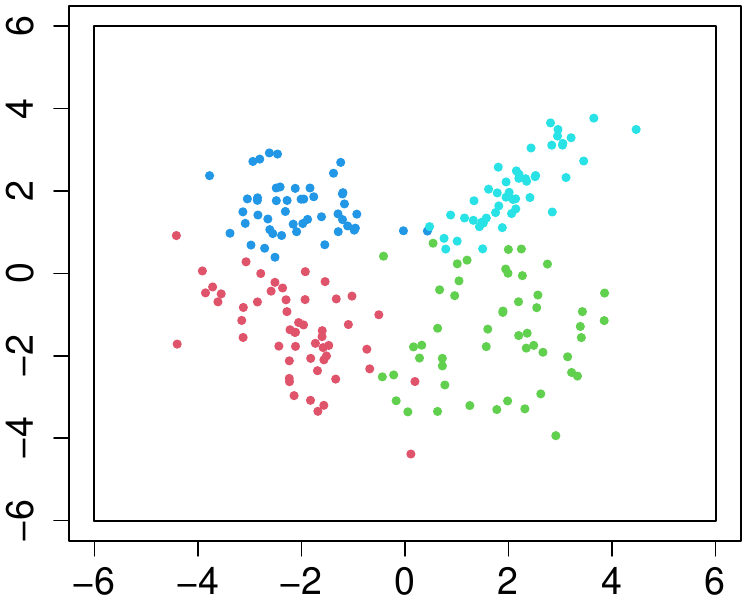}
\caption{Sample}
\end{subfigure}
\hfill
\begin{subfigure}[c]{.3\linewidth}
\centering
\includegraphics[height = 1.6in, width = \linewidth]{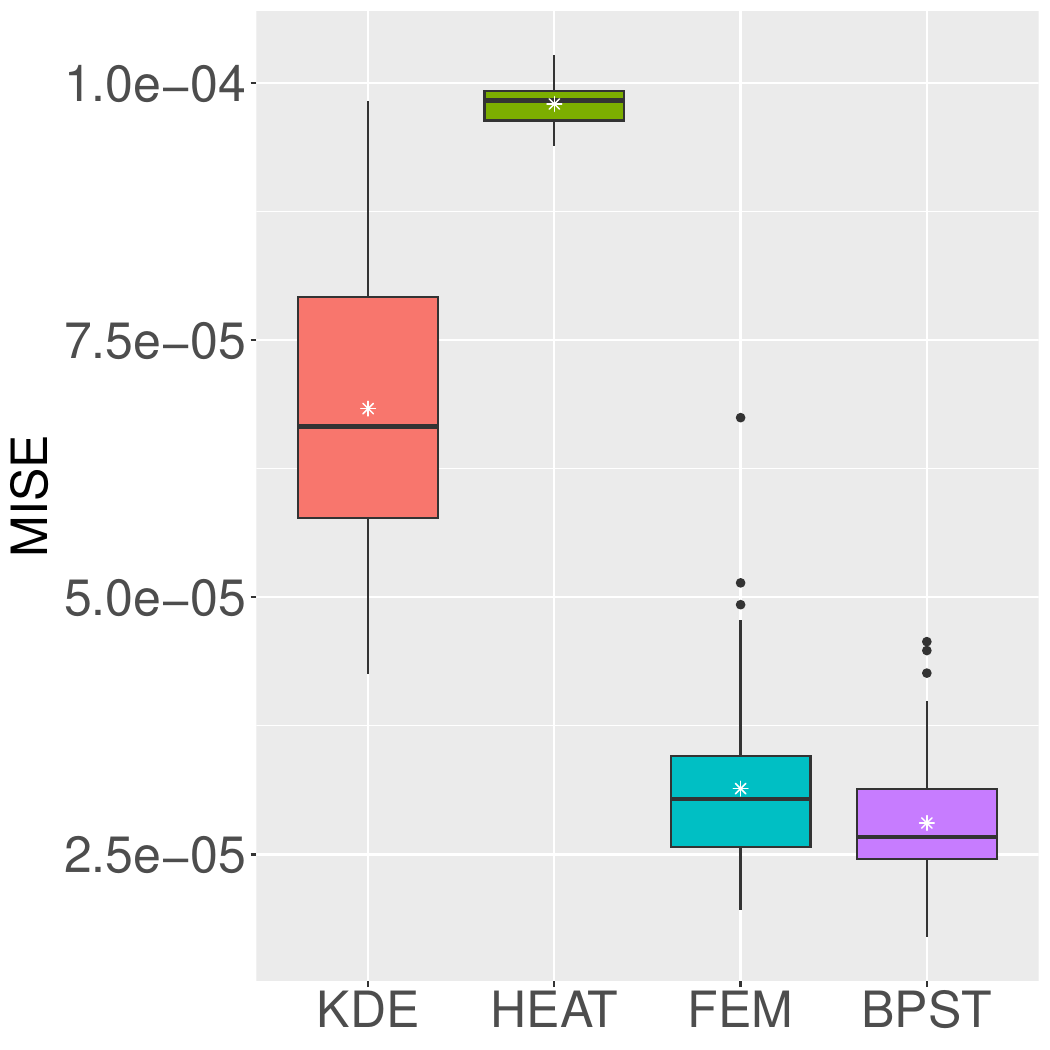}
\caption{MISEs}
\end{subfigure}
\hfill
\begin{subfigure}[c]{0.27\linewidth}
\centering
\includegraphics[height = 1.6in, width = 1.65in]{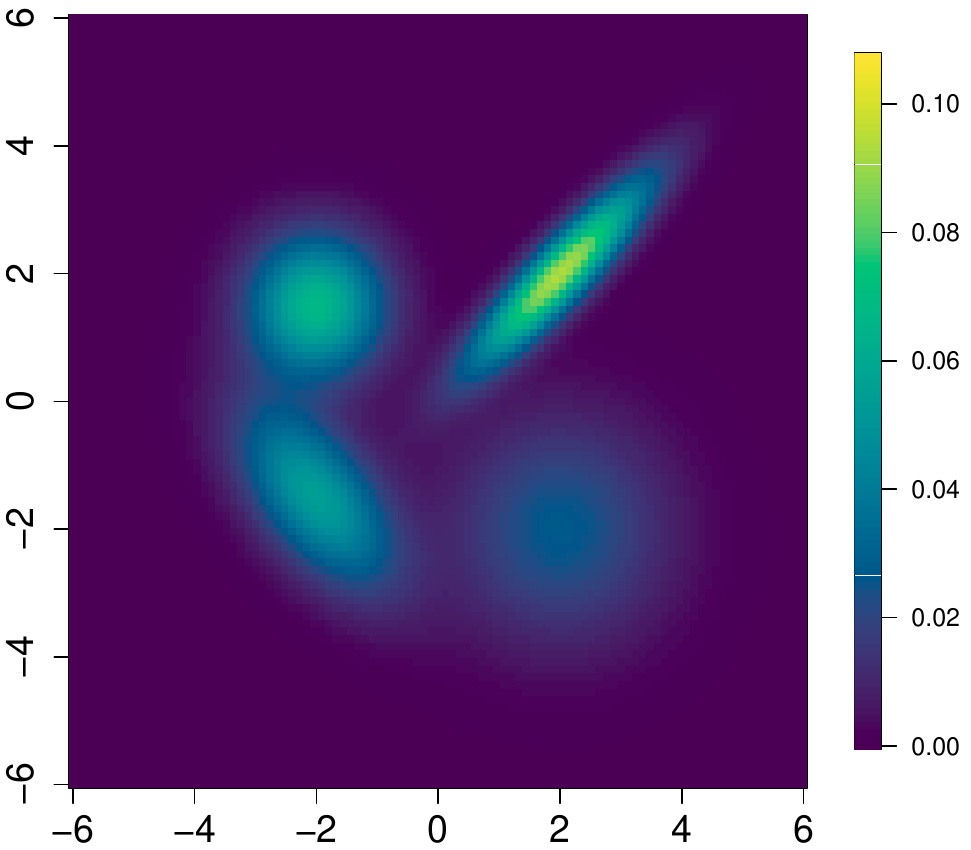}
\caption{TRUE}
\end{subfigure}
\newline
\begin{subfigure}[c]{0.23\linewidth}
\centering
\includegraphics[height = 1.6in, width = 1.45in]{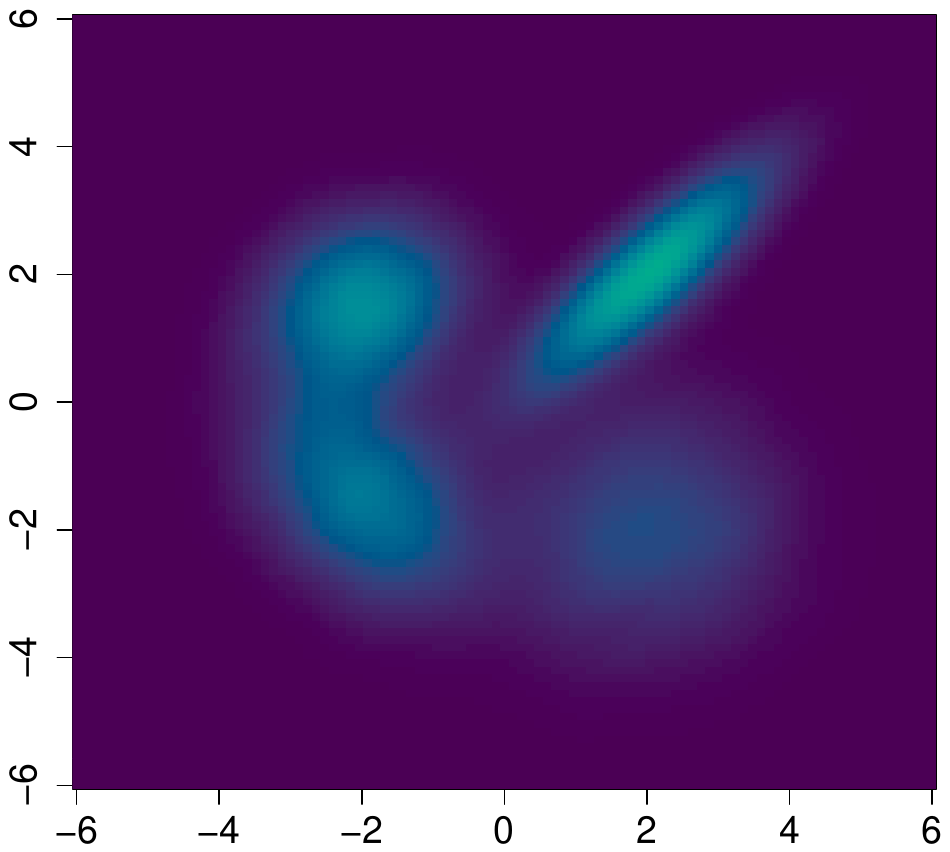}
\caption{KDE}
\end{subfigure}
\begin{subfigure}[c]{.23\linewidth}
\centering
\includegraphics[height = 1.6in, width = 1.45in]{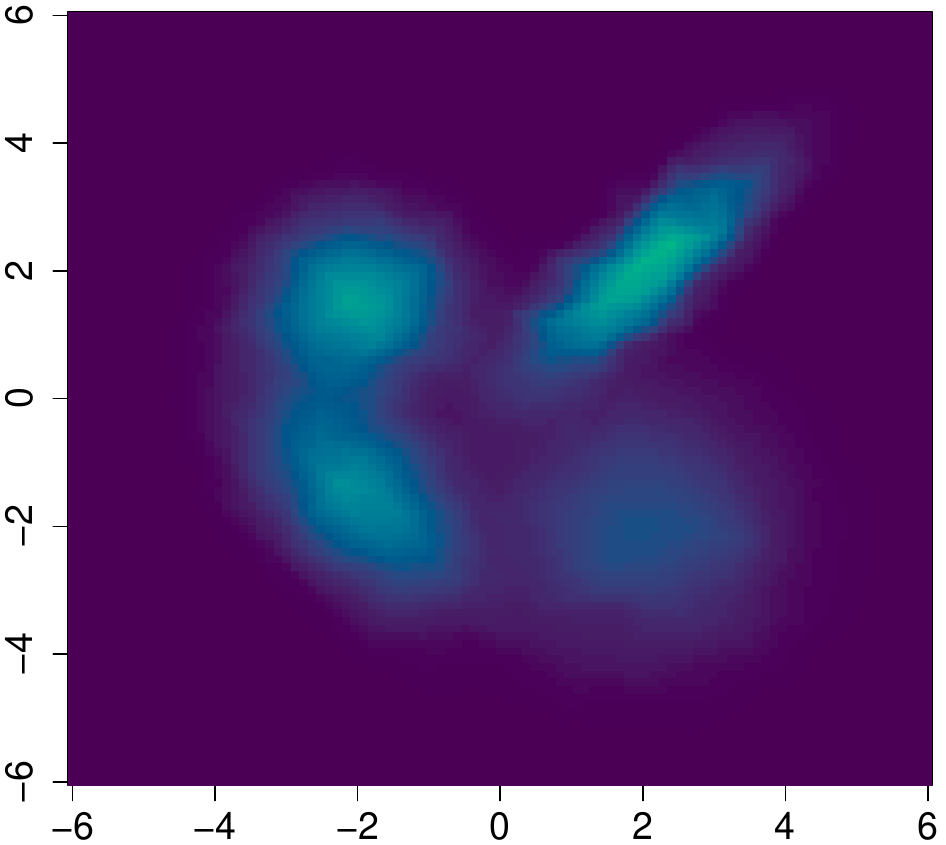}
\caption{HEAT}
\end{subfigure}
\hfill
\begin{subfigure}[c]{.23\linewidth}
\centering
\includegraphics[height = 1.6in, width = 1.45in]{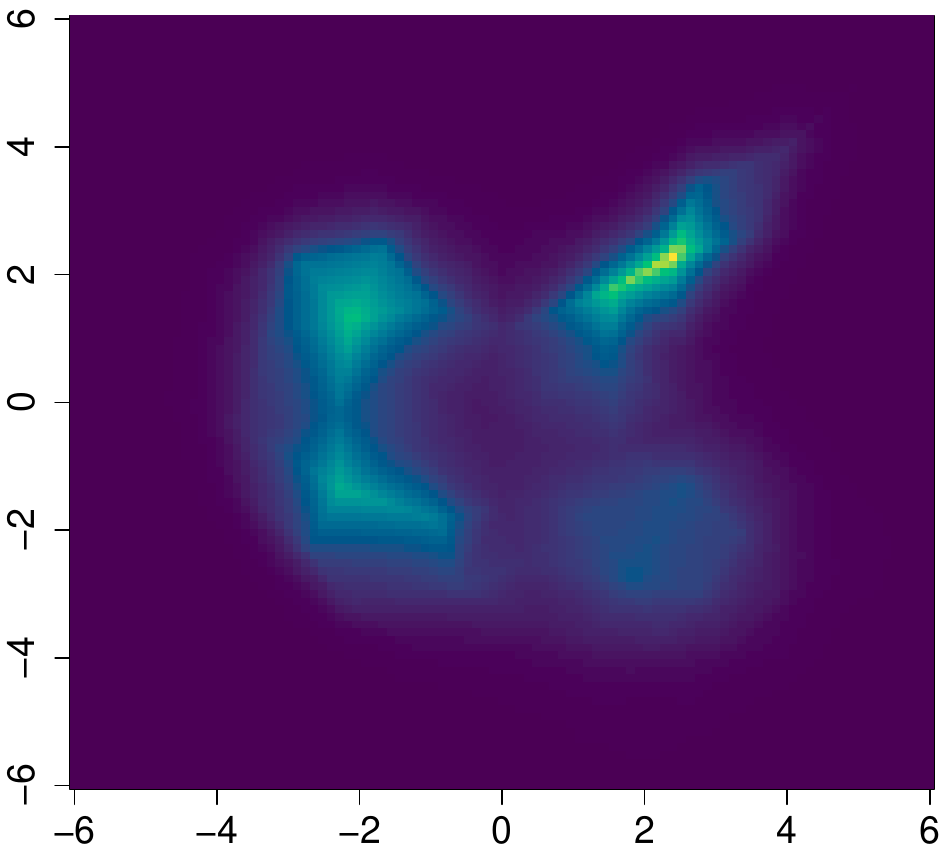}
\caption{FEM}
\end{subfigure}
\hfill
\begin{subfigure}[c]{0.27\linewidth}
\centering
\includegraphics[height = 1.6in, width = 1.65in]{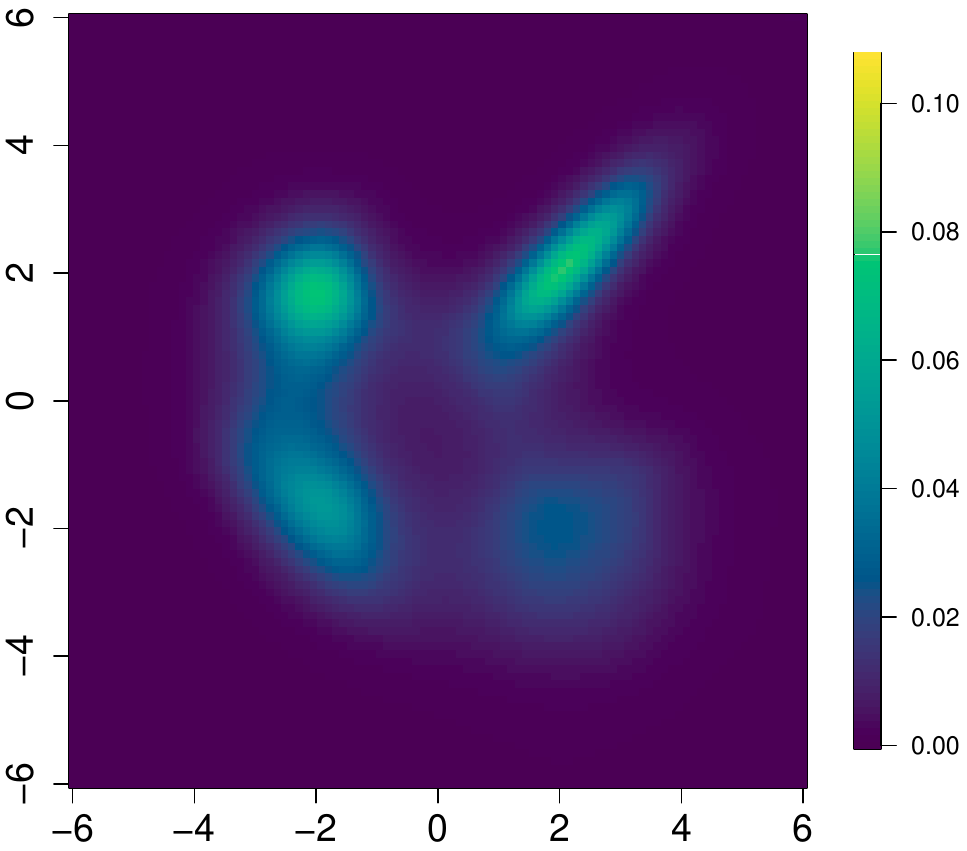}
\caption{BPST}
\end{subfigure}
\caption{(a) A single realization of the data; (b) comparison of  methods using MISE; (c) true density; (d)-(G) estimated surface of density using KDE, HEAT, FEM, and BPST.} 
\label{fig:sim1}
\end{figure}
\paragraph{}In this simulation, the KDEs are obtained using the optimal bandwidth matrix automatically selected by the \texttt{ks::kde()} function using the $5$-fold CV. The DE-PDE and HEAT initial estimates are estimated using \texttt{fdaPDE::DE.FEM()} and \texttt{fdaPDE::DE.heat.FEM()} respectively, after discretizing the rectangular domain into a triangular mesh, with the maximum area of the triangles being 1 and the minimum triangle angle equal to 30 degrees, eventually generating 133 nodes and 228 triangles in total over the entire domain. The smoothing parameter for FEM is chosen using a $5$-fold CV. For BPST, the triangle size is selected according to the requirements for theoretical convergence (e.g., $\beta$-quasi-uniform triangulation) and the discussions in Section \ref{ssec:parameter-selction}. Eventually, 1568 and 50 triangles are generated for the initial and final optimized estimators of BPST, respectively. Comparatively, a larger triangulation is generated for the initial estimate than the final estimate to have a better initial approximation from the algorithm that depends on the triangle areas and the proportion of data points that fall into the triangles. 

\begin{figure}[!ht]
\centering
\begin{subfigure}[c]{.48\linewidth}
\centering
\includegraphics[height = 0.5\linewidth]{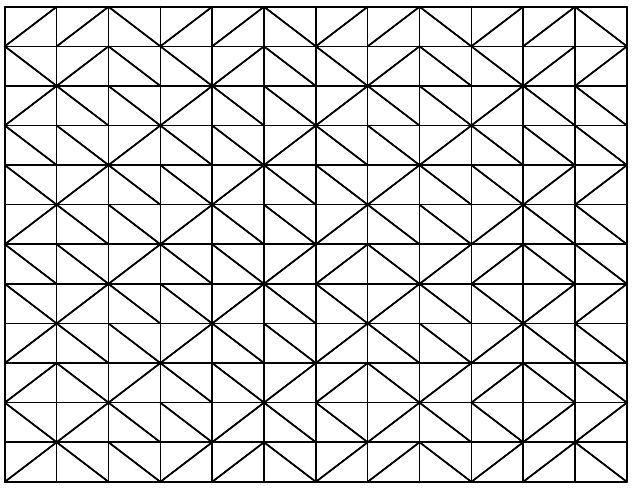}
\caption{}
\end{subfigure}
\hfill
\begin{subfigure}[c]{.48\linewidth}
\centering
\includegraphics[height = 0.5\linewidth]{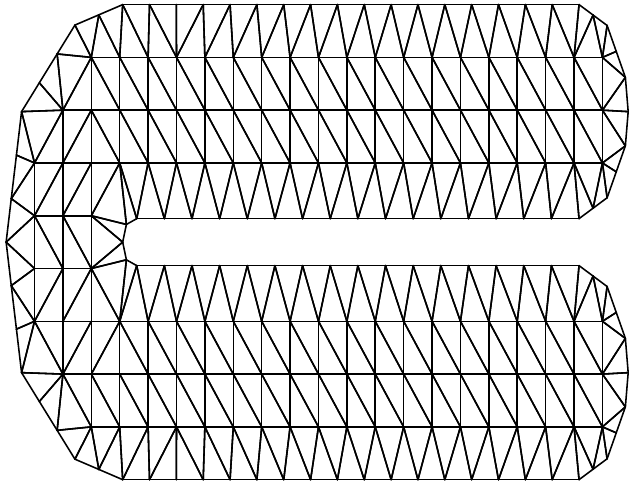}
\caption{}
\end{subfigure}
\caption{Triangulation plot for the domains used in (a) Simulation 1, and (b) Simulation 2.} 
\label{fig:sim12triang}
\end{figure}

\paragraph{}Figure \ref{fig:sim1} shows that the comparative methods, including BPST, closely approximate the true density form by detecting all four modes of the mixture distribution. KDE, depicted in Figure \ref{fig:sim1} (d), identifies the four modes but tends to oversmooth, leading to underestimated density values in various parts of the domain. This issue arises from the modes' differing orientations and degrees of anisotropy in the Gaussian distribution. Initial estimates for FEM i.e., HEAT, shown in Figure \ref{fig:sim1} (e), serve as foundational elements for the estimation of the density through FEM. It shows marked improvements in smoothness and accuracy. FEM, as shown in Figure \ref{fig:sim1} (f), effectively captures the structural details of the four modes by using its locally discretized triangles and Finite Element basis functions to maintain the inherent heterogeneous structure of the function. BPST, presented in Figure \ref{fig:sim1} (f), excels in representing the true density, particularly in capturing abrupt changes and the varying spread and anisotropy of the components of the Gaussian mixture. It provides precise estimations of high- and low-density regions near the modes, making it the most effective method based on visual qualitative assessment. Figure \ref{fig:sim1} (b) features boxplots that compare the MISE values of these methods over 100 replications, with BPST and FEM displaying nearly equivalent and the lowest MISE scores, though BPST shows less variability in MISE values compared to FEM.

\subsection{Simulation study 2}
\label{ssec:sim2}
In this scenario, we analyze a simplified density in a complex Horseshoe domain \citep{Ramsay2002}, using the test function from Section 5.1 of \cite{WoodBravingtonHedley2008}. To maintain non-negativity, the function's value is increased by five and subsequently normalized by dividing by its integral, ensuring the formation of a proper density. The domain is characterized by a significant concavity that nearly bifurcates it into two regions, with one region exhibiting considerably higher values than the other. This simulation mimics practical scenarios, as exemplified by the analysis of motor vehicle theft in Portland presented in Section \ref{SEC:application}, where two proximate but distinct regions display contrasting high- and low-density values indicative of the likelihood of event occurrences.   

\paragraph{}The implementation details for KDE in this study remain consistent with those outlined in Simulation 1. For FEM and HEAT, the smoothing parameter is determined using a 2-fold CV. The horseshoe domain is discretized based on specific mesh settings: a maximum triangle area of 0.012 and a minimum triangle angle of 30 degrees, resulting in a mesh comprising 520 nodes and 930 triangles. In the case of BPST, the initial and final optimized estimates are generated on a triangulated horseshoe domain, as in Figure \ref{fig:sim12triang}(b), featuring 356 and 112 triangles, respectively. It is important to note that the number of triangles utilized in BPST is significantly higher than those used for a rectangular domain in Simulation 1. This increase is due to the unique shape constraints of the Horseshoe domain and the need for a uniform triangulation that adapts to the prominent concave structure of the domain.  

\begin{figure}[!ht]
\centering
\begin{subfigure}[c]{.25\linewidth}
\centering
\includegraphics[height = 1.6in, width = 1.55in]{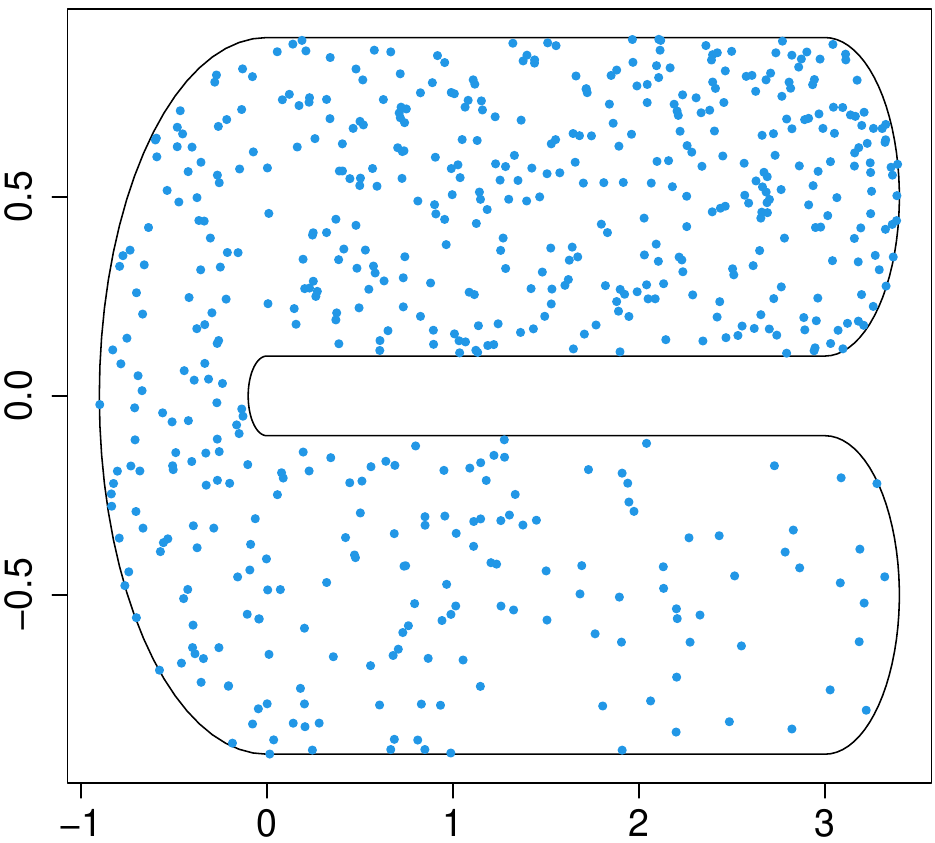}
\caption{Sample}
\end{subfigure}
\hfill
\begin{subfigure}[c]{.3\linewidth}
\centering
\includegraphics[height = 1.7in, width = \linewidth]{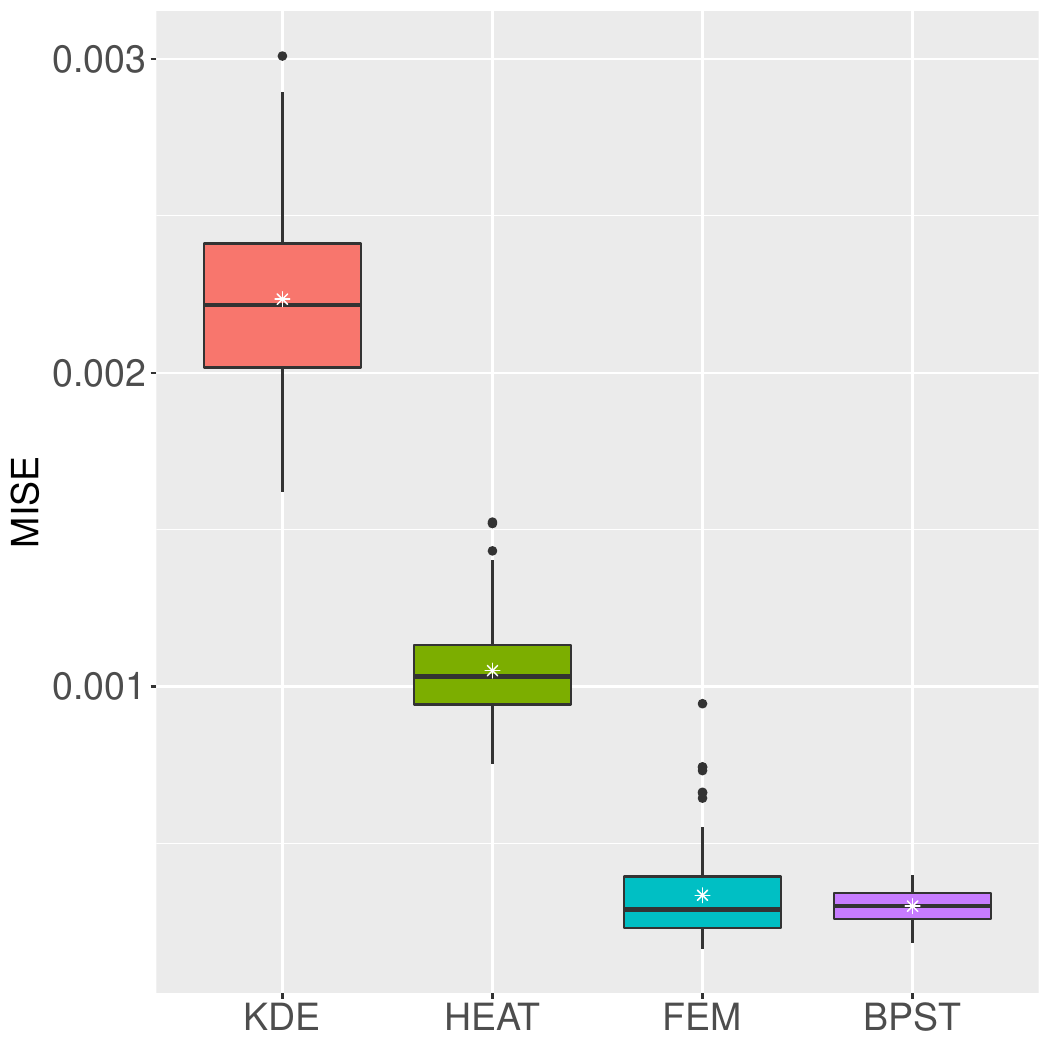}
\caption{MISEs}
\end{subfigure}
\hfill
\begin{subfigure}[c]{0.24\linewidth}
\centering
\includegraphics[height = 1.7in, width = 1.75in]{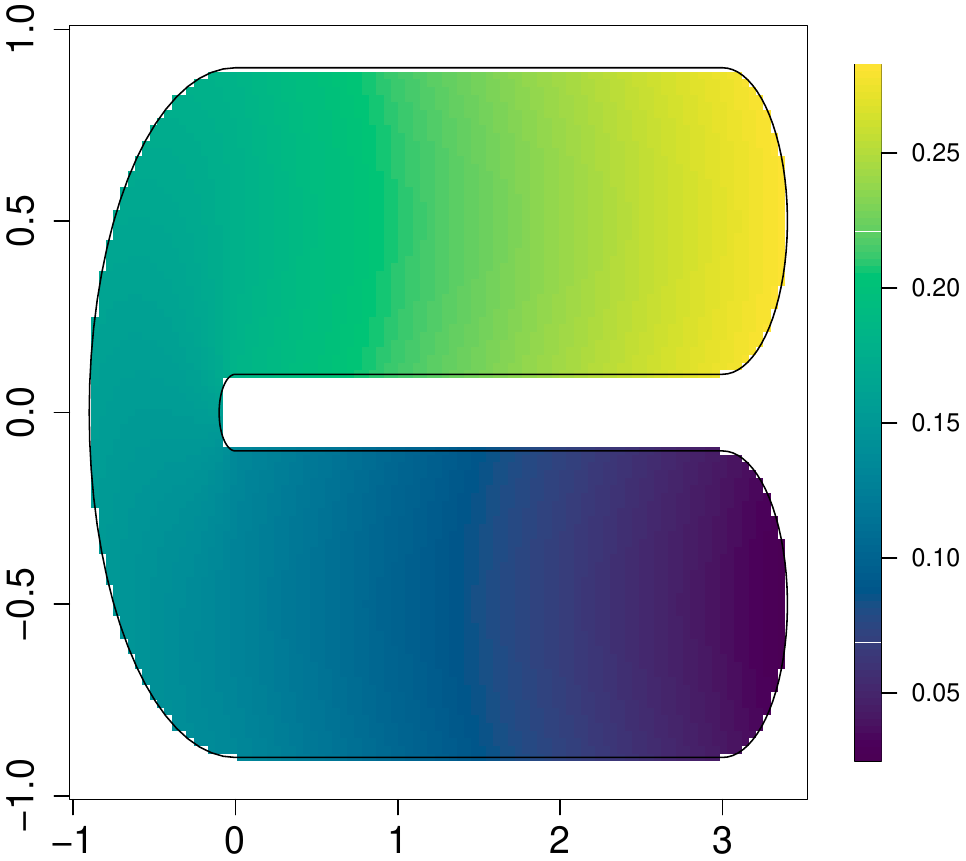}
\caption{TRUE}
\end{subfigure}
\newline
\begin{subfigure}[c]{0.24\linewidth}
\centering
\includegraphics[height = 1.7in, width = \linewidth]{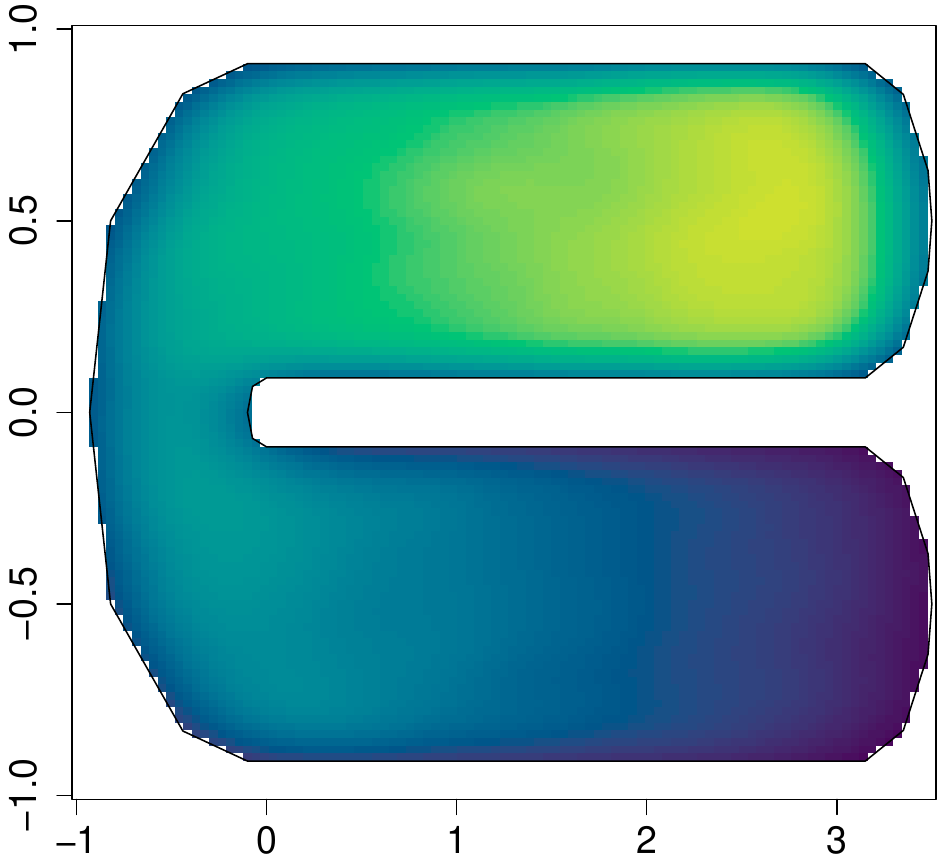}
\caption{KDE}
\end{subfigure}
\begin{subfigure}[c]{.24\linewidth}
\centering
\includegraphics[height = 1.7in, width = \linewidth]{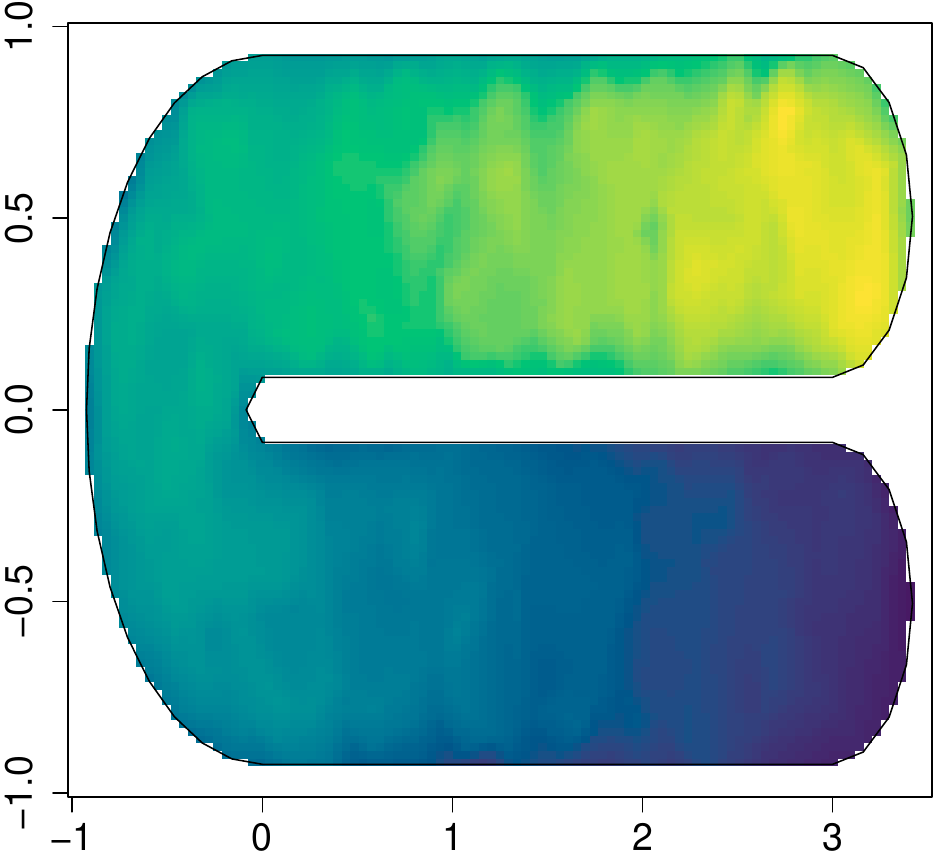}
\caption{HEAT}
\end{subfigure}
\hfill
\begin{subfigure}[c]{.24\linewidth}
\centering
\includegraphics[height = 1.7in, width = \linewidth]{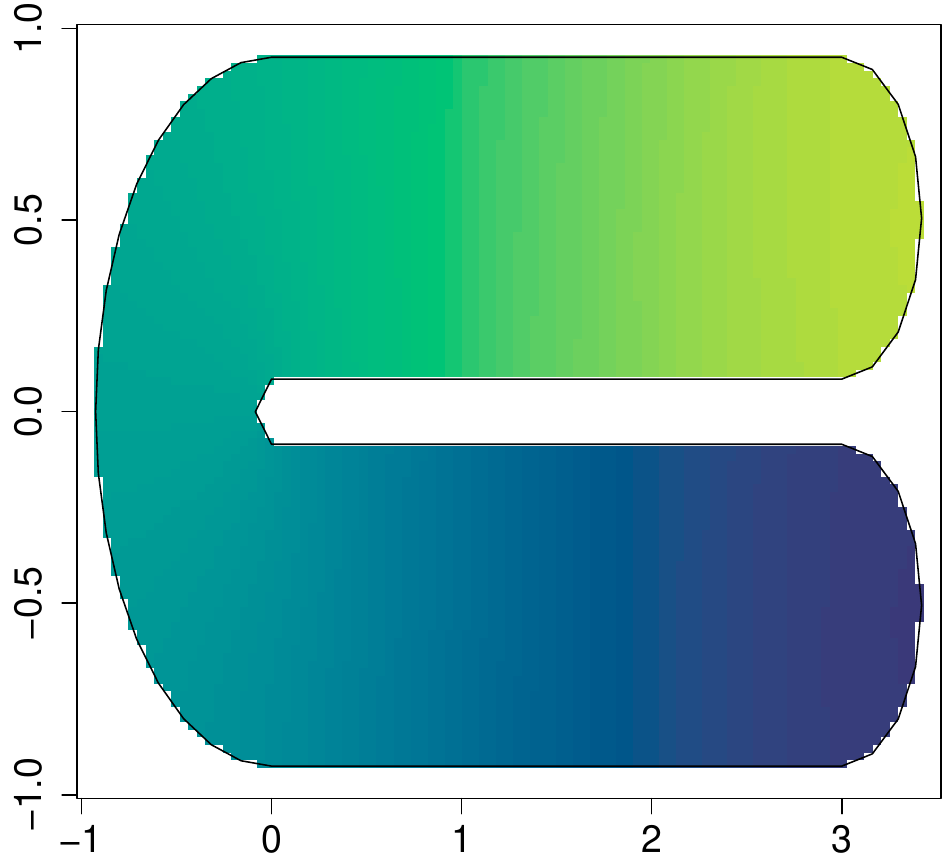}
\caption{FEM}
\end{subfigure}
\hfill
\begin{subfigure}[c]{0.24\linewidth}
\centering
\includegraphics[height = 1.7in, width = 1.75in]{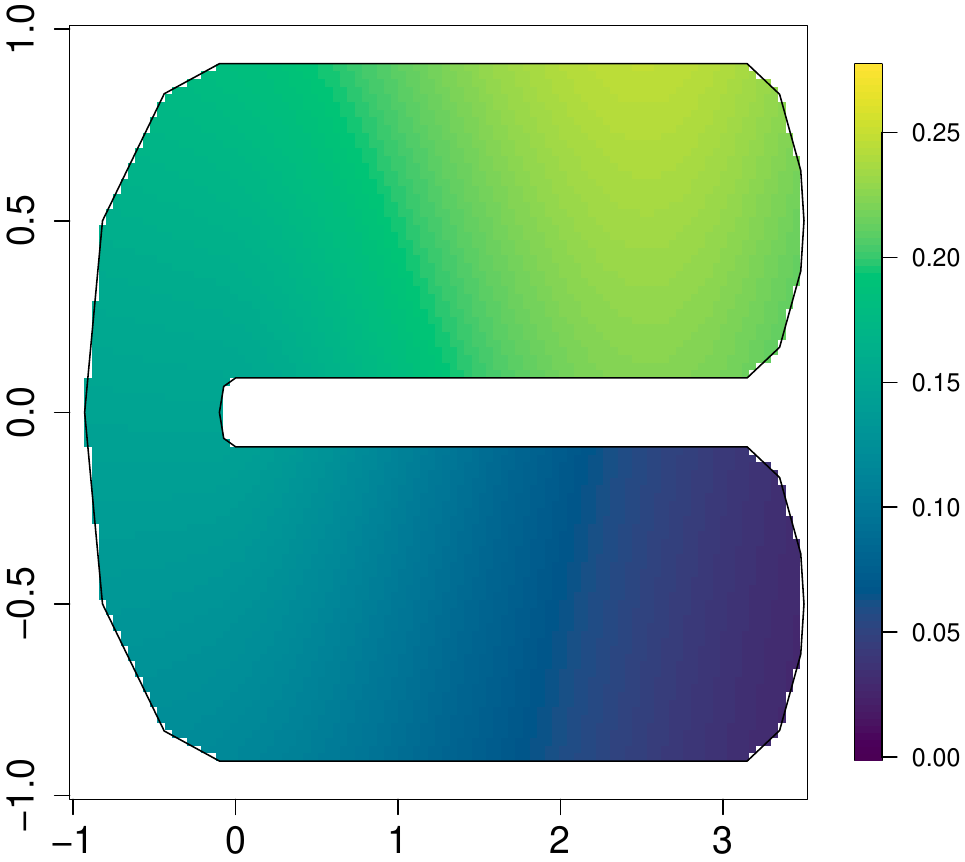}
\caption{BPST}
\end{subfigure}
\caption{(a) A single data realization; (b) comparison of methods using MISE; (c) true density; (d)--(g) estimated density surfaces for KDE, HEAT, FEM, and BPST.}
\label{fig:sim2}
\end{figure}

\paragraph{}Figure \ref{fig:sim2} illustrates that KDE does not accurately capture the structure of the underlying density, concentrating most density estimates in the center of the upper arm of the horseshoe domain. This outcome highlights the limitations of Euclidean distance-based methods in density estimation, as discussed in the Introduction, particularly regarding edge effects near the domain boundary due to a scarcity of observations. In contrast, both FEM and BPST adhere to the shape constraints of the domain and acknowledge that Euclidean proximity between points does not necessarily correlate with the closeness in actual density magnitudes. Although the FEM closely approximates the true form, it consistently underestimates the density values across the domain. BPST, on the other hand, not only closely mirrors the true density form but also provides nearly perfect estimates of the values across the domain's two distinct parts. From the MISE plot (Figure \ref{fig:sim2} (b)), it is evident that FEM and BPST enjoy the smallest MISE values, while BPST has an advantage over FEM. BPST, on the other hand, also enjoys the smallest variance in MISE values compared to all other competing methods.

\subsection{Simulation study 3}
\label{ssec:sim3}
This simulation study examines a more complex density configuration on the same intricate horseshoe domain as in Simulation 2. Unlike Simulation 2, the true density is now defined as a mixture that incorporates the density profile from Simulation 2 along with two additional Gaussian distributions and a skewed Gaussian distribution. The parameters for these distributions are detailed as follows:
\begin{itemize}
    \item Means: 
      \(\mu_1 = (0.9, -0.5)'\),
      \(\mu_2 = (2, -0.5)'\);
    \item Variances:
      \(\Sigma_1 = \begin{bmatrix}
      0.04 & 0 \\
      0 & 0.01
      \end{bmatrix}\),
      \(\Sigma_2 = \begin{bmatrix}
      0.02 & 0 \\
      0 & 0.01
      \end{bmatrix}\);
    \item Skewed Gaussian distribution (from the \texttt{R} package \texttt{sn}):\\
      \(\xi = \begin{pmatrix}
      1.3\\
      0
      \end{pmatrix}\),
      \(\Omega = \begin{bmatrix}
      0.5 & 0 \\
      0 & 0.1
      \end{bmatrix}\),
      \(\alpha = \begin{pmatrix}
      0\\
      6
      \end{pmatrix}\), 
      \(\tau = 0\);
    \item Mixture weights: \(\pi = (0.2, 0.05, 0.05, 0.7)\).
\end{itemize}

\begin{figure}[!ht]
\centering
\begin{subfigure}[l]{.34\linewidth}
\centering
\includegraphics[height = 1.75in, width = 0.8\linewidth]{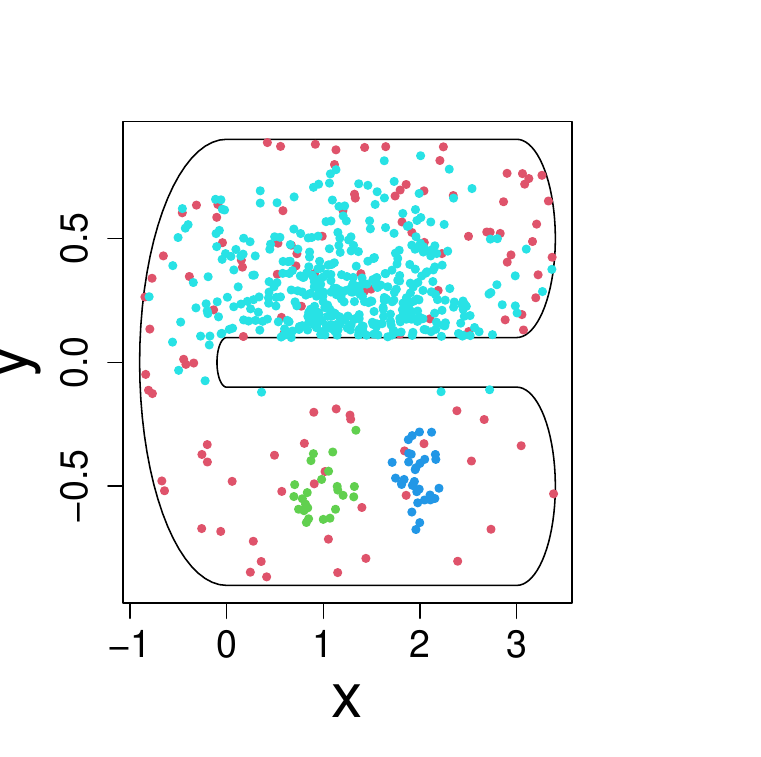}
\caption{Sample}
\end{subfigure}
\begin{subfigure}[c]{.3\linewidth}
\centering
\includegraphics[height = 1.75in, width = \linewidth]{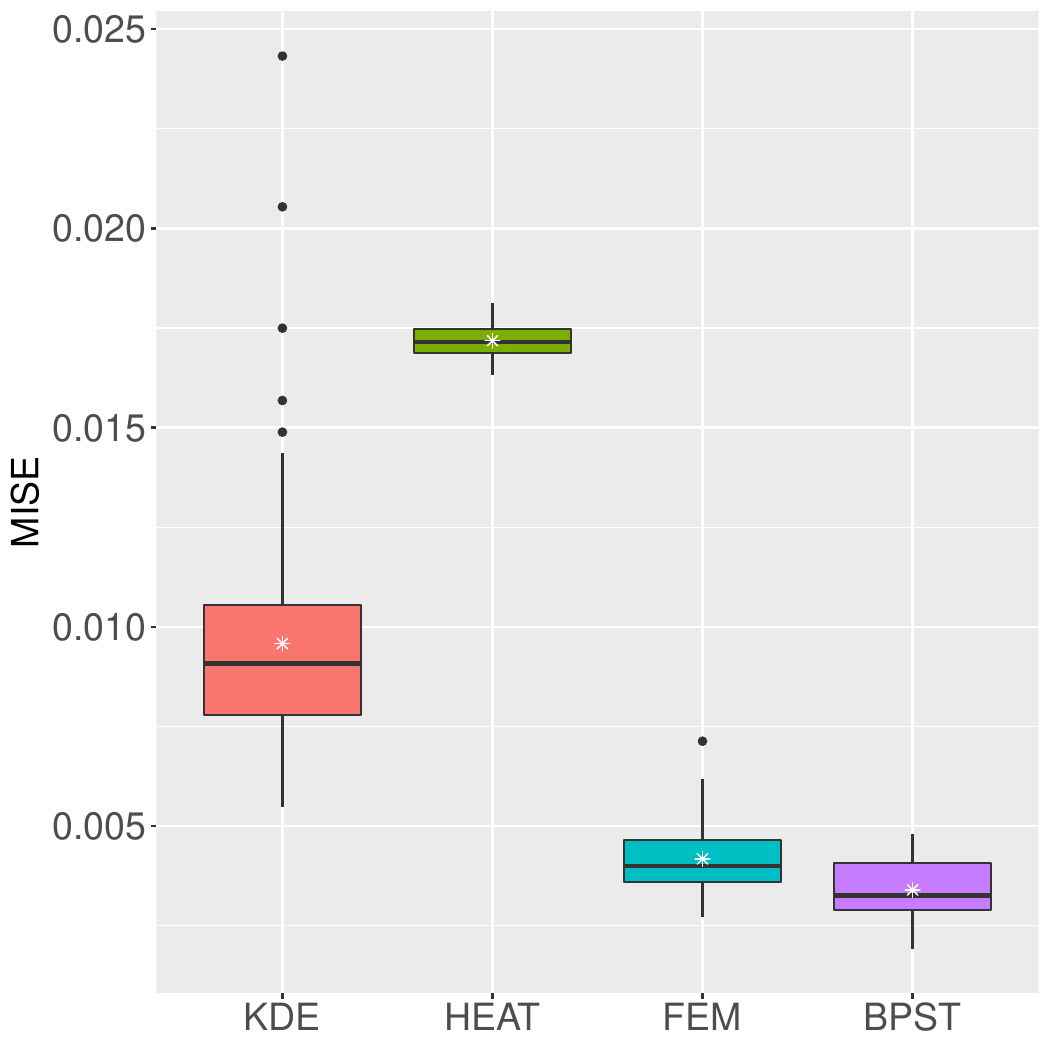}
\caption{MISEs}
\end{subfigure}
\hfill
\begin{subfigure}[c]{0.27\linewidth}
\centering
\includegraphics[height = 1.75in, width = 1.7in]{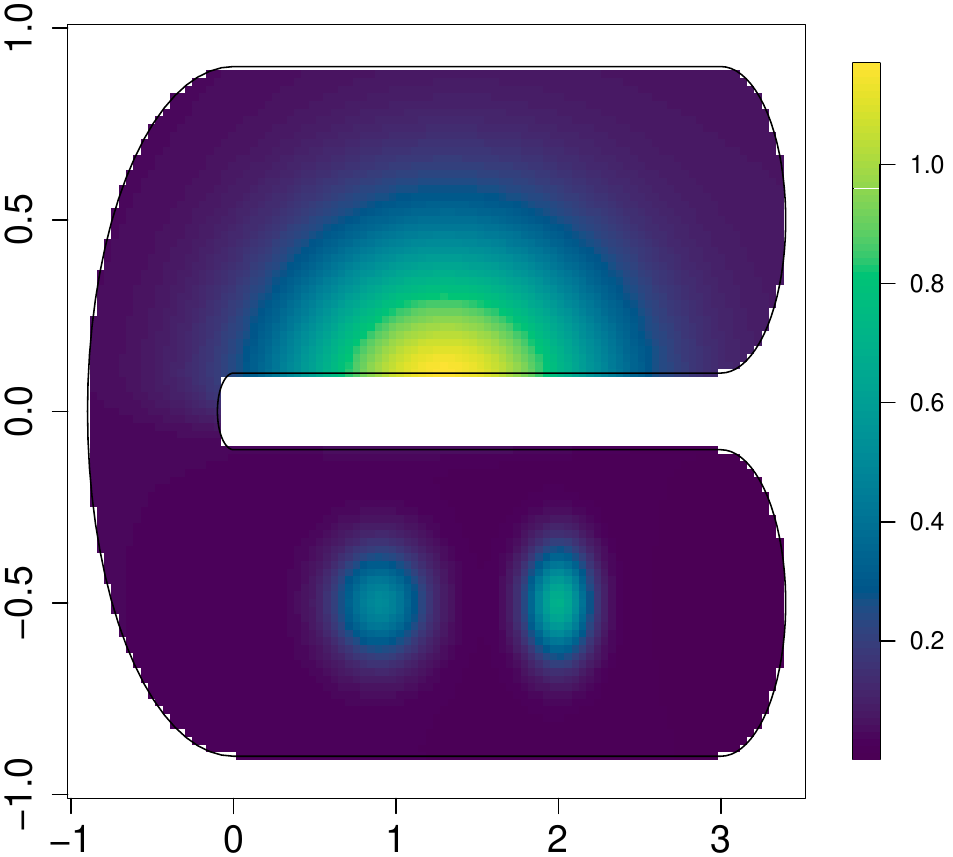}
\caption{TRUE}
\end{subfigure}
\newline
\begin{subfigure}[c]{0.23\linewidth}
\centering
\includegraphics[height = 1.75in, width = 1.45in]{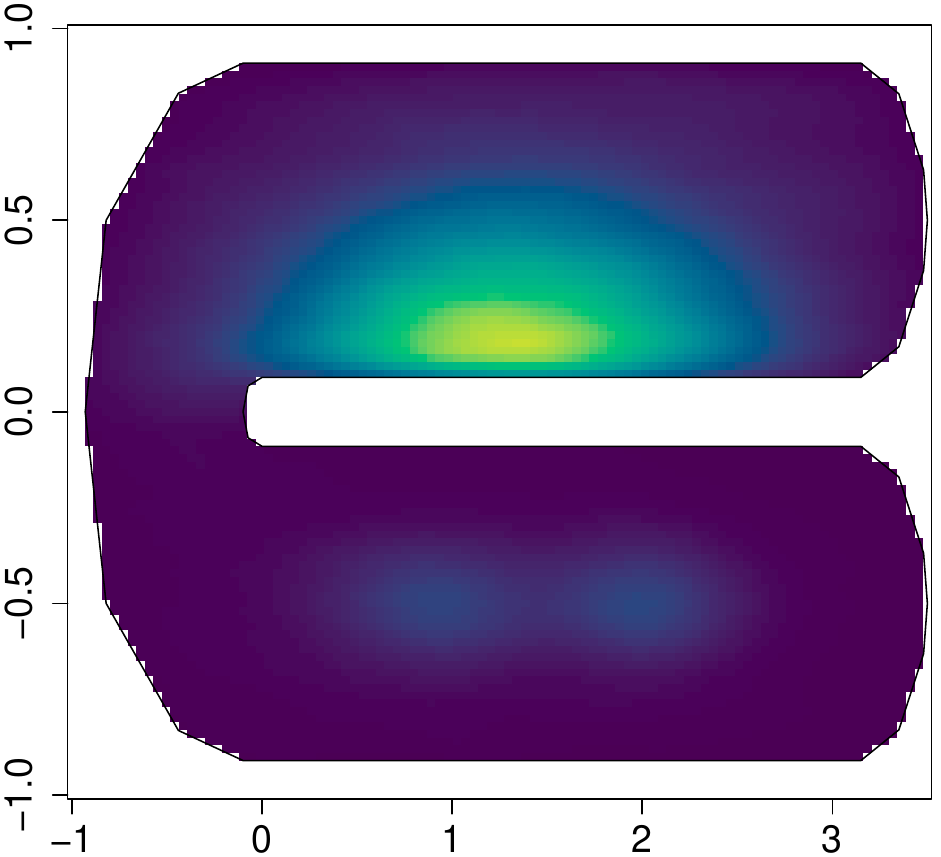}
\caption{KDE}
\end{subfigure}
\hfill
\begin{subfigure}[c]{.23\linewidth}
\centering
\includegraphics[height = 1.75in, width = 1.45in]{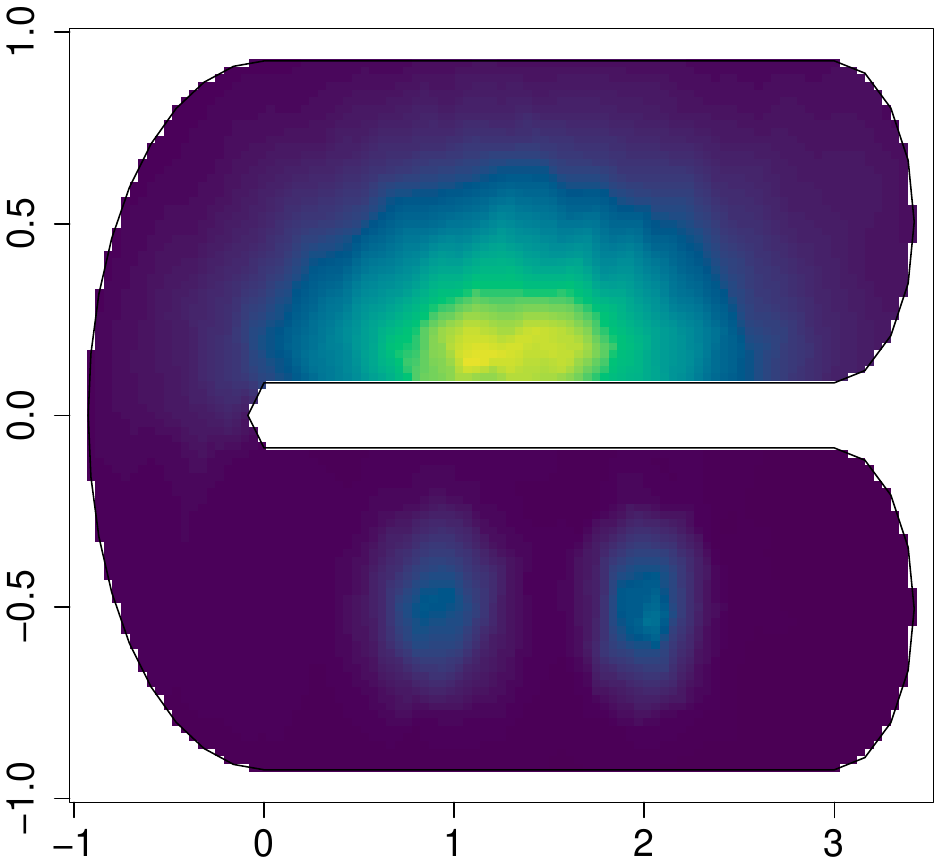}
\caption{HEAT}
\end{subfigure}
\hfill
\begin{subfigure}[c]{.23\linewidth}
\centering
\includegraphics[height = 1.75in, width = 1.45in]{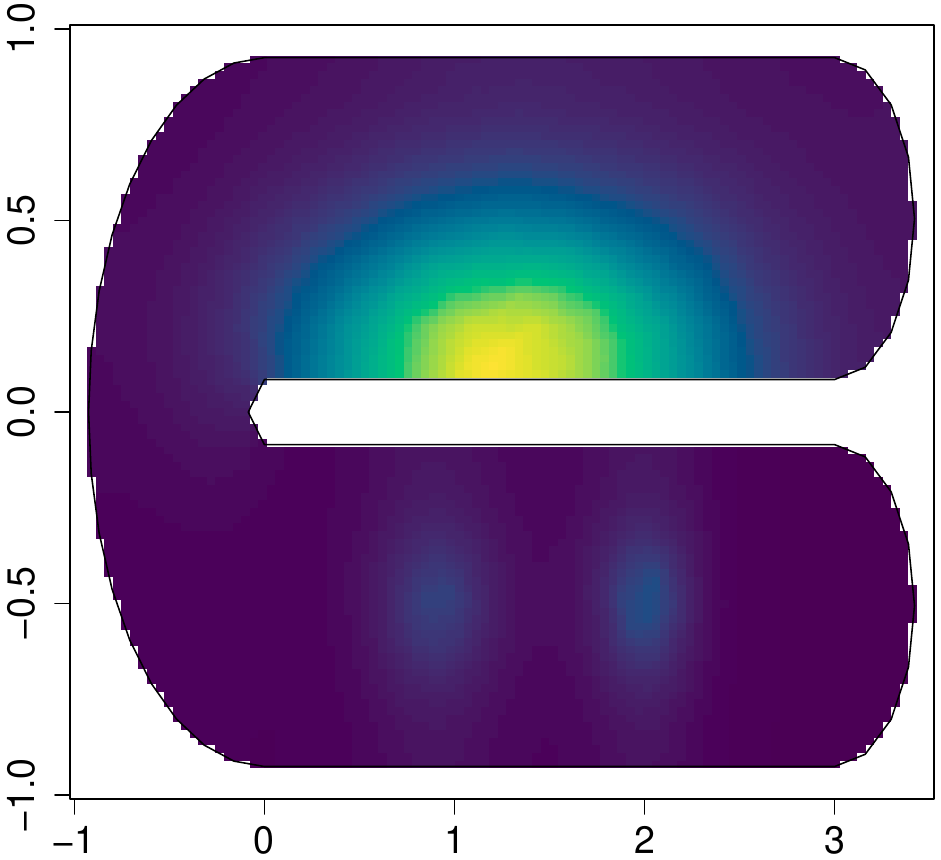}
\caption{FEM}
\end{subfigure}
\hfill
\begin{subfigure}[c]{0.27\linewidth}
\centering
\includegraphics[height = 1.75in, width = 1.7in]{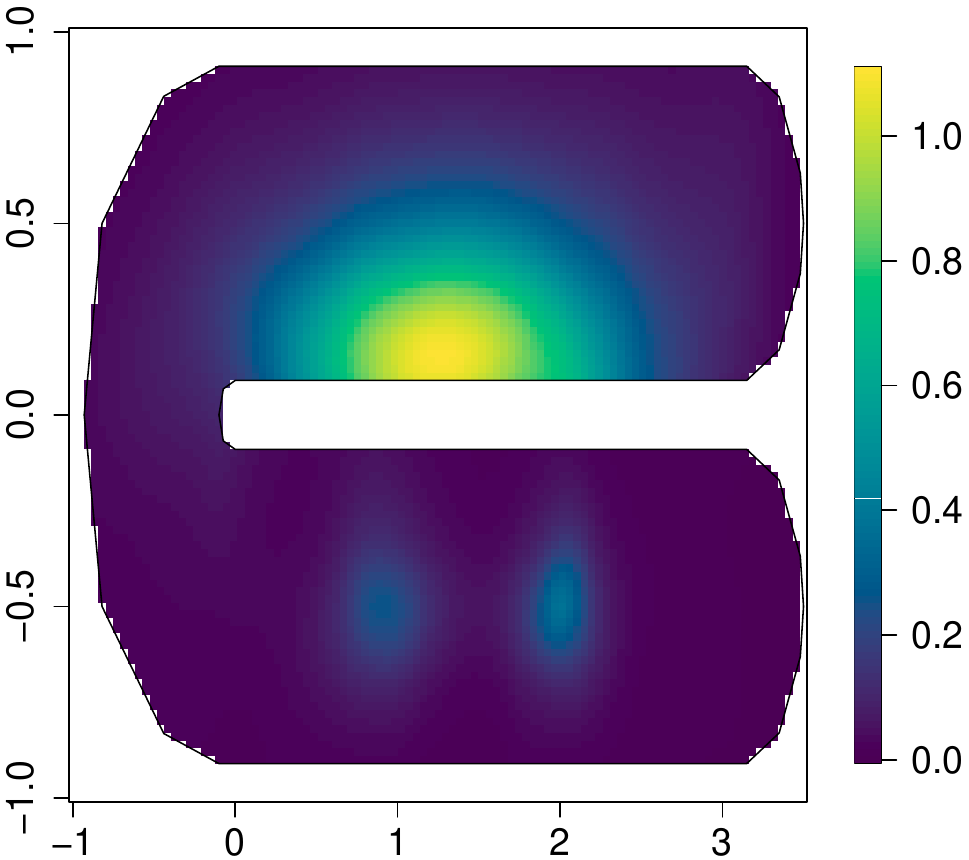}
\caption{BPST}
\end{subfigure}
\hfill
\caption{(a) A single data realization; (b) comparison of methods using MISE; (c) true density; (d)-(g) estimated density surfaces for KDE, HEAT, FEM, and BPST.}
\label{fig:sim3}
\end{figure}

\paragraph{}The KDE and FEM estimates are derived in the same manner as in Simulation 2. BPST is applied to a triangulated domain using $356$ triangles for the initial estimator and $214$ triangles for the optimized estimator. As evident from Figure \ref{fig:sim3}, while the KDE estimate accurately identifies the modal points of the density, it struggles to constrain the elongation of the two Gaussian modes in the lower horseshoe arm. Both HEAT and FEM tend to oversmooth the density of the mixture components in this area, resulting in underestimated density values. Whereas, BPST accurately captures the high-density region on the upper arm of the domain to the correct magnitude. Although BPST slightly underestimates the modes of the Gaussian distribution on the lower arm, it compensates by precisely delineating the shape and scatterness of those modes, outperforming other methods. The boxplots of MISEs in Figure \ref{fig:sim3}(b) further highlight that BPST surpasses competing methods, achieving the smallest MISE values and variance.

\section{Real Data Application} 
\label{SEC:application}


\begin{figure}[!ht]
\centering
\begin{subfigure}[c]{.3\linewidth}
\centering
\includegraphics[height = 1.75in, width = \linewidth]{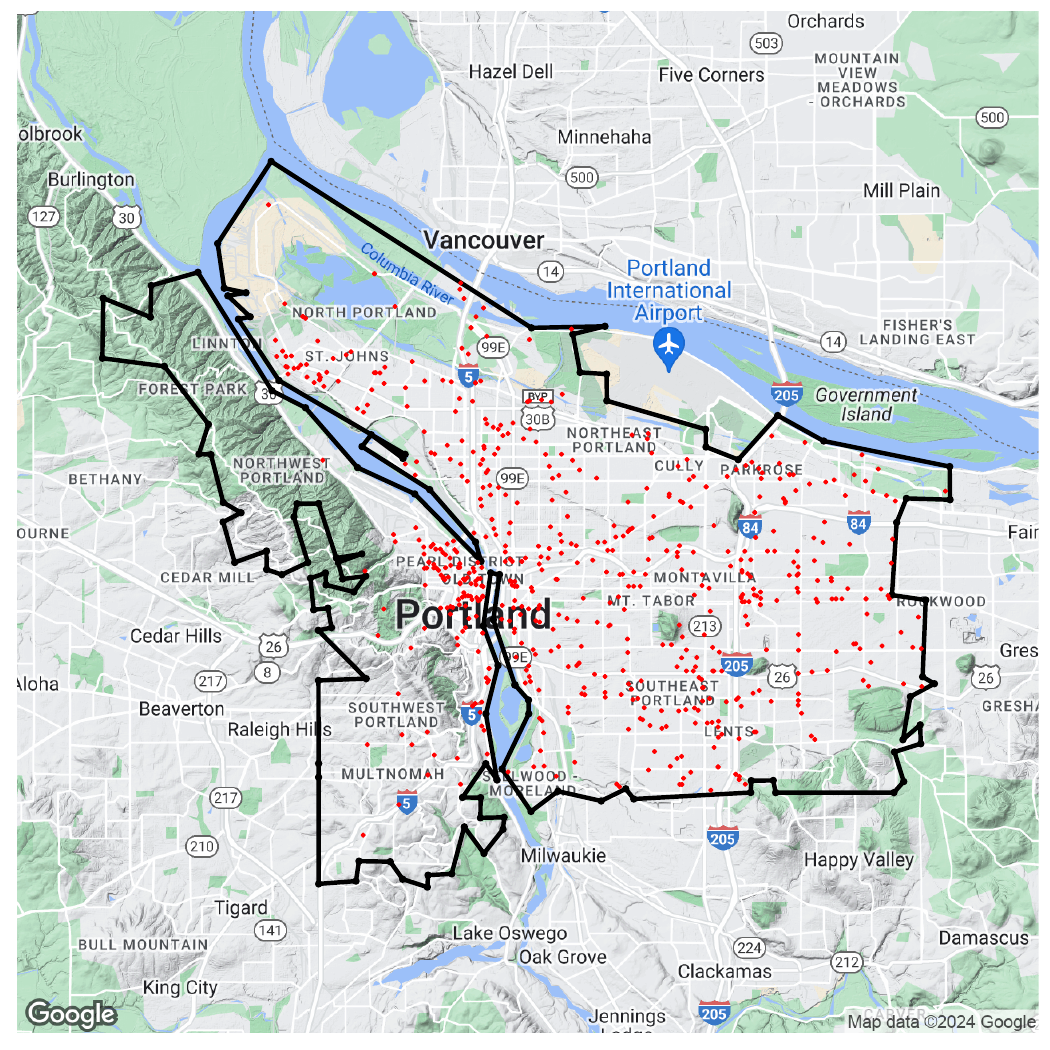}
\caption{}
\end{subfigure}
\hfill
\hfill
\begin{subfigure}[c]{.3\linewidth}
\centering
\includegraphics[height = 1.75in, width = \linewidth]{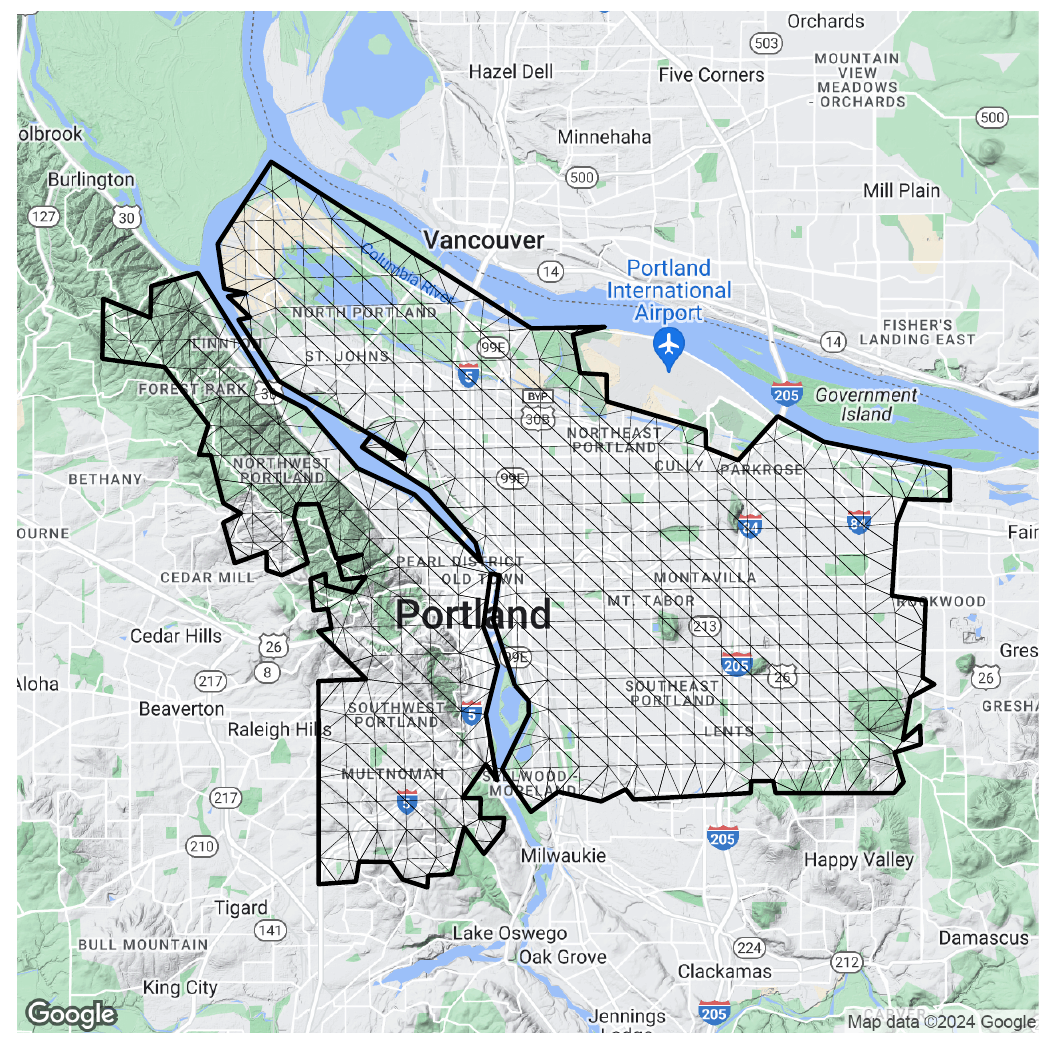}
\caption{}
\end{subfigure}
\hfill
\begin{subfigure}[c]{0.34\linewidth}
\centering
\includegraphics[height = 1.75in, width = \linewidth]{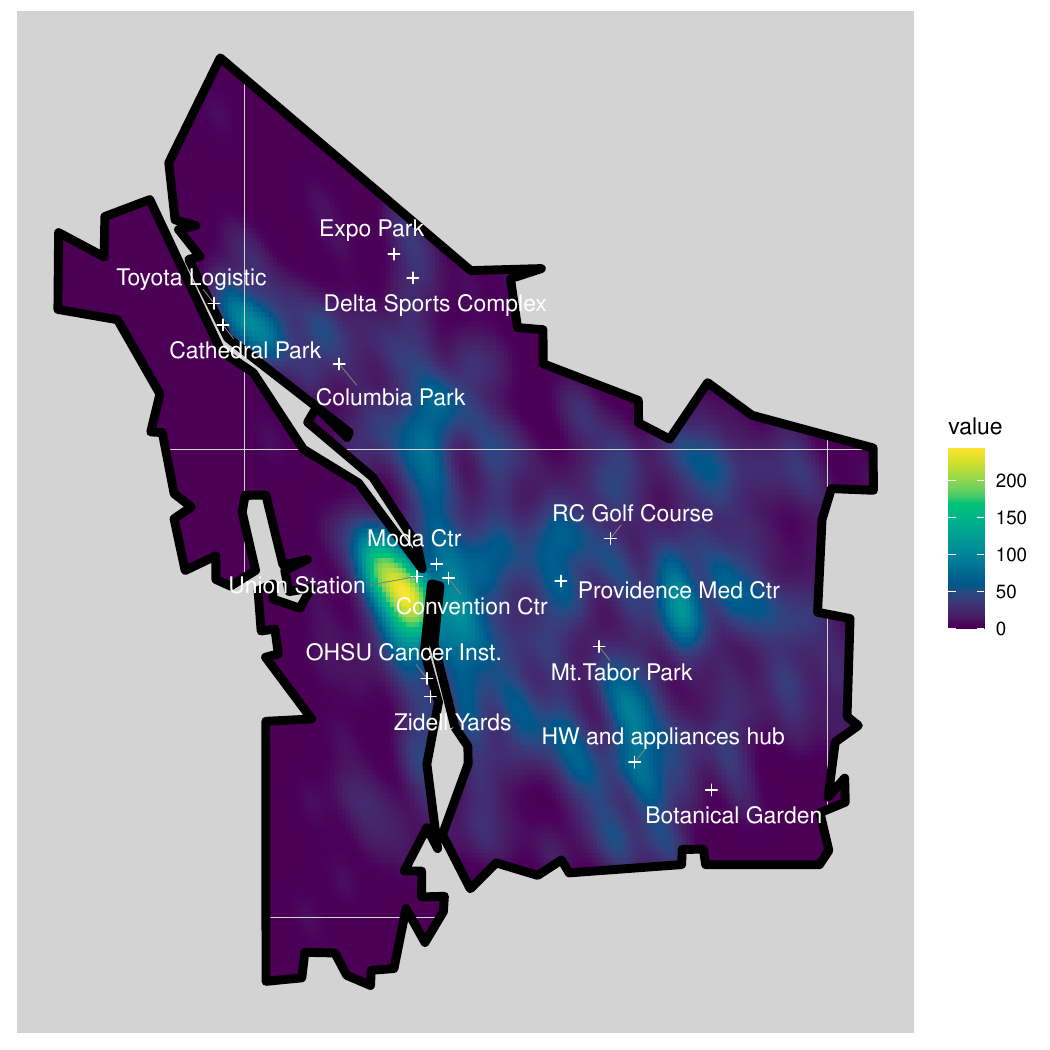}
\caption{}
\end{subfigure}
\hfill
\newline
\begin{subfigure}[c]{.34\linewidth}
\centering
\includegraphics[height = 1.75in, width = \linewidth]{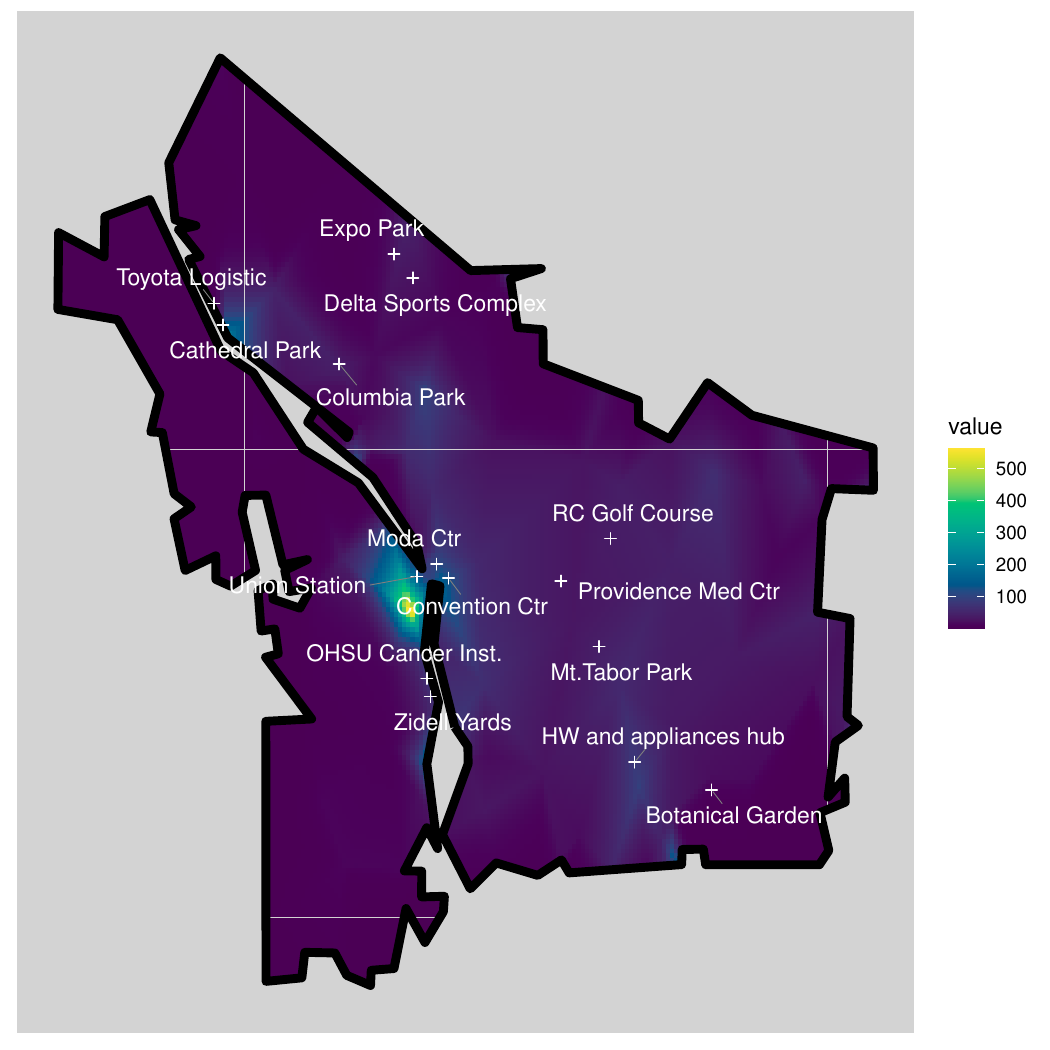}
\caption{}
\end{subfigure}
\hfill
\begin{subfigure}[c]{.34\linewidth}
\centering
\includegraphics[height = 1.75in, width = \linewidth]{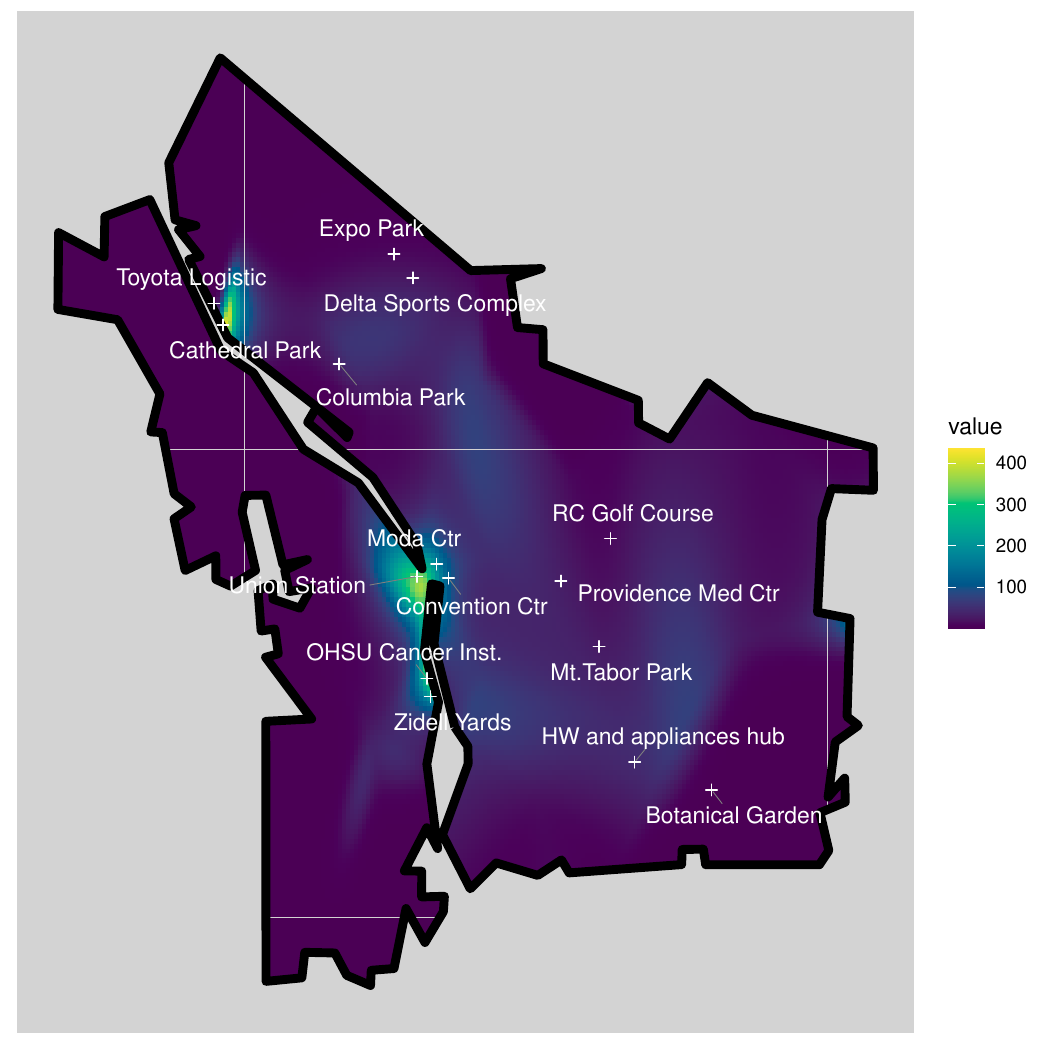}
\caption{}
\end{subfigure}
\hfill
\begin{subfigure}[c]{0.3\linewidth}
\centering
\includegraphics[height = 1.75in, width = \linewidth]{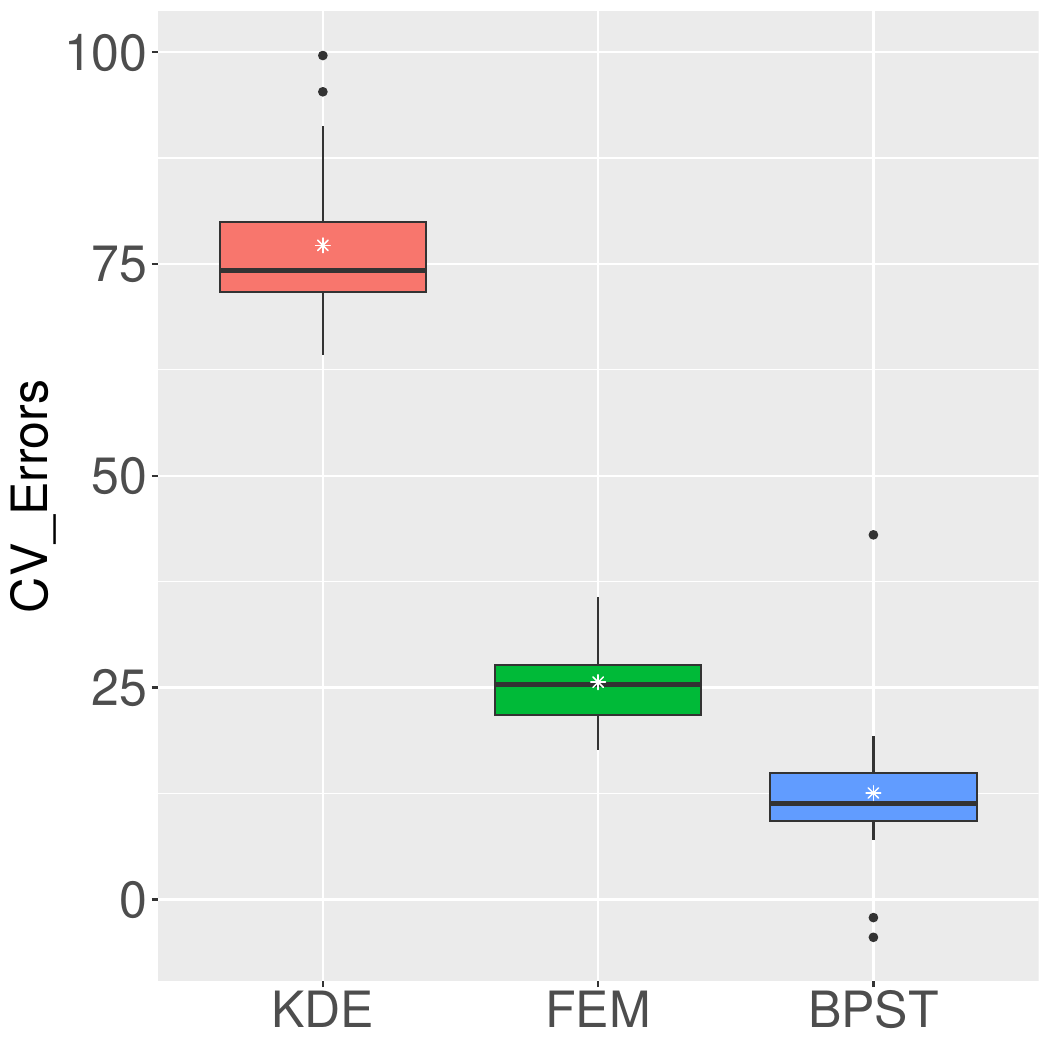}
\caption{}
\end{subfigure}
\hfill
\caption{(a) Motor vehicle theft data from Portland, Oregon, of June 2023; (b) triangulation of the domain(Portland city, Oregon); (c) -- (e) the estimated surface of density using KDE, FEM, and BPST; (f) comparison of methods using transformed CV values.}
\label{fig:realdata}
\end{figure}

\paragraph{}For the real-world data application, we analyzed motor vehicle theft reports from Portland City, Oregon, collected in June 2023. The dataset includes geographical information for 602 reported theft locations, presented in both latitude-longitude and UTM (Easting-Northing) coordinates. For additional details on monthly neighborhood offense statistics, please refer to the website of \href{https://public.tableau.com/app/profile/portlandpolicebureau/viz/New_Monthly_Neighborhood/MonthlyOffenseTotals}{Portland Police Bureau}.

\paragraph{}The data depicted in Figure \ref{fig:realdata}(a) highlight two important aspects: i) the complex topology of the domain influenced by the tributaries of the Willamette River, which divides Portland into two distinct parts, and ii) the significant variation in motor vehicle theft reports throughout the city. The city's geography is further complicated by numerous water bodies, primarily formed by the Columbia River. The variation in crime reports is evident from the starkly different numbers of theft incidents on the two opposite river banks. The eastern side exhibits significantly higher occurrences than the western bank, both in the north and south directions, resembling the situations considered in Simulations 2 and 3. The main hotspots for theft reports are the city center, especially the north and northeast parts, which house shopping and sporting complexes, nature parks, golf courses, and the largest of the city's car logistics center with ample parking areas where cars are parked for extended periods during sports events, concerts, and movie screenings or car repairing and logistic services. On the southeast side, the abundance of waterfront parks, trails, and shipyards and the existence of the OHSU cancer research institute hosts large parking lots with heavy occupancy of cars for the large part of the day. These two aspects make these data well-suited to validate the usefulness of the proposed method for density estimation, where the shape constraints of the domain drive the occurrences of events. Accurate density estimation is crucial for sensitive analyses like this, as an inaccurate estimate may lead to incorrect inferences or decision-making processes. For instance, if a method estimates comparatively high nonnegative density in a region on the western side of the river where no points are observed, it could unfairly tarnish that neighborhood's reputation, resulting in a drastic decline in housing market prices and other potential downgrades.

\paragraph{}KDE and DE-PDE methods use the same computational resources as those used in simulation studies. For KDE, the bandwidth matrix is optimized using 5-fold CV. For DE-PDE, the triangular mesh is constructed with a minimum triangle angle of 25 degrees, and the smoothing parameter is also determined through the 5-fold CV. In the case of BPST, the analysis uses the same triangular mesh used in FEM, with a total of $169$ triangles, to ensure methodological comparability. The optimal smoothing parameter for BPST is selected by minimizing the CV error, as detailed in Section \ref{ssec:penalty}. Since from the simulation studies, the performance of the final FEM and BPST estimates are proven to be significantly better compared to their corresponding initial estimates, for conciseness, we only discuss results for the three main methods KDE, FEM, and BPST here for the real data analysis.

\paragraph{}A qualitative comparison of the fits of various methods is illustrated in Figure \ref{fig:realdata} (c)--(e). The KDE method, while capturing the actual structure of the density, tends to underestimate and oversmooth areas where density changes abruptly, particularly in small city neighborhoods. This limitation is inherent to Euclidean distance-based methods, where points near high-occurrence events are likely to receive high estimated values and vice versa for low-occurrence areas. The DE-PDE estimate shows considerable improvement in tracking the dynamic density patterns across the domain, albeit with an overall overestimation. The limitations of linear finite element basis functions are evident as they fail to capture nonlinear pattern changes across the domain. The BPST method emerges as the most accurate in estimating the underlying density, surpassing the limitations of the comparative methods. It adeptly captures all abrupt changes in density, conforms to the domain's shape constraints, and provides significant flexibility in estimation by accurately delineating the complex nonlinear structure of the density across the domain using bivariate penalized splines.

 
\paragraph{}A more quantitative comparison can be made using the CV errors described in Section \ref{ssec:penalty}. Note that, unlike MISEs, this quantity is no longer guaranteed nonnegative. The methods with lower transformed values are indicative of better performance. The BPST method exhibits the smallest CV values in Figure \ref{fig:realdata}(f), significantly outperforming competing methods.

\section{Conclusions and discussions} 
\label{SEC:conclusions}

The BPST method introduced in this paper demonstrates consistent performance in various complex domains and successfully addresses the structural complexities of unknown underlying density functions. It marks a significant enhancement over existing density estimation methods. The discretization approach that we have detailed offers two primary advantages. First, domain triangulation effectively meets the challenges posed by complex domain boundaries. Secondly, the use of bivariate penalized spline basis functions allows for precise capture of abrupt local changes in the density, while preserving the inherent global smoothness typical of spline functions. Moreover, the method simplifies parameter tuning by requiring the selection of a single smoothing parameter, $\lambda_n$, which can be estimated through CV, making the approach more flexible and less dependent on the intricacies of the estimation of several tuning parameters.

\paragraph{}An exciting avenue to look into would be to find some concrete thumb rule regarding the choice of the degree of the spline basis polynomials and the number of triangles. Although the exploration approach worked pretty well in terms of results, a definitive approach or rule would be much more time-efficient in decision-making for the implementors. Another fascinating direction would be to extend the current approach to higher dimensions and non-Euclidean geometries, such as general manifolds. Density estimation in these multidimensional domains demands some flexible method that can overcome the limitations of Euclidean distances, and our proposed method using BPST can open new potential avenues for such demands in higher dimensions. The BPST method can be extended to 3D domains using trivariate splines on tetrahedral partitions of the space following \cite{lai_schumaker_2007}. However, this presents a nontrivial task in terms of computational challenges compared to 2D cases and requires a more thorough investigation. 

\paragraph{}Considering the example of real data analysis, an intriguing path to explore would be to find out how the proposed method can be extended to the analysis of intensity estimation in the spatial point process models and, if possible, to the modeling of intensity in the joint spatiotemporal point processes over complicated domains. Modeling spatiotemporal intensity function would help understand the evolution of the underlying process over time, which might be very important in some fields of study like sociology and biomedical sciences. 

\paragraph{}The requirement of uncertainty quantification in nonparametric density estimation also holds a significant position in future directions. The recent approach proposed by \cite{GineNickl2010} based on Rademacher symmetrization provides a potential extension to the proposed BPST setting.

\section{Acknowledgement} 
\label{SEC:acknowledgement}

Guannan Wang and Li Wang gratefully acknowledge the partial support for this research from the National Institutes of Health under grant 1R01AG085616. Shan Yu's research was partially supported by the 3Cavaliers program at the University of Virginia. Guannan Wang's research also received partial support from the Simons Foundation under Grant 963447.  Li Wang's research was additionally supported by the National Science Foundation under grant 2203207.

\bibliographystyle{apalike}
\bibliography{references}

\begin{thebibliography}{}

\bibitem[Arnone et~al., 2023]{fdaPDE}
Arnone, E., Clemente, A., Sangalli, L.~M., Lila, E., Ramsay, J., and Formaggia,
  L. (2023).
\newblock {\em fdaPDE: Physics-Informed Spatial and Functional Data Analysis}.
\newblock R package version 1.1-16.

\bibitem[Carando et~al., 2009]{carando2009nonparametric}
Carando, D., Fraiman, R., and Groisman, P. (2009).
\newblock Nonparametric likelihood based estimation for a multivariate
  lipschitz density.
\newblock {\em Journal of Multivariate Analysis}, 100(5):981--992.

\bibitem[Chac{\'o}n and Duong, 2018]{chacon2018multivariate}
Chac{\'o}n, J.~E. and Duong, T. (2018).
\newblock {\em Multivariate kernel smoothing and its applications}.
\newblock CRC Press.

\bibitem[Chaudhuri and Marron, 1999]{chaudhuri1999sizer}
Chaudhuri, P. and Marron, J.~S. (1999).
\newblock Sizer for exploration of structures in curves.
\newblock {\em Journal of the American Statistical Association},
  94(447):807--823.

\bibitem[Cox and O'Sullivan, 1990]{cox1990asymptotic}
Cox, D.~D. and O'Sullivan, F. (1990).
\newblock Asymptotic analysis of penalized likelihood and related estimators.
\newblock {\em The Annals of Statistics}, pages 1676--1695.

\bibitem[Cule et~al., 2010]{cule}
Cule, M., Samworth, R., and Stewart, M. (2010).
\newblock Maximum likelihood estimation of a multi-dimensional log-concave
  density.
\newblock {\em Journal of the Royal Statistical Society: Series B (Statistical
  Methodology)}, 72(5):545--607.

\bibitem[Ferraccioli et~al., 2021]{Ferraccioli2021}
Ferraccioli, F., Arnone, E., Finos, L., Ramsay, O., and Sangalli, L.~M. (2021).
\newblock Nonparametric density estimation over complicated domains.
\newblock {\em Journal of the Royal Statistical Society: Series B (Statistical
  Methodology)}.

\bibitem[GINÉ and NICKL, 2010]{GineNickl2010}
GINÉ, E. and NICKL, R. (2010).
\newblock Adaptive estimation of a distribution function and its density in
  sup-norm loss by wavelet and spline projections.
\newblock {\em Bernoulli}, 16(4):1137--1163.

\bibitem[Good and Gaskins, 1971]{goodd1971nonparametric}
Good, I. and Gaskins, R.~A. (1971).
\newblock Nonparametric roughness penalties for probability densities.
\newblock {\em Biometrika}, 58(2):255--277.

\bibitem[Gu, 1993]{Gu1993a}
Gu, C. (1993).
\newblock Smoothing spline density estimation: A dimensionless automatic
  algorithm.
\newblock {\em Journal of the American Statistical Association},
  88(422):495--504.

\bibitem[Gu, 2013]{gu2013smoothing}
Gu, C. (2013).
\newblock {\em Smoothing spline ANOVA models}, volume 297.
\newblock Springer.

\bibitem[Gu and Qiu, 1993]{Gu1993}
Gu, C. and Qiu, C. (1993).
\newblock Smoothing spline density estimation: Theory.
\newblock {\em The Annals of Statistics}, 21(1):217--234.

\bibitem[Huang, 1998]{huang1998projection}
Huang, J.~Z. (1998).
\newblock Projection estimation in multiple regression with application to
  functional anova models.
\newblock {\em The annals of statistics}, 26(1):242--272.

\bibitem[Huang, 2003]{huang2003asymptotics}
Huang, J.~Z. (2003).
\newblock Asymptotics for polynomial spline regression under weak conditions.
\newblock {\em Statistics \& probability letters}, 65(3):207--216.

\bibitem[Huang and Su, 2021]{huang2021asymptotic}
Huang, J.~Z. and Su, Y. (2021).
\newblock Asymptotic properties of penalized spline estimators in concave
  extended linear models: Rates of convergence.
\newblock {\em The Annals of Statistics}, 49(6):3383--3407.

\bibitem[Lai and Schumaker, 2007]{lai_schumaker_2007}
Lai, M.-J. and Schumaker, L.~L. (2007).
\newblock {\em Spline Functions on Triangulations}.
\newblock Encyclopedia of Mathematics and its Applications. Cambridge
  University Press.

\bibitem[Lai and Wang, 2013]{lai2013bivariate}
Lai, M.-J. and Wang, L. (2013).
\newblock Bivariate penalized splines for regression.
\newblock {\em Statistica Sinica}, pages 1399--1417.

\bibitem[Leonard, 1978]{leonard1978density}
Leonard, T. (1978).
\newblock Density estimation, stochastic processes and prior information.
\newblock {\em Journal of the Royal Statistical Society: Series B
  (Methodological)}, 40(2):113--132.

\bibitem[Lindgren et~al., 2011]{lindgren2011explicit}
Lindgren, F., Rue, H., and Lindstr{\"o}m, J. (2011).
\newblock An explicit link between gaussian fields and gaussian markov random
  fields: the stochastic partial differential equation approach.
\newblock {\em Journal of the Royal Statistical Society: Series B (Statistical
  Methodology)}, 73(4):423--498.

\bibitem[Loftsgaarden and Quesenberry, 1965]{Loftsgaarden:Quesenberry:65}
Loftsgaarden, D.~O. and Quesenberry, C.~P. (1965).
\newblock {A Nonparametric Estimate of a Multivariate Density Function}.
\newblock {\em The Annals of Mathematical Statistics}, 36(3):1049 -- 1051.

\bibitem[Parzen, 1962]{Parzen}
Parzen, E. (1962).
\newblock {On Estimation of a Probability Density Function and Mode}.
\newblock {\em The Annals of Mathematical Statistics}, 33(3):1065 -- 1076.

\bibitem[Ramsay, 2002]{Ramsay2002}
Ramsay, T. (2002).
\newblock Spline smoothing over difficult regions.
\newblock {\em Journal of the Royal Statistical Society. Series B (Statistical
  Methodology)}, 64(2):307--319.

\bibitem[Sain et~al., 1994]{sain1994cross}
Sain, S.~R., Baggerly, K.~A., and Scott, D.~W. (1994).
\newblock Cross-validation of multivariate densities.
\newblock {\em Journal of the American Statistical Association},
  89(427):807--817.

\bibitem[Scott, 2015]{scott2015multivariate}
Scott, D.~W. (2015).
\newblock {\em Multivariate density estimation: theory, practice, and
  visualization}.
\newblock John Wiley \& Sons.

\bibitem[Silverman, 1982]{Silverman1982}
Silverman, B.~W. (1982).
\newblock On the estimation of a probability density function by the maximum
  penalized likelihood method.
\newblock {\em The Annals of Statistics}, 10(3):795--810.

\bibitem[Stone, 1982]{stone1982optimal}
Stone, C.~J. (1982).
\newblock Optimal global rates of convergence for nonparametric regression.
\newblock {\em The annals of statistics}, pages 1040--1053.

\bibitem[Utreras, 1988]{utreras1988convergence}
Utreras, F.~I. (1988).
\newblock Convergence rates for multivariate smoothing spline functions.
\newblock {\em Journal of approximation theory}, 52(1):1--27.

\bibitem[Wand and Jones, 1994]{wand1994kernel}
Wand, M.~P. and Jones, M.~C. (1994).
\newblock {\em Kernel smoothing}.
\newblock CRC press.

\bibitem[Wang et~al., 2019]{BPST}
Wang, G., Wang, L., Lai, M.~J., Kim, M., Li, X., Mu, J., Wang, Y., and Yu, S.
  (2019).
\newblock Bpst: Bivariate spline over triangulation.
\newblock R package version 1.0. Available at
  \url{https://github.com/FIRST-Data-Lab/BPST}.

\bibitem[Weinberger, 1974]{weinberger1974variational}
Weinberger, H.~F. (1974).
\newblock {\em Variational methods for eigenvalue approximation}.
\newblock SIAM.

\bibitem[Wood et~al., 2008]{WoodBravingtonHedley2008}
Wood, S.~N., Bravington, M.~V., and Hedley, S.~L. (2008).
\newblock Soap film smoothing.
\newblock {\em Journal of the Royal Statistical Society: Series B (Statistical
  Methodology)}, 70(5):931--955.

\bibitem[Yu et~al., 2021]{yu_wang}
Yu, S., Wang, G., Wang, L., and Yang, L. (2021).
\newblock Multivariate spline estimation and inference for image-on-scalar
  regression.
\newblock {\em Statistica Sinica}, 31(3):1463--1487.

\end{thebibliography}

\newpage
\appendix
\section{Appendix}\label{Appendix}
 \renewcommand{\theequation}{\thesection.\arabic{equation}}
 
\subsection{Technical lemmas}\label{App1}

In this section, we introduce some technical lemmas that will be used to prove the main results.

\begin{lemma}[Theorem 10.10 of \cite{lai_schumaker_2007}]
\label{lemm1}
Suppose that Assumption \ref{assmp2} holds, and let $1 \leq p \leq \infty$. Then for every $h \in W_{p}^{m + 1}(\Omega)$ with $m \geq 3r + 2$, there exists a spline $s \in S_{m}^{r}(\triangle_n)$, such that
\[
\|D_{z_1}^{\alpha_1}D_{z_2}^{\alpha_2}(h - s)\|_{p, \Omega} \leq K |\triangle_n|^{m + 1 - \alpha_1 - \alpha_2} |h|_{m + 1, p, \Omega},
\]
for all $0 \leq \alpha_1 + \alpha_2 \leq m$, with the constant $K$ depending only on $r$, $m$ and the shape parameter $\beta$, if $\Omega$ is convex and also depends on the Lipschitz constant of the boundary of $\Omega$, when $\Omega$ is not convex.
\end{lemma}


\begin{lemma}[Lemma 1, Section S1 of \cite{lai2013bivariate}]
\label{lemm2}
Following the notation of Section \ref{SEC:estimation}, if we assume $\{B_h\}_{h \in \mathcal{H}}$ to be the set of basis functions for the spline space $\mathcal{S}^r_m(\triangle_n)$, then there exist constants $C_1, C_2 > 0$, depending on $m$ and $\beta$, such that
\begin{equation*}
C_1|\triangle_n|^2\sum\limits_{h \in \mathcal{H}}\gamma_h^2 \leq \left\|\sum\limits_{h \in \mathcal{H}}\gamma_h B_h\right\|^2_2 \leq C_2|\triangle_n|^2\sum\limits_{h \in \mathcal{H}}\gamma_h^2,
\end{equation*}
for all $\gamma_h, h \in \mathcal{H}$.
\end{lemma}

\paragraph{}The following lemma determines the uniform rate at which the empirical inner product (or, equivalently, the norm) approximates the theoretical inner product (or the norms).

\begin{lemma}[Lemma 2, Section S.1 of \cite{lai2013bivariate}]
\label{lemm3}
Under Assumption \ref{assmp5}, if $s_1 = \sum_{h \in \mathcal{H}}\gamma^*_h B_h$ and $s_2 = \sum_{h \in \mathcal{H}}\tilde{\gamma}_h B_h$ be any two spline functions from $\mathcal{S}^r_m(\triangle_n)$, then,
\[
\sup\limits_{\substack{s _1, s_2\in \mathcal{S}^r_m(\triangle_n)\\\|s_1\|_2, \|s_2\|_2 \neq 0}} \left|\frac{\langle s_1, s_2\rangle_n - \langle s_1, s_2\rangle_2}{\|s_1\|_2\|s_2\|_2}\right| = O_p\left(\sqrt{N_n \log(n)/n}\right).
\]
\end{lemma}

As a direct consequence of Lemma \ref{lemm3}, we have
\[
\sup\limits_{\substack{s \in \mathcal{S}^r_m(\triangle_n)\\\|s\|_2 \neq 0}} \left|\frac{\|s\|_n^2}{\|s\|^2_2} - 1\right| = O_p\left(\sqrt{N_n \log(n)/n}\right).
\]

\paragraph{}The following lemma from \cite{huang2021asymptotic} plays a pivotal role in establishing the two main theorems of this paper.

\begin{lemma}[Lemma 4.1 (Convexity Lemma) of \cite{huang2021asymptotic}]
\label{lemm4}
Consider a convex functional $\mathcal{Q}(\cdot)$ and a continuous functional $\mathcal{R}(\cdot)$, both defined on a convex set of functions, say $\widetilde{\mathcal{C}}$. Now for all functions $h \in \widetilde{\mathcal{C}}$, satisfying $\mathcal{R}(h) = a$(where, $a \in \mathbb{R}$), if there exists a function $h^* \in \widetilde{\mathcal{C}}$ with $\mathcal{R}(h^*) < a$, such that
$\mathcal{Q}(h^*) < \mathcal{Q}(h)$ or $\frac{\partial}{\partial \zeta} \mathcal{Q}(h^* + \zeta (h - h^*))\bigg\rvert_{\zeta = 1^{+}} > 0$,
then any minimizer $h_{min}$ of $\mathcal{Q}(\cdot)$ in $\widetilde{\mathcal{C}}$ satisfies $\mathcal{R}(h_{min}) \leq a$.
\end{lemma}

\begin{lemma}
\label{prop1}
Under Assumption \ref{assmp3}, that the true log density $g_0 \in W_{\infty}^{m + 1}(\Omega)$ and $m \geq 3r + 2$, there exists a function or, more precisely, a bivariate spline function $\tilde{s} \in \mathcal{S}^r_m(\triangle_n)$ and constants $C_1, C_2, C_3, C_4$, depending on $m, d$ and $g_0$ such that $\|\tilde{s} - g_0\|_{\infty} \leq C_1|\triangle_n|^{m + 1}|g_0|_{m + 1, \infty, \Omega}$
and $\|\tilde{s} - g_0\|_2 \leq C_3|\triangle_n|^{m + 1}$. Moreover, if $q \leq m$, then $\mathcal{E}_q(\tilde{s}) \leq C_4 |\triangle_n|^{2(m - q + 1)}$.
\end{lemma}

\begin{proof} Note that
\[
\|\tilde{s} - g_0\|_2^2 = \int_{\bs{x} \in \Omega} \{(\tilde{s} - g_0)(\bs{x})\}^2 \,d\bs{x} \leq \int_{\bs{x} \in \Omega} \{\sup\limits_{\bs{x}\in \Omega}(\tilde{s} - g_0)(\bs{x})\}^2\,d\bs{x} = \|\tilde{s} - g_0\|_{\infty}^2 |\Omega|.
\]
As a direct consequence of Lemma \ref{lemm1} and since $g_0 \in W^{m+1}_{\infty}$, substituting $\alpha_1 = \alpha_2 = 0$,
\[
\|\tilde{s} - g_0\|_{\infty} \leq C_1|\triangle_n|^{m + 1} |g_0|_{m + 1, \infty, \Omega} \leq C_2|\triangle_n|^{m + 1}. 
\]
Hence,
\begin{align*}
\|\tilde{s} - g_0\|_2 \leq |\Omega|^{1/2}\|\tilde{s} - g_0\|_{\infty} \leq C_2|\Omega|^{1/2}|\triangle_n|^{m + 1} \equiv C_3 |\triangle_n|^{m + 1}.
\end{align*}
Now, it is left to show that for $q \leq m,  \mathcal{E}_q(\tilde{s}) \leq C_4 |\triangle_n|^{2(m + 1 - q)}$. Note that
\begin{align*}
\mathcal{E}_q(\tilde{s}) &=  \sum\limits_{|\bs{\alpha}| = q}c_{\bs{\alpha}} \int_{\Omega}|D_{x_1}^{\alpha_1} D_{x_2}^{\alpha_2}(\tilde{s})|^2\,d\bs{x}= \sum\limits_{|\bs{\alpha}| = q}c_{\bs{\alpha}}\left\{\sum\limits_{T \in \triangle_n}\int_{T}\left|D_{x_1}^{\alpha_1} D_{x_2}^{\alpha_2}(\tilde{s})\right|^2\,d\bs{x}\right\}\\
&= \sum\limits_{|\bs{\alpha}| = q}c_{\bs{\alpha}}\left\{\sum\limits_{T \in \triangle_n}\int_{T}\left|D_{x_1}^{\alpha_1} D_{x_2}^{\alpha_2}(\tilde{s} - g_0 + g_0)\right|^2\,d\bs{x}\right\}\\
&= \sum\limits_{|\bs{\alpha}| = q}c_{\bs{\alpha}}\left\{\sum\limits_{T \in \triangle_n}\int_{T}\left|D_{x_1}^{\alpha_1} D_{x_2}^{\alpha_2}(\tilde{s} - g_0) + D_{x_1}^{\alpha_1} D_{x_2}^{\alpha_2}(g_0 - l) \right|^2\,d\bs{x}\right\},
\end{align*}
where $l \in \mathbb{P}_{m'}$, where $m' < q$, implying that $l \in \mathbb{P}_q$ as well, since $\mathbb{P}_{m'} \subseteq \mathbb{P}_q$. Note that $D_{x_1}^{\alpha_1} D_{x_2}^{\alpha_2}g_0$ is replaced by $D_{x_1}^{\alpha_1} D_{x_2}^{\alpha_2}(g_0 - l)$, since $D_{x_1}^{\alpha_1} D_{x_2}^{\alpha_2}l = 0$ as $\alpha_1 + \alpha_2 = q > m'$.

Then,
\begin{align}
\mathcal{E}_q(\tilde{s}) &= \sum\limits_{|\bs{\alpha}| = q}c_{\bs{\alpha}}\left\{\sum\limits_{T \in \triangle_n}\int_{T}\left|D_{x_1}^{\alpha_1} D_{x_2}^{\alpha_2}(\tilde{s} - g_0) + D_{x_1}^{\alpha_1} D_{x_2}^{\alpha_2}(g_0 - l) \right|^2\,d\bs{x}\right\} \notag\\
    &\leq \sum\limits_{|\bs{\alpha}| =q}2c_{\bs{\alpha}}\left[\sum\limits_{T \in \triangle_n}\int_{T}\left\{\left|D_{x_1}^{\alpha_1} D_{x_2}^{\alpha_2}(\tilde{s} - g_0)\right|^2 + \left|D_{x_1}^{\alpha_1} D_{x_2}^{\alpha_2}(g_0 - l) \right|^2\right\}\,d\bs{x}\right]  \notag\\
    &= \sum\limits_{|\bs{\alpha}| =q}2c_{\bs{\alpha}}\left[\sum\limits_{T \in \triangle_n}\left\{\|D_{x_1}^{\alpha_1} D_{x_2}^{\alpha_2}(\tilde{s} - g_0)\|_{2, T}^2 + \|D_{x_1}^{\alpha_1} D_{x_2}^{\alpha_2}(g_0 - l) \|_{2, T}^2\right\}\right] \notag\\
    &\leq C |\triangle_n|^{2(m+1-q)}|g_0|_{m+1, 2, \Omega} \hspace{2.6 in} \label{**} \\
    &\leq C_4 |\triangle_n|^{2(m+1-q)}, \notag
    \end{align}
where $C, C_4 > 0$ are constants and (\ref{**}) follows from the proofs and discussions of Theorem 5.18, Equation (5.18) and Theorem 5.19 of \cite{lai_schumaker_2007}. The last inequality holds because $g_0 \in W^{m+1}_{\infty}(\Omega)$ according to Assumption \ref{assmp3}.

\paragraph{}A brief support for the last line of inequality can be obtained as follows:

\paragraph{}From (\ref{**}), we have, $\mathcal{E}_q(\tilde{s}) \leq C |\triangle_n|^{2(m+1-q)}|g_0|_{m+1, 2, \Omega}$. Since $g_0 \in W_{\infty}^{m+1}(\Omega)$, , it implies: $g_0 \in \left\{g : \sum_{k = 0}^{m+1}|g|_{k, \infty, \Omega} < \infty\right\}$, which is equivalent to: 
\[
g_0 \in \left\{g: \sum_{k = 0}^{m+1}\left(\max\limits_{\alpha_1 + \alpha_2 = k}\|D_{\alpha_1}^{x_1}D_{\alpha_2}^{x_2}g\|_{\infty, \Omega}\right) < \infty\right\}.
\]
Hence,
\begin{align*}
    |g_0|_{m+1, 2, \Omega} &= \left(\sum\limits_{\alpha_1 + \alpha_2 = m+1}\|D_{\alpha_1}^{x_1}D_{\alpha_2}^{x_2}g_0\|^2_{2, \Omega}\right)^{1/2} \leq |\Omega|^{1/2} \left(\sum\limits_{\alpha_1 + \alpha_2 = m+1}\|D_{\alpha_1}^{x_1}D_{\alpha_2}^{x_2}g_0\|^2_{\infty, \Omega}\right)^{1/2}\\
    &\leq |\Omega|^{1/2} M_1 = M,
\end{align*}
since, $g_0 \in W^{m + 1}_{\infty}(\Omega)$, where $M_1, M > 0$ are constants. 
\end{proof}

\begin{lemma}
\label{prop2}
Given that Assumption \ref{assmp5} holds, under Assumption \ref{assmp2}, the empirical and theoretical norms are equivalent, i.e.,
\[
\sup\limits_{\substack{s \in \mathcal{S}^r_m(\triangle_n)\\ \|s\| \neq 0} }\left|\frac{\|s\|_n}{\|s\|} - 1\right| = o_p(1).
\]
\end{lemma}

\begin{proof}
Under Assumption \ref{assmp5}, it is evident that
\[
\lim_{n \xrightarrow{}\infty}\frac{N_n \log(n)}{n} =\lim_{n \xrightarrow{}\infty} \frac{Cn^{\eta}\log(n)}{n} = 0.
\]
Therefore, using Lemma \ref{lemm3}, we have
\[\sup\limits_{\substack{s \in \mathcal{S}^r_m(\triangle_n)\\\|s\|_2 \neq 0}} \left|\frac{\|s\|_n^2}{\|s\|^2_2} - 1\right| = O_p\left(\sqrt{N_n \log(n)/n}\right) = o_p(1),
\]
and from the equivalence of theoretical and $L_2$ norm under Assumption \ref{assmp4}, we have,
\[
\sup\limits_{\substack{s \in \mathcal{S}^r_m(\triangle_n)\\\|s\| \neq 0}} \left|\frac{\|s\|_n^2}{\|s\|^2} - 1\right| = O_p\left(\sqrt{N_n \log(n)/n}\right) = o_p(1).
\]
\end{proof}
\paragraph{}Now, suppose $V(h)$ is a quadratic function defining a metric such that $V(\hat{h} - h)$ indicates a good estimate $\hat{h}$ of $h$. The asymptotic analysis of the BPST estimator heavily depends on the eigenanalysis of the quadratic penalty functional $V(h) = \|h\|^2 = \int_{\bs{x} \in \Omega}h^2(\bs{x})\omega(\bs{x}) \,d\bs{x}$, where $\omega(\cdot)$ is a non-negative weight function bounded away from zero and infinity. 

\paragraph{}Following \cite{weinberger1974variational}, Chapter 3, a quadratic functional $Q_1$ is completely continuous concerning another quadratic functional $Q_2$ if, for any $\epsilon > 0$, there exists a finite set of linear functionals $L_1, L_2, \cdots, L_k$ such that $L_j(h) = 0, j = 1, \cdots, k$, implies $Q_1(h) \leq \epsilon Q_2(h)$. For an illustrative example, see Section 9.1 of \cite{gu2013smoothing}. 

\paragraph{}Theorem 3.1 on page 52 of \cite{weinberger1974variational} states that if $V$ is completely continuous concerning $\mathcal{E}_q$ and $\mathcal{E}_q + V$, there exists eigenvalues $e_\nu$ and corresponding eigenfunctions $\psi_\nu$ such that 
\[
V(\psi_\nu, \psi_\mu) = e_\nu \delta_{\nu, \mu} \quad \text{ and }\quad (\mathcal{E}_q + V)(\psi_\nu, \psi_\mu) = \delta_{\nu, \mu},
\]
with $1 \geq e_\nu \downarrow 0$ and $\delta_{\nu, \mu}$ being the Kronecker delta. If we define $\phi_\nu = e_\nu^{-1/2}\psi_\nu$, then $V$ and $\mathcal{E}_q$ can be diagonalized as
\[
V(\phi_\nu, \phi_\mu) = \delta_{\nu, \mu} \quad \text{ and }\quad \mathcal{E}_q(\phi_\nu, \phi_\mu) = \rho_\nu \delta_{\nu, \mu},
\]
where $0 \leq \rho_\nu = e_\nu^{-1} - 1 \uparrow \infty$, and $\rho_\nu$ and $\phi_\nu$ are the eigenvalues and eigenfunctions of $\mathcal{E}_q$ with respect to $V$, respectively. A function $h$ satisfying $\mathcal{E}_q(h) < \infty$ can be expressed as a Fourier series $h = \sum_\nu h_\nu \phi_\nu$, with $h_\nu = V(h, \phi_\nu)$ being the Fourier coefficients, leading to
\[
V(h) = \sum\limits_\nu h_\nu^2 \quad \text{ and }\quad \mathcal{E}_q(h) = \sum\limits_\nu\rho_\nu h_\nu^2,
\]
which implies
\[
\|h\|^2 + \lambda_n\mathcal{E}_q(h) = (V + \lambda_n \mathcal{E}_q)(h)= \sum\limits_\nu(1 + \lambda_n\rho_\nu)h_\nu^2.
\] 

\paragraph{}The following lemma provides the rate at which the eigenvalues of the quadratic functionals diverge to infinity.

\begin{lemma}\label{prop3}
Considering the form $V(h) = \|h\|^2 = \int_{\bs{x} \in \Omega}h^2(\bs{x})\omega(\bs{x})d\bs{x}$ for a weight function $\omega(\cdot)$ bounded away from zero and infinity, $V$ is completely continuous with respect to $\mathcal{E}_q$. Moreover, $\rho_\nu \uparrow \infty$ and $\rho_\nu \asymp \nu^{2q/d}$ for all sufficiently large $\nu$.
\end{lemma}

\begin{proof}
The result is directly derived from Theorem 5.3 in \cite{utreras1988convergence}.
\end{proof}

\begin{lemma}\label{prop4}
Suppose that there exists a positive constant $K$ such that $\rho_\nu \geq K\nu^{2q/d}$ $(q > d/2)$, for all large $\nu$. Then as $\lambda_n \xrightarrow{} 0$, with $n \xrightarrow{}\infty$, it holds that
\[
\sum_\nu \frac{1}{1 + \lambda_n \rho_\nu} = O(\lambda_n^{-1/(2q/d)}).
\]
\end{lemma}

\begin{proof}
The proof can be found in Lemma 9.1 of \cite{gu2013smoothing}.
\end{proof}

\subsection{Lemmas providing sufficient conditions for Assumptions  \ref{cond1}--\ref{cond3}}
\label{App3}

This section describes lemmas that provide sufficient conditions for Assumptions \ref{cond1}--\ref{cond3} in Section \ref{SEC:results}.

\begin{lemma}
[Sufficient condition for Assumption \ref{cond1}]
\label{lemm5}

Suppose Assumption \ref{assmp3} holds and $\|h_1\|_{\infty} \leq C$ for some constant $C > 0$. If there are constants $C_1, C_2, C_3 > 0$, such that
\[
-C_1\|h_2\|^2 \leq \frac{\,d^2}{\,d\alpha^2}\Lambda(h_1 + \alpha h_2) \leq -C_2 \|h_2\|^2, ~
0 \leq \alpha \leq 1,
\]
whenever $\|h_2\|_{\infty} \leq C_3$, then Assumption \ref{cond1} holds if $\|g_0\|_{\infty} \leq C$.
\end{lemma}

\begin{proof}
Consider a function $h$ with $\|h\|_{\infty}\leq C,$ for some constant $C > 0$. From Lemma \ref{prop1}, it is evident that $\tilde{s}$ would be the minimizer of the unpenalized log-likelihood $\Lambda(.)$. Therefore, a Taylor expansion of the function $\Lambda(g_0 + h) - \Lambda(g_0)$ at the minimal point $\tilde{s}$ would be,
$$\Lambda(g_0 + h) - \Lambda(g_0) \approx \Lambda(\tilde{s} + h) - \Lambda(\tilde{s}),$$since the first-order derivative term of the Taylor polynomial would be zero. Now since, $\tilde{s}$ minimizes $\Lambda(.)$, then 
$$\frac{\,d}{\,d\alpha}\Lambda\{(1 - \alpha)\tilde{s} + \alpha(\tilde{s} + h)\}\bigg\rvert_{\alpha=0} = 0.$$

\paragraph{}Now, considering an integration by parts,
\begin{align*}
    \Lambda(\tilde{s} + h) - \Lambda(\tilde{s}) &= \int_{0}^1 (1 - \alpha)\frac{\,d^2}{\,d\alpha^2}\Lambda\{(1 - \alpha)\tilde{s} + \alpha(\tilde{s} + h)\}\,d\alpha\\
    &= \int_{0}^1(1-\alpha) \frac{\,d^2}{\,d\alpha^2}\Lambda(h_1 + \alpha h_2)\,d\alpha
\end{align*}
where, $h_1 = \tilde{s}$ and $h_2 = h$.

\paragraph{}Note that, $\|h_1\|_{\infty} = \|\tilde{s}\|_{\infty} = \|(\tilde{s} + g_0 - g_0)\|_{\infty}\leq \|g_0\|_{\infty} + \|\tilde{s} - g_0\|_{\infty} \leq C^*$, for some constant $C^* > 0$, as from the statement of the Lemma we have $\|g_0\|_{\infty} \leq C$ and from Lemma \ref{prop1}, $\|\tilde{s} - g_0\|_{\infty}\leq C_2 |\triangle_n|^{m+1}\equiv C_4$, with $C_4$ being a constant greater than $0$, for large $n$ and under Assumption \ref{assmp2}. In addition, $\|h_2\|_{\infty} = \|h\|_{\infty} \leq C$ for constant $C > 0$.

\paragraph{}Hence it directly follows from the statement of the Lemma that,
$$C_1\|h\|^2 \leq \{\Lambda(\tilde{s}+h) - \Lambda(\tilde{s})\} = \int_{0}^1 (1 - \alpha)\frac{\,d^2}{\,d\alpha^2}\Lambda(h_1 + \alpha h_2)\,d\alpha \approx \{\Lambda(g_0+h) - \Lambda(g_0)\} \leq C_2\|h\|^2.$$
\end{proof}

\begin{lemma}
[Sufficient condition for Assumption \ref{cond2}] 
\label{lemm6}

If $\mathrm{Var}[L^{\prime}(\tilde{g}_n; h)] \leq C$, for some constant $C > 0$ and any $h$ with $\|h\|^2 = 1$, then Assumption \ref{cond2} holds.
\end{lemma}

\begin{proof}
Given a set of basis functions $\left\{B_1, \cdots, B_H\right\}$ for $\mathcal{S}_m^r(\triangle_n)$, with $H = {N_n(m+1)(m+2)}/2$, an orthonormal basis can be constructed as say $\left\{\xi_k : k = 1, \cdots, H\right\}$ using Gram-Schimdt orthogonalization.

\paragraph{}Note that any $s \in \mathcal{S}_m^r(\triangle_n)$ can be written as $s = \sum_{k = 1}^H s_k\xi_k$, with $s_k = \langle s, \xi_k \rangle$. From the definition, $L^{\prime}(\tilde{g}_n; s) = \sum_k s_kL^{\prime}(\tilde{g}_n; \xi_k)$.
Then, we have
\begin{align}\label{A.1}
\frac{|({\rm E}_n - {\rm E})L^{\prime}(\tilde{g}_n; s)|^2}{\|s\|^2 + \lambda_n\mathcal{E}_q(s)} &\leq \frac{|({\rm E}_n - {\rm E})L^{\prime}(\tilde{g}_n; s)|^2}{\|s\|^2}  = \sum\limits_k s_k^2\nonumber = \sum\limits_k \left|({\rm E}_n - {\rm E})L^{\prime}(\tilde{g}_n; \xi_k)\right|^2\nonumber\\
&\leq \sum\limits_k \frac{C}{n} \asymp \frac{C_1}{n|\triangle_n|^2} = O_p\left(\frac{1}{n|\triangle_n|^2}\right).
\end{align}

\paragraph{}Now consider the eigenanalysis from discussions related to Lemmas \ref{prop3} and \ref{prop4}. Since $V(s) = \|s\|^2$ is completely continuous with respect to $\mathcal{E}_q(s)$,it implies an eigen decomposition of $s$ as $s = \sum_{\nu} s_{\nu}\phi_{\nu}$ and $L^{\prime}(\tilde{g}_n; s) = \sum_\nu s_\nu L^{\prime}(\tilde{g}_n; \phi_\nu)$. Then, using $\|s\|^2 = \sum_\nu s_\nu^2$, and the diagonalised form of $\mathcal{E}_q(S)$, along with Cauchy-Schwartz inequality, 
\begin{align}\label{A.2}
\frac{|({\rm E}_n - {\rm E})L^{\prime}(\tilde{g}_n; s)|^2}{\|s\|^2 + \lambda_n\mathcal{E}_q(s)} &= \frac{|({\rm E}_n - {\rm E})\sum\limits_{\nu}s_{\nu}L'(\tilde{g}_n; \phi_{\nu})|^2}{\sum\limits_{\nu} s_\nu^2 (1+ \lambda_n \rho_\nu)}= \frac{|\sum\limits_{\nu}s_{\nu}({\rm E}_n - {\rm E})L'(\tilde{g}_n; \phi_{\nu})|^2}{\sum\limits_{\nu} s_\nu^2 (1+ \lambda_n \rho_\nu)}\nonumber\\
&\leq \frac{\sum\limits_\nu s_\nu^2\sum\limits_\nu \left|({\rm E}_n - {\rm E})L^{\prime}(\tilde{g}_n; \phi_\nu)\right|^2}{\sum\limits_{\nu} s_\nu^2 (1+ \lambda_n \rho_\nu)}\nonumber\\
&\leq  \frac{(C/n)\sum\limits_\nu s_\nu^2}{\sum\limits_k s_\nu^2 (1+ \lambda_n \rho_\nu)} = \frac{C}{n}\sum_\nu\frac{s_\nu^2}{\sum\limits_\nu s_\nu^2 (1+ \lambda_n \rho_\nu)} \leq \frac{C}{n}\sum_\nu\frac{s_\nu^2}{s_\nu^2 (1+ \lambda_n \rho_\nu)}\nonumber\\
&= \frac{C}{n}\sum_\nu\frac{1}{1+ \lambda_n \rho_\nu} =O_p\left(\frac{1}{n\lambda_n^{d/2q}}\right).
\end{align}
Combining both \eqref{A.1} and \eqref{A.2}, it follows that
\[
\sup\limits_{s \in \mathcal{S}^r_m(\triangle_n)}\frac{|({\rm E}_n - {\rm E})L^{\prime}(\tilde{g}_n; s)|^2}{\|s\|^2 + \lambda_n\mathcal{E}_q(s)} = O_p\left(\frac{1}{n|\triangle_n|^d} \wedge \frac{1}{n\lambda_n^{d/2q}}\right).
\]
\end{proof}

\begin{lemma}
[Sufficient condition for Assumption \ref{cond3}]
\label{lemm7}
Suppose Assumptions \ref{assmp3} hold, then we have
\begin{enumerate}[label=\normalfont(\roman*)]
\item $\|\tilde{g}_n\|_{\infty} = O(1),$
\item $L(\tilde{g}_n + \alpha s)$, as a function of $\alpha$(for $ s \in \mathcal{S}^r_m(\triangle_n)$), is twice continuously differentiable and in addition to that there exist constants $C_1, C_2 > 0$ such that 
\[
\frac{\, d^2}{\,d \alpha^2}L(\tilde{g}_n + \alpha s) \geq C_2\|s\|^2, ~ 0 \leq \alpha \leq 1
\]
holds for $s \in \mathcal{S}^r_m(\triangle_n)$, with $\|s\|_{\infty} \leq C_1$, with probability tending to one as $n \xrightarrow{} \infty$.  
\end{enumerate}
\end{lemma}

\begin{proof}
Part (i) follows directly from Theorem \ref{theo1}, and what remains is to prove Part (ii). Note that
\[
\frac{\,d}{\,d\alpha}L(\tilde{g}_n + \alpha s)\bigg\rvert_{\alpha = 1^{+}} - \frac{\,d}{\,d\alpha}L(\tilde{g}_n + \alpha s)\bigg\rvert_{\alpha = 0^{+}} = \int_{0}^1 \frac{\,d^2}{\,d\alpha^2}L(\tilde{g}_n + \alpha s)\,d\alpha.
\]

\paragraph{}Given the fact that for $s \in \mathcal{S}_m^r(\triangle_n), \|s\|_{\infty} \leq \|s - g_0\|_{\infty} + \|g_0\|_{\infty} \leq C_1$(for large $n$), for some constant $C_1 > 0$, [from Lemma \ref{prop1} and since $g_0 \in W^m_{\infty}(\Omega)$], the result follows from Lemma \ref{lemm7}, i.e.,
\[
\frac{\,d}{\,d\alpha}L(\tilde{g}_n + \alpha s)\bigg\rvert_{\alpha = 1^{+}} - \frac{\,d}{\,d\alpha}L(\tilde{g}_n + \alpha s)\bigg\rvert_{\alpha = 0^{+}} = \int_{0}^1 \frac{\,d^2}{\,d\alpha^2}L(\tilde{g}_n + \alpha s)\,d\alpha \geq C_2 \|s\|^2,
\]
for $0 \leq \alpha \leq 1$.
\end{proof}

\subsection{Validation of the assumptions}
\label{App4}

\subsubsection{Verification of Assumption \ref{cond1}}
\label{sssec:verif_cond1}

Note that
\begin{align*}
\frac{\,d^2}{\,d\alpha^2}\Lambda(h_1 + \alpha h_2) &= \frac{\,d^2}{\,d\alpha^2}{\rm E}[L(h_1 + \alpha h_2; X_1, \cdots, X_n)]\\
&=  \frac{\,d^2}{\,d\alpha^2} {\rm E}\left\{-\frac{1}{n}\sum\limits_{i=1}^{n}(h_1 + \alpha h_2)(X_i) + \int_{\bs{x} \in \Omega} \exp{(h_1 + \alpha h_2)(\bs{x})}\,d\bs{x}\right\}\\
&= \int_{\bs{x} \in \Omega} \{h_2(\bs{x})\}^2 \exp\{(h_1 + \alpha h_2)(\bs{x})\}\,d\bs{x}\\
&= \mathrm{Var}\{h_2(X_{\alpha})\},
\end{align*}
where $X_{\alpha}$ is a random variable having log density $g_{X_{\alpha}}(\bs{x}) = (h_1 + \alpha h_2)(\bs{x}), \bs{x} \in \Omega$. Thus, it is now enough to establish that $C_1\|h_2\|^2 \leq \mathrm{Var}(h_2(X_{\alpha})) \leq C_2 \|h_2\|^2$, for $0 \leq \alpha \leq 1$, to satisfy Assumption \ref{cond1} through Lemma \ref{lemm5}. Now, $\|h_1 + \alpha h_2\|_{\infty} \leq C_4$, for some positive constant $C_4$, since from Lemma \ref{lemm5}, $\|h_1\|_{\infty} \leq C$ and $\|h_2\|_{\infty} \leq C_3$. This implies that $g_{X_{\alpha}}(\cdot)$ is also bounded on $\Omega$, i.e. there exist some constants $c_1, c_2 \in \mathbb{R}$, such that $c_1 \leq g_{X_{\alpha}}(\bs{x}) \leq c_2$, for all $\bs{x} \in \Omega$. Then, we have,
\begin{align*}
&\mathrm{Var}(h_2(X_{\alpha})) = \int_{\bs{x} \in \Omega} \{h_2(\bs{x})\}^2 \exp(g_{X_{\alpha}}(\bs{x}))\,d\bs{x},
\end{align*}
which implies that $\exp(c_1) \int_{\bs{x} \in \Omega}\{h_2(\bs{x})\}^2\,d\bs{x} \leq \mathrm{Var}\{h_2(X_{\alpha})\} \leq \exp(c_2) \int_{\bs{x} \in \Omega}\{h_2(\bs{x})\}^2\,d\bs{x}$ and $C_1 \|h_2\|_2^2 \leq \mathrm{Var}\{h_2(X_{\alpha})\} \leq C_2 \|h_2\|_2^2$. This verifies that Assumption \ref{cond1} is satisfied by $\Lambda(\cdot)$, as under Assumption \ref{assmp4}, the theoretical norm $\|.\|$ and the $L_2$ norm $\|.\|_2$ are equivalent.

\subsubsection{Verification of Assumption \ref{cond1}}
\label{sssec:verif_cond2}

Note that 
\begin{align*}
L^{\prime}[(\tilde{g}_n; s)(\bs{x})] &= \frac{\,d}{\,d\alpha}L[(\tilde{g}_n + \alpha s)(\bs{x})]\\
&= \frac{\,d}{\,d\alpha} \left[-\tilde{g}_n(\bs{x}) - \alpha s(\bs{x}) + \int_{\bs{x} \in \Omega} \exp(\tilde{g}_n(\bs{x})) \exp\left(\alpha  s(\bs{x})\right)\right]\bigg\rvert_{\alpha = 0^{+}}\\
&= -s(\bs{x}) + {\rm E}_{\tilde{g}_n}(s(\bs{X})),
\end{align*}
where $\bs{X}$ has density $\exp(\tilde{g}_n(\bs{x}))$. Then, $\mathrm{Var}(L^{\prime}[(\tilde{g}_n; s)(\bs{x})]) = \mathrm{Var}(s(\bs{x})) \leq \|s\|^2$, from the definition of theoretical norm $\|.\|^2$. Hence, Assumption \ref{cond2} holds for the present scenario through Lemma \ref{lemm6}. 

\subsubsection{Verification of Assumption \ref{cond3}}\label{sssec:verif_cond3}

To verify the Assumption \ref{cond3}, note that
\begin{align*}
\frac{\,d^2}{\,d \alpha^2} & L[(\tilde{g}_n + \alpha s)(\bs{x_1}, \cdots, \bs{x}_n)] \\
&= \frac{\,d^2}{\,d \alpha^2}\left[-\frac{1}{n}\sum\limits_{i = 1}^{n}\tilde{g}_n(\bs{X}_i) - \frac{1}{n}\sum\limits_{i = 1}^{n}\alpha s(\bs{X}_i) + \int_{\bs{x} \in \Omega} \exp((\tilde{g}_n + \alpha s)(\bs{x}))\,d\bs{x}\right]\\
&= \frac{\,d^2}{\,d \alpha^2}\left[{\rm E}\left\{-\frac{1}{n}\sum\limits_{i = 1}^{n}\tilde{g}_n(\bs{X}_i) -\frac{1}{n}\sum\limits_{i = 1}^{n}\alpha s(\bs{X}_i)\right\} + \int_{\bs{x} \in \Omega} \exp\left\{(\tilde{g}_n + \alpha s)(\bs{x})\right\}\,d\bs{x}\right]\\
&= \frac{\,d^2}{\,d \alpha^2} \Lambda(\tilde{g}_n + \alpha s) = \mathrm{Var}\{s(\bs{X}'_{\alpha})\},
\end{align*}
where $\bs{X}'_{\alpha}$ is a random variable having log density $g_{\bs{X}'_{\alpha}} = (\tilde{g}_n + \alpha s)(\bs{x}), \bs{x} \in \Omega$. Following the arguments in Section \ref{sssec:verif_cond1}, we have
$C_1\|s\|^2 \leq \mathrm{Var}(s(\bs{X}'_{\alpha})) \leq C_2\|s\|^2$, for some constants $C_1, C_2 > 0$, which implies,
\[
\frac{\,d^2}{\,d \alpha^2}L[(\tilde{g}_n + \alpha s)(\bs{x_1}, \cdots, \bs{x}_n)] = \mathrm{Var}(s(\bs{X}'_{\alpha})) \geq C_1 \|s\|^2.
\]
Therefore, Assumption \ref{cond3} is verified through Lemma \ref{lemm7}.

\subsection{Proof of Theorems \ref{theo1} and \ref{theo2}}
\label{App5}

The proofs for Theorems \ref{theo1} and \ref{theo2} follow directions and analogies from the proof of master theorems in Section 4 of \cite{huang2021asymptotic}. However, in the present scenario, the generalization to the 2D BPST method can be considered as an extension of the univariate scenarios from \cite{huang2021asymptotic}.

\begin{proof}[\unskip\nopunct]\paragraph{Proof of Theorem \ref{theo1}}
Note that, for $\tilde{s} \in \mathcal{S}_m^r(\triangle_n)$ from Lemma \ref{prop1},
\begin{equation}
\label{A.3}
\|\tilde{s} - g_0\| + \lambda_n^{1/2}\mathcal{E}_q^{1/2}(\tilde{s}) \leq C_2|\triangle_n|^{m + 1} + C_3^{1/2}\lambda_n^{1/2}|\triangle_n|^{(m + 1 - q)}.
\end{equation}
To apply the Convexity Lemma (Lemma \ref{lemm4}) in the present scenario, consider a convex functional $\mathcal{Q}(h) = \Lambda(h) + \lambda_n \mathcal{E}_q(h)$, and the continuous function as
$\mathcal{R}(h) = \|\tilde{s} - h\| + \lambda_n^{1/2}\mathcal{E}_q^{1/2}(\tilde{s} - h)$. Defined on $\widetilde{\mathcal{C}} = \mathcal{S}_m^r(\triangle_n)$. The convexity of $\mathcal{Q}(\cdot)$ follows from the convexity of $\Lambda(\cdot) = {\rm E}[L(\cdot)]$ and the penalty functional itself. The continuity of $\mathcal{R}(\cdot)$ follows from the reverse triangle inequality as
$|\mathcal{R}(h_1) - \mathcal{R}(h_2)| \leq \|h_1 - h_2\| + \lambda_n^{1/2}\mathcal{E}_q^{1/2}(h_1 - h_2).$

\paragraph{}For applying the Lemma, $a \in \mathbb{R}$ can be taken as $a = c(|\triangle_n|^{m + 1} + \lambda_n^{1/2}|\triangle_n|^{m + 1 -q})$, with $c > 0$ being a constant. Now in Lemma \ref{lemm4}, consider $h^* = \tilde{s}$ and then $\mathcal{R}(\tilde{s}) = 0 < a$. Now if it can be shown that $\mathcal{Q}(\tilde{s}) < \mathcal{Q}(h)$, with $h \in \mathcal{S}_m^r(\triangle_n)$ and $\mathcal{R}(h) = a$, then Lemma \ref{lemm4} implies that $\tilde{g}_n$, the minimizer of $\mathcal{Q}(h)$ over $\mathcal{S}_m^r(\triangle_n)$, satisfies $\mathcal{R}(\tilde{g}_n) < a$. 

Given that, as a direct consequence,
\begin{align*}
\|\tilde{g}_n - g_0\| + \lambda_n^{1/2}\mathcal{E}_q^{1/2}(\tilde{g}_n) &\leq \|\tilde{g}_n - \tilde{s}\| + \lambda_n^{1/2}\mathcal{E}_q^{1/2}(\tilde{g}_n - \tilde{s}) + \|\tilde{s} - g_0\| + \lambda_n^{1/2}\mathcal{E}_q^{1/2}(\tilde{s})\\
&=\mathcal{R}(\tilde{g}_n) + \|\tilde{s} - g_0\| + \lambda_n^{1/2}\mathcal{E}_q^{1/2}(\tilde{s})\\
&\leq a + C_1 |\triangle_n|^{m+1} + C_3^{1/2}\lambda_n^{1/2}|\triangle_n|^{(m+1 - q)},
\end{align*}
where the last inequality is a result of \eqref{A.3}. To bound the terms involved in computations related to $a$, such as $(|\triangle_n|^{m+1} + \lambda_n^{1/2}|\triangle_n|^{m + 1 - q})$ and $|\triangle_n|^{2(m+1)} + \lambda_n|\triangle_n|^{2(m + 1 - q)}$, by each other, the following inequality will be of extreme use 
\renewcommand{\theequation}{\thesection.\arabic{equation}}
\begin{equation}\label{A.4}
\frac{1}{2}(w_1 + w_2)^2 < w_1^2 + w_2^2 < (w_1 + w_2)^2, \hspace{0.3 in} w_1, w_2 > 0
\end{equation}

\paragraph{}Now, using \eqref{A.4}, 
\begin{align*}
\|\tilde{g}_n - g_0\| + \lambda_n^{1/2}\mathcal{E}_q^{1/2}(\tilde{g}_n) &\leq c(|\triangle_n|^{m+1} + \lambda_n^{1/2}|\triangle_n|^{m + 1 -q}) + C_1 |\triangle_n|^{m + 1} + C_3^{1/2}\lambda_n^{1/2}|\triangle_n|^{(m + 1 - q)}\\
&= O_p\left(|\triangle_n|^{m+1} \vee \lambda_n |\triangle_n|^{m + 1 - q}\right),
\end{align*}
which implies $\|\tilde{g}_n - g_0\|^2 + \lambda_n \mathcal{E}_q(\tilde{g}_n) \leq O_p\left(|\triangle_n|^{2(m+1)} \vee \lambda_n^{1/2} |\triangle_n|^{2(m + 1 - q)}\right),$ and hence the desired result will be proved.

\paragraph{}What remains to prove is that $\mathcal{Q}(\tilde{s}) < \mathcal{Q}(h)$, given $\mathcal{R}(\tilde{s}) = 0 < a$,  and having $\mathcal{R}(h) = a$ for $h \in \mathcal{S}^r_m(\triangle_n)$.

\paragraph{}Note that, for any $h \in \mathcal{S}^r_m(\triangle_n)$ with $\mathcal{R}(h) \leq a$, using the form of $\mathcal{R}(.)$, 
\begin{align}
\label{A.5}
\|h - \tilde{s}\|_{\infty} &\leq U_n \|h - \tilde{s}\| \leq U_n \mathcal{R}(h) \leq U_n a,
\end{align}

\paragraph{}Also, we have
\begin{align}\label{A.6}
\|h - g_0\|_{\infty} &\leq \|h - \tilde{s}\|_{\infty} + \|\tilde{s} - g_0\|_{\infty} \leq U_n a + C_2 |\triangle_n|^{m + 1} = o(1), \hspace{0.2 in} \text{as} \hspace{0.05 in}n \xrightarrow{}\infty,
\end{align}
since, for large $n$, using Lemma \ref{prop1} and the conditions on Theorem \ref{theo1} regarding the joint behavior of $U_n, \lambda_n$, and $|\triangle_n|$ values, $\|h - g_0\|_{\infty} \leq C_6$, for some constant $C_6 > 0$.

\paragraph{}Now, under Assumption \ref{cond1}, we can obtain that
\renewcommand{\theequation}{\thesection.\arabic{equation}}
\begin{align}\label{A.7}
\mathcal{Q}(h) - \Lambda(g_0) &= \Lambda(h)
- \Lambda(g_0) + \lambda_n\mathcal{E}_q(h)\nonumber\\
&\geq C_1 \|h - g_0\|^2 + \lambda_n\mathcal{E}_q(h)\\
&\geq \frac{1}{2}(C_1\wedge 1) \left\{\|h - g_0\| + \lambda_n^{1/2}\mathcal{E}^{1/2}_q(h)\right\}^2,\nonumber
\end{align}
and
\begin{align}\label{A.8}
\mathcal{Q}(\tilde{s}) - \Lambda(g_0) &= \Lambda(\tilde{s}) - \Lambda(g_0) + \lambda_n \mathcal{E}_q(\tilde{s})\nonumber\\
&\leq C_2 \|\tilde{s} - g_0\|^2 + \lambda_n\mathcal{E}_q(\tilde{s})\\
&\leq \frac{1}{2}(C_2 \vee 1) \left\{\|\tilde{s} - g_0\| + \lambda_n^{1/2}\mathcal{E}^{1/2}_q(\tilde{s})\right\}^2. \nonumber
\end{align}

\paragraph{}For $h \in \mathcal{S}^r_m(\triangle_n)$ with $\mathcal{R}(h) = a$, using (\ref{A.3}), we have
\begin{align*}
a = c(|\triangle_n|^{m+1} + \lambda_n^{1/2}|\triangle_n|^{m + 1 -q}) &= \|h - \tilde{s}\| + \lambda_n^{1/2}\mathcal{E}^{1/2}_q(h - \tilde{s}) = \mathcal{R}(h)\\
&\leq\|\tilde{s} - g_0\| + \|h - g_0\| +\\
& \hspace{0.2 in}\lambda_n^{1/2}\mathcal{E}_q^{1/2}(h) + \lambda_n^{1/2}\mathcal{E}_q^{1/2}(\tilde{s})\\
&\leq \|h - g_0\| + \lambda_n^{1/2}\mathcal{E}_q^{1/2}(h) + \\
&\hspace{0.2 in}C_2|\triangle_n|^{m + 1} + C_3^{1/2}\lambda_n^{1/2}|\triangle_n|^{(m + 1 - q)}.
\end{align*}

\paragraph{}Therefore, we have 
\[
\left\{(c - C_2)|\triangle_n|^{m + 1} + (c - C_3^{1/2})\lambda_n^{1/2}|\triangle_n|^{m + 1 -q}\right\} \leq  \|h - g_0\| + \lambda_n^{1/2}\mathcal{E}_q^{1/2}(h).
\]

\paragraph{}By considering the constant $c$ to be large enough (to obtain a strict inequality), using the bound of $\|\tilde{s} - g_0\| + \lambda_n^{1/2}\mathcal{E}^{1/2}_q(\tilde{s})$ from \eqref{A.3}, we have
\[
\frac{1}{2}(C_2 \vee 1)\left\{\|\tilde{s} - g_0\| + \lambda_n^{1/2}\mathcal{E}^{1/2}_q(\tilde{s})\right\}^2 < \frac{1}{2}(C_1\wedge1) \left\{\|h - g_0\| + \lambda_n^{1/2}\mathcal{E}^{1/2}_q(h)\right\}^2,
\]
that is $\mathcal{Q}(\tilde{s}) < \mathcal{Q}(h)$ considering the forms in the RHS of \eqref{A.7} and \eqref{A.8}. Hence, the proof for this part is complete.

\paragraph{}Now, from \eqref{A.6},
\[
\|h\|_{\infty}\leq \|h - g_0\|_{\infty} + \|g_0\|_{\infty} < C_7\|g_0\|_{\infty},
\] 
for some constant $C_7 > 0$ and large $n$, and note that it holds for any $h \in \mathcal{S}^r_m(\triangle_n)$ with $\mathcal{R}(h) \leq a$. Since $\mathcal{R}(\tilde{g}_n) = 0 < a$, it is evident that
\[
\|\tilde{g}_n\|_{\infty} \leq C_7 \|g_0\|_{\infty} < \infty, ~\text{and}~
\|\tilde{g}_n - g_0\|_{\infty} \leq \|\tilde{g}_n\|_{\infty} + \|g_0\|_{\infty} < \infty,
\]
that is, $\|\tilde{g}_n\|_{\infty} = o(1)$ and $\|\tilde{g}_n - g_0\|_{\infty} = o(1)$. Hence, the proof of Theorem \ref{theo1} is now complete.
\end{proof}

\begin{proof}[\unskip\nopunct]\paragraph{Proof of Theorem \ref{theo2}} 
Analogous to the previous theorem, let the convex and continuous functional defined on $\widetilde{\mathcal{C}} = \mathcal{S}^r_m(\triangle_n)$ be $\mathcal{Q}(h) = L(h) + \lambda_n\mathcal{E}_q(h)$, and $\mathcal{R}(h) = \|h - \tilde{g}_n\| + \lambda_n^{1/2}\mathcal{E}^{1/2}_q(h - \tilde{g}_n)$, respectively. The constant $a \in \mathbb{R}$ can be taken as
\[
a^2 = c^2\left(\frac{1}{n|\triangle_n|^d} \wedge \frac{1}{n\lambda_n^{d/2q}}\right),
\]
with $c > 0$ being a constant. Now in Lemma \ref{lemm4}, consider $h^* = \tilde{g}_n$, which results in $\mathcal{R}(\tilde{g}_n) = 0 < a$. Thus, if it can be shown that 
\begin{equation}\label{A.9}
\frac{\partial}{\partial \alpha}\mathcal{Q}(\tilde{g}_n + \alpha (h - 
\tilde{g}_n))\bigg\rvert_{\alpha = 1^{+}} > 0,
\end{equation}
for $h \in \mathcal{S}^r_m(\triangle_n)$ with $\mathcal{R}(h) = a$, then Lemma \ref{lemm4} implies that $\hat{g}_n$, the minimizer of $\mathcal{Q}(h)$ over the spline space $\mathcal{S}^r_m(\triangle_n)$, satisfies $\mathcal{R}(\hat{g}_n) \leq a$, i.e.,
\[
\|\hat{g}_n - \tilde{g}_n\|^2 + \lambda_n\mathcal{E}_q(\hat{g}_n - \tilde{g}_n)\leq a^2 = c^2\left(\frac{1}{n|\triangle_n|^d} \wedge \frac{1}{n\lambda_n^{d/2q}}\right),
\]
and Theorem \ref{theo2} can be proved.

\paragraph{}However, it still remains to prove \eqref{A.9}. Note that
\begin{align*}
&\frac{\partial}{\partial \alpha} \mathcal{Q}(\tilde{g}_n + \alpha (h - 
\tilde{g}_n))\bigg\rvert_{\alpha = 1^{+}} \\
&= \frac{\partial}{\partial \alpha}\left\{L(\tilde{g}_n + \alpha (h - 
\tilde{g}_n)) + \lambda_n \mathcal{E}_q(\tilde{g}_n + \alpha (h - 
\tilde{g}_n))\right\}\bigg\rvert_{\alpha = 1^{+}}\\
&= \frac{\partial}{\partial \alpha}L(\tilde{g}_n + \alpha (h - 
\tilde{g}_n))\bigg\rvert_{\alpha = 1^{+}} + \lambda_n \frac{\partial}{\partial \alpha}\mathcal{E}_q(\tilde{g}_n + \alpha (h - 
\tilde{g}_n))\bigg\rvert_{\alpha = 1^{+}}\\
&= \frac{\partial}{\partial \alpha}L(\tilde{g}_n + \alpha (h - 
\tilde{g}_n))\bigg\rvert_{\alpha = 1^{+}} +\\
&\hspace{0.2 in} \lambda_n \frac{\partial}{\partial \alpha}\left\{\mathcal{E}_q(\tilde{g}_n) + 2\alpha\mathcal{E}_q(\tilde{g}_n, h - \tilde{g}_n) + \alpha^2 \mathcal{E}_q(h - \tilde{g}_n)\right\}\bigg\rvert_{\alpha = 1^{+}}\\
&= \frac{\partial}{\partial \alpha}L(\tilde{g}_n + \alpha (h - 
\tilde{g}_n))\bigg\rvert_{\alpha = 1^{+}} + 2\lambda_n\mathcal{E}_q(\tilde{g}_n, h - \tilde{g}_n) +\\
&\hspace{0.2 in} 2\lambda_n\mathcal{E}_q(h - \tilde{g}_n) + \frac{\partial}{\partial \alpha}L(\tilde{g}_n + \alpha (h - 
\tilde{g}_n))\bigg\rvert_{\alpha = 0^{+}} - \frac{\partial}{\partial \alpha}L(\tilde{g}_n + \alpha (h - 
\tilde{g}_n))\bigg\rvert_{\alpha = 0^{+}}\\
&\equiv A + B
\end{align*}
with
\[
A = \frac{\partial}{\partial \alpha}L(\tilde{g}_n + \alpha (h - \tilde{g}_n))\bigg\rvert_{\alpha = 0^{+}} + 2\lambda_n\mathcal{E}_q(\tilde{g}_n, h - \tilde{g}_n),
\]
and 
\[
B = 2\lambda_n\mathcal{E}_q(h - \tilde{g}_n) -\frac{\partial}{\partial \alpha}L(\tilde{g}_n + \alpha (h - \tilde{g}_n))\bigg\rvert_{\alpha = 0^{+}} + \frac{\partial}{\partial \alpha}L(\tilde{g}_n + \alpha (h - \tilde{g}_n))\bigg\rvert_{\alpha = 1^{+}}.
\]

\paragraph{}Note that $A = ({\rm E}_n - {\rm E})L^{\prime}(\tilde{g}_n, h - \tilde{g}_n)$, from the definition of $L^{\prime}(\cdot)$. Then using Assumption \ref{cond2}, for any $h \in \mathcal{S}^r_m(\triangle_n)$ with $\mathcal{R}(h) \leq a$
\begin{align*}
|A| &= \left\{\|h - \tilde{g}_n\| + \lambda_n^{1/2}\mathcal{E}^{1/2}_q(h - \tilde{g}_n)\right\}^{1/2} \times O_p\left\{\left(\frac{1}{n|\triangle_n|^d} \wedge \frac{1}{n\lambda_n^{d/2q}}\right)^{1/2}\right\}\\
&\leq a\hspace{0.03 in} O_p\left(\frac{a}{c}\right)=O_p\left(\frac{a^2}{c}\right).
\end{align*}

\paragraph{}And, from the definition of $U_n$,
\[\|h - \tilde{g}_n\|_{\infty} \leq U_n\|h - \tilde{g}_n\| \leq U_n \mathcal{R}(h) \leq U_n a = U_n \times c \times O_p\left\{\left(\frac{1}{n|\triangle_n|^d} \wedge \frac{1}{n\lambda_n^{d/2q}}\right)^{1/2}\right\} = o(1),\]
for large $n$, that is, $\|h - \tilde{g}_n\|_{\infty}\leq C_8$, for some constant $C_8 > 0$. 
  
\paragraph{}Then, for any $h \in \mathcal{S}^r_m(\triangle_n)$ with $\mathcal{R}(h) = a$, using Assumption \ref{cond3},
\begin{align*}
B &= 2\lambda_n\mathcal{E}_q(h - \tilde{g}_n) - \frac{\partial}{\partial \alpha}L(\tilde{g}_n + \alpha (h - 
\tilde{g}_n))\bigg\rvert_{\alpha = 0^{+}} + \frac{\partial}{\partial \alpha}L(\tilde{g}_n + \alpha (h - 
\tilde{g}_n))\bigg\rvert_{\alpha = 1^{+}}\nonumber\\
&\geq 2\lambda_n\mathcal{E}_q(h - \tilde{g}_n) + C_2\|h - \tilde{g}_n\|^2\nonumber\\
&\geq \frac{1}{2}(C_2\wedge2)\left\{\|h - \tilde{g}_n\| + \lambda_n^{1/2}\mathcal{E}_q^{1/2}(h - \tilde{g}_n)\right\}^2 =  \frac{1}{2}(C_2\wedge2)\mathcal{R}^2(h)\nonumber\\
&= \frac{1}{2}(C_2\wedge2) a^2,
\end{align*} 
and 
\[
\frac{\partial}{\partial \alpha}\mathcal{Q}(\tilde{g}_n + \alpha (h - 
\tilde{g}_n))\bigg\rvert_{\alpha = 1^{+}} = A + B \geq - O_p\left(\frac{a^2}{c}\right) + \frac{1}{2}(C_2\wedge2)a^2> 0,
\]
with sufficiently large  $c$. Hence, it completes the proof of the theorem.
\end{proof}
\end{document}